\documentclass[reprint,amsmath,amssymb,aps,prd]{revtex4-2}

\usepackage{dcolumn}
\usepackage{bm}
\usepackage{subcaption}
\usepackage{hyperref}
\usepackage[pdftex]{graphicx}
\usepackage{amsmath, amssymb, amsthm, float, braket}
\usepackage{bbold}
\usepackage{bm}

\usepackage{tikz}
\usepackage{tikz-3dplot}
\usetikzlibrary{calc}

\DeclareMathOperator{\GL}{GL}
\DeclareMathOperator{\SL}{SL}

\begin{document}

\preprint{APS/123-QED}

\title{Baby Skyrmion crystals}
\author{Paul Leask}
\affiliation{School of Mathematics, University of Leeds, Leeds, LS2 9JT, England, UK}
\email{mmpnl@leeds.ac.uk}

\date{\today}

\begin{abstract}
This paper describes a model for baby Skyrme crystal chunks with arbitrary potential by considering energy contributions from the bulk and surface of a crystal chunk.
We focus on two potentials which yield distinct Skyrme lattices: the standard potential $V=m^2(1-\varphi^3)$ and the easy plane potential $V=\frac{1}{2}m^2 (\varphi^1)^2$.
In both models, the static energy functional is minimized over all $2$-dimensional period lattices, yielding the minimal energy crystal structure(s).
For the standard potential, the Skyrmions form a hexagonal crystal structure, whereas, for the easy plane potential, the minimal energy crystal structure is a square lattice of half-charge lumps.
We find that square crystal chunks are the global minima in the easy plane model for charges $B>6$ with $2B$ a perfect square ($m^2=1$).
In contrast, we observe that hexagonal crystal chunks in the standard model become the global minima for surprisingly large charges, $B>954$ ($m^2=0.1$).
\end{abstract}

\maketitle



\section{Introduction}

The Skyrme model \cite{Skyrme_1961} is a nonlinear field theory of pions which possesses topological solitons that describe baryons.
It has been derived as a low-energy effective field theory of quantum chromodynamics (QCD) in the large colour limit \cite{Witten_1983_1,Witten_1983_2} and, more recently, from holographic QCD models such as the Sakai-Sugimoto model \cite{Sugimoto_2005}.
One of the outstanding problems in the Skyrme model is the correct prediction of nuclear binding energies.
One would like to be able to predict correct binding energies using the Bethe--Weizs\"acker semi empirical mass formula, which is composed of five terms: the volume term, the surface term, the Coulomb term, the asymmetry term, and the pairing term.
The coefficients in each term are normally determined empirically, and the problem at hand is: can the coefficients be estimated by using Skyrmions?

In the Skyrme model, the classical mass of a Skyrmion roughly plays the same role as the volume and surface terms \cite{Ma_2019}.
To be able to address these first two terms, we need to understand the phases of nuclear matter in the Skyrme model.
An important question arises when studying phases of nuclear matter regarding the nature of high-density and low-density phases, and the transition between these phases.
At high densities the Skyrmions form a crystal, whereas at low densities the Skyrmions are localised to their corresponding lattice points.
As the ground state of nuclear matter has a crystalline structure in the classical approximation \cite{Kugler_1988}, understanding the infinite crystalline structure is key.

The baby Skyrme model \cite{Piette_1995} is a $(2+1)$-dimensional analogue of the $(3+1)$-dimensional Skyrme model, where interest in the baby Skyrme model has peaked again with the apparent prevalence of baby Skyrmions in condensed matter systems \cite{Yu_2010}, quantum hall systems \cite{Kovalev_2018,Sondhi_1993}, chiral magnetic systems \cite{Schroers_2021} and nematic liquid crystals \cite{Ackerman_2017}.
In this paper we investigate the crystalline structure for baby Skyrmions and formulate our method in terms of an arbitrary potential.
The choice of potential is crucial for baby Skyrmions as it determines the behaviour of the solitons and, thus, the underlying Skyrmion crystalline structure.
For the first time, we propose a method to determine the surface energy contribution of a crystal chunk, once the minimal energy infinite crystal structure is determined.
In order to predict the minimal energy of a charge-$B$ crystal chunk with a fixed area, we then study isoperimetric problems for particular crystal symmetries.

For the standard baby Skyrme model \cite{Piette_1995} we find that the solitons form a hexagonal crystal structure with $D_6$ symmetry, which was first proposed by Hen and Karliner \cite{Hen_2008,Hen_2009}.
This hexagonal crystal structure is not unique to the baby Skyrme model; it also arises in quantum hall systems \cite{Sondhi_1993}, chiral magnetic Skyrmion systems \cite{Rybakov_2019,Bogdanov_1994}, Ginzburg-Landau vortices \cite{Kleiner_1964} and $3$D Skyrmions in analogy with fullerene shells in carbon chemistry \cite{Battye_1998}.
However, in the easy plane model \cite{Jaykka_2010,Nitta_2013,Nitta_2014} the optimal crystal structure is found to be a square lattice of half solitons, similar to that of the conjectured cubic crystal of half Skyrmions in the Skyrme model.
This square crystal structure also arises in chiral magnets with easy plane anisotropy \cite{Batista_2015}.

This paper is laid out as follows.
We begin by discussing the general baby Skyrme model.
From here, we introduce our numerical minimisation procedure and define the initial configurations that we use to initialise our algorithm.
Then static multisoliton solutions are considered on the plane and a discussion of the possible global minima is presented.
In Sec~\ref{sec: Baby Skyrmions on a lattice} we investigate the lattice structure of baby Skyrmions and formulate a method to determine the optimal soliton crystal.
Once these minimal energy infinite crystals are known, we construct a crystal slab model to numerically determine the surface energy of a crystal chunk.
Finally, we study chunks of the infinite crystal in a bid to predict the classical energies of baby Skyrmion crystals.


\section{Baby Skyrme model}
\label{sec: Baby Skyrme model}

The general static baby Skyrme model consists of a single scalar field $\varphi: \Sigma \rightarrow S^2$ where $(\Sigma,g)$ is a $2$-dimensional Riemannian manifold, and $(S^2,h,\omega)$ is the $2$-sphere embedded in $\mathbb{R}^3$ with the induced flat Euclidean metric $h$ and area $2$-form $\omega$.
We will often write the baby Skyrme field as the $3$-vector $\varphi=(\varphi^1,\varphi^2,\varphi^3)$.
The static energy functional of this model on $\Sigma$ is given by
\begin{equation}
    E[\varphi] = \int_{\Sigma} \left\{ \frac{1}{2} |\textrm{d}\varphi|^2 + \frac{\kappa^2}{2}|\varphi^*\omega|^2 + V(\varphi) \right\} \textrm{vol}_g,
\label{eq: Baby Skyrme model - Energy functional}
\end{equation}
where $V: S^2 \rightarrow \mathbb{R}$ is the potential of the baby Skyrme model, $|\cdot|$ denotes the Hilbert-Schmidt norm and $\textrm{vol}_g$ is the volume form on $\Sigma$ associated with its metric $g$.
The parameter $\kappa$ is a standard coupling constant for which we will set $\kappa=1$ for our numerical analysis.
Note that the differential form $\textrm{d}\varphi \in \Omega^1(\Sigma)$ is a linear map $\textrm{d}\varphi_x: T_x\Sigma \rightarrow T_{\varphi(x)} S^2$ and so the Hilbert-Schmidt norm $|\textrm{d}\varphi_x|$ depends on both the domain metric $g$ and target metric $h$.
However, the pullback of the area form $\omega \in \Omega^2(S^2)$ is $\varphi^*\omega\in\Omega^2(\Sigma)$ and so its norm only depends on the metric $g$.

We will follow the terminology of harmonic map theory and refer to the first term in~\eqref{eq: Baby Skyrme model - Energy functional} as the Dirichlet energy.
The Dirichlet term is also commonly referred to as the $\sigma$-model term.
The second term is known as the Skyrme energy.
It is conventional to label the three terms in~\eqref{eq: Baby Skyrme model - Energy functional} as $E_2, E_4$, and $E_0$ respectively, where each term is thought of as a polynomial in spatial derivatives with the subscript denoting the degree.

Let us introduce oriented local coordinates $(x^1,x^2)$ on the domain $\Sigma$ and let $\{\partial_1,\partial_2\}$ be a local orthonormal basis for the tangent space $T_x\Sigma$ at $x\in\Sigma$.
Then the Dirichlet energy in local coordinates is given by \cite{Baird_Wood_2003}
\begin{align}
    E_2 & = \frac{1}{2}\int_{\Sigma} |\textrm{d}\varphi|^2 \, \textrm{vol}_g \nonumber \\
    & = \frac{1}{2} \int_{\Sigma} g^{\mu\nu}\partial_\mu\varphi^\alpha\partial_\nu\varphi^\beta h_{\alpha\beta} \, \sqrt{\det g} \, \textrm{d}x^1 \textrm{d}x^2,
\label{eq: Baby Skyrme model - Dirichlet energy}
\end{align}
where $\partial_i:=\frac{\partial}{\partial x^i}$.
The Skyrme energy in local coordinates is
\begin{align}
    E_4 & = \frac{\kappa^2}{2}\int_{\Sigma} |\varphi^*\omega|^2 \, \textrm{vol}_g \nonumber \\
    & = \frac{\kappa^2}{4} \int_{\Sigma} g^{\alpha\beta}g^{\mu\nu}(\varphi^*\omega)_{\alpha\mu}(\varphi^*\omega)_{\beta\nu} \, \sqrt{\det g} \, \textrm{d}x^1 \textrm{d}x^2.
\label{eq: Baby Skyrme model - Skyrme energy}
\end{align}
It is easy to show that the pullback of the area $2$-form $\omega$ on the $2$-sphere to $\Sigma$ is given by
\begin{equation}
    \varphi^* \omega = \varphi \cdot \left( \partial_1\varphi \times \partial_2\varphi \right) \textrm{d}x^1 \wedge \textrm{d}x^2 \in \Omega^2(\Sigma).
\label{eq: Baby Skyrme model - Area 2-form}
\end{equation}

If the domain $\Sigma$ is compact, then the baby Skyrme map $\varphi: \Sigma \rightarrow S^2$ has an associated topological degree given by the pullback of the normalised area $2$-form of the target space $S^2$,
\begin{equation}
    B[\varphi] = -\frac{1}{4\pi}\int_{\Sigma} \varphi^*\omega \in \mathbb{Z}.
\label{eq: Baby Skyrme model - Geometric charge}
\end{equation}
In terms of the local coordinates $(x^1,x^2)$ on $\Sigma$, the topological degree is explicitly
\begin{equation}
    B[\varphi] = -\frac{1}{4\pi} \int_{\Sigma} \varphi \cdot \left( \partial_1 \varphi \times \partial_2 \varphi \right) \, \textrm{d}x^1 \textrm{d}x^2.
\label{eq: Baby Skyrme model - Explicit charge}
\end{equation}
We refer to minimisers of the static energy functional $E$ for fixed degree $B$ as baby Skyrmions.
The topological degree $B$ is also referred to as the topological charge, or just charge, which we adopt throughout.
Finding baby Skyrmions involves numerically solving partial differential equations.
We do this using an accelerated gradient descent algorithm for second order dynamics, detailed in Sec~\ref{subsec: Numerical minimisation procedure}.

In order for static (multi)soliton solutions to exist in the baby Skyrme system, we must evade Derrick's non-existence Theorem.
Consider a variation $\varphi_\lambda: \Sigma \times \mathbb{R} \rightarrow S^2$ of the baby Skyrme field $\varphi$ such that $\varphi_{\lambda=0}=\varphi$.
This has infinitesimal generator $\partial_\lambda \varphi_\lambda|_{\lambda=0} \in \Gamma(\varphi^{-1}T S^2)$, where $\varphi^{-1}T S^2$ is the vector bundle  over $\Sigma$ with fibre $T_{\varphi(x)}S^2$ over $x\in\Sigma$.
Explicitly, if we consider the spatial rescaling $x \mapsto e^\lambda x$, then we have a one-parameter family of maps $\varphi_\lambda=\varphi(e^\lambda x)$ such that $\varphi_{\lambda=0}=\varphi$.
The rescaled static energy functional is then
\begin{equation}
    E_\lambda = E[\varphi_\lambda]  = E_2 + e^{2\lambda} E_4 + e^{-2\lambda} E_0.
\label{eq: Baby Skyrme model - Derricks theorem}
\end{equation}
If the baby Skyrme field configuration $\varphi$ is a minimiser of the energy $E$, then we require
\begin{equation}
    \left.\frac{d}{d\lambda}\right|_{\lambda=0} E[\varphi_\lambda] = E_4 - E_0 = 0,
\end{equation}
which yields the familiar virial constraint $E_4=E_0$.
Unlike the $(3+1)$-dimensional Skyrme model, the potential $E_0=\int_{\Sigma}V(\varphi) \textrm{vol}_g$ is necessary in the baby Skyrme model otherwise the energy $E[\varphi]$ can be lowered by spatial rescaling and thus cannot have minima.
So the baby Skyrmions have a preferred size determined by the ratio $\sqrt{\kappa/m}$.
There also exists a lower topological Bogomol'nyi bound on the (static) energy given by \cite{Speight_2010}
\begin{equation}
    E \geq \pm \left( 1 + \kappa\braket{V}\right)\int_\Sigma \varphi^* \omega = 4 \pi |B| \left( 1 + \kappa\braket{V} \right),
\end{equation}
where $\braket{V}$ is the average value of $V: S^2 \rightarrow \mathbb{R}$ on $S^2$.

In comparison to the $\sigma$-model, the addition of the Skyrme term stabilizes the $\sigma$-model to spatial rescalings.
The addition of any term that is cubic, or more, in spatial derivatives would stabilise the model (for example, the $O(3)$ $\sigma$-model coupled to a massive vector meson \cite{Foster_2009}), however the Skyrme term is the lowest order expression that retains the second order nature of the equations of motion in terms of time derivatives.

Throughout this paper, there are three choices of the physical space $\Sigma$ that we will consider.
The first physical space we will consider is the plane $\Sigma=\mathbb{R}^2$.
For the solitons to have finite energy, it is necessary to impose the boundary conditions
\begin{equation}
    \lim_{|x| \rightarrow \infty} \varphi(x) \equiv \varphi_\infty = \textrm{constant},
\label{eq: Baby Skyrme model - Finite energy BCs}
\end{equation}
and select $\varphi_\infty$ from the vacuum manifold of the model, i.e. such that $V[\varphi_\infty] = 0$.
Without loss of generality, we choose the vacuum $\varphi_\infty=(0,0,1)$ throughout.
This gives us the one-point compactification of space $\mathbb{R}^2 \cup \{\infty\} \cong S^2$.
The baby Skyrme field can then be viewed as the map $\varphi: S^2 \rightarrow S^2$, which has a conserved topological charge $B \in \pi_2(S^2)=\mathbb{Z}$, characterized as the winding number of the map and given explicitly by~\eqref{eq: Baby Skyrme model - Explicit charge}.

The second physical space we will consider is that of the $2$-torus $\Sigma=\mathbb{R}^2/\Lambda$, in which our field satisfies the doubly-periodic boundary conditions $\varphi(x)=\varphi(x +n_1 X^1 + n_2 X^2)$.
Here $n_1,n_2\in\mathbb{Z}$ and $X^1,X^2\in\mathbb{R}^2$ are a fundamental pair of periods that generate the lattice $\Lambda$.
The maps $\varphi: \mathbb{R}^2/\Lambda \rightarrow S^2$ have an associated integer degree, and so admit topological solitons.

Finally, the third physical space we consider is the infinite cylinder $\Sigma=S^1\times\mathbb{R}$.
This corresponds to a Dirichlet boundary condition in the $x^2$-direction, $\lim_{|x^2| \rightarrow \infty}=\varphi_\infty$, and a periodic boundary condition in the $x^1$-direction, $\varphi(x)=\varphi(x +n_1 X^1)$, where $n_1\in\mathbb{Z}$ and $X^1\in\mathbb{R}^2$ is a vector in the $x^1$-direction.
The maps $\varphi: S^1\times\mathbb{R} \rightarrow S^2$ also have a conserved integer topological degree and admit topological solitons.


\subsection{Initial configurations}
\label{subsec: Initial configurations}

To initialise the numerical minimisation procedure, the gradient descent algorithm requires an initial configuration, or approximation to the static soliton.
Consider the axially symmetric field configuration
\begin{equation}
    \varphi= \left( \sin f(r) \cos B\theta, \sin f(r) \sin B\theta, \cos f(r) \right),
\label{eq: Initial configurations - Radial ansatz}
\end{equation}
with the monotonically decreasing radial profile function $f(r)$ satisfying $f(0)=\pi$ and $f(\infty)=0$.
Equivalently, the profile function $f$ vanishes at the boundary of the grid.
Here, $r$ and $\theta$ are polar coordinates in the plane and there exists an internal phase that has been set to zero by applying the global symmetry that rotates the $\varphi^1, \varphi^2$ field components \cite{Salmi_2014}.
For our numerical minimisation procedure, the profile function is taken to be \cite{Karliner_2008}
\begin{equation}
    f(r) = \pi \exp(-r).
\label{eq: Initial configurations - Profile function}
\end{equation}

The initial field configuration is a linear superposition of static solutions; typically we use a set-up of $N$ charge-$1$ skyrmions with a favourable relative phase-shift between each other (for maximal attraction).
This is known as the attractive channel, and is dependent upon the choice of potential.
The superposition is justified because the profile function decays exponentially.
The superposition is done in the complex field formalism, i.e. where $W: S^2 \rightarrow \mathbb{C}P^1$ is the stereographic projection of the $\varphi$ field of $S^2$.
We use the profile function of a static solution (typically of topological charge one) to obtain
\begin{equation}
    W[\varphi] = \frac{\varphi^1 + i\,\varphi^2}{1 + \varphi^3}.
\label{eq: Initial configurations - Riemann sphere coordinate}
\end{equation}
Using the radial ansatz~\eqref{eq: Initial configurations - Radial ansatz}, this is
\begin{equation}
    W = \tan \left( \frac{f(r)}{2} \right) e^{iB\theta}.
\label{eq: Initial configurations - Complex coordinate}
\end{equation}
We can then assume that if the solitons are well separated in relation to their size, then we can approximate the resulting solution as $W = \sum_{i}^N W_i$, where $N$ is the total number of solitons in the system, and $B=\sum_{i}^N B_i$ is the total baryon number of the system.
In terms of the stereographic coordinate $W$, the baby Skyrme field is
\begin{equation}
    \varphi= \left( \frac{W + \bar{W}}{1 + |W|^2}, -i\frac{(W - \bar{W})}{1 + |W|^2}, \frac{1 - |W|^2}{1 + |W|^2} \right).
\label{eq: Initial configurations - Skyrme field}
\end{equation}


\subsection{Numerical minimisation procedure}
\label{subsec: Numerical minimisation procedure}

In order to find local minima of the static energy, we must numerically relax the baby Skyrme field.
The numerical methods are carried out on a $N_1 \times N_2$ grid with lattice spacings $\Delta x^1, \Delta x^2$.
The baby Skyrme energy is then discretised using a 4th order central finite-difference scheme.
This yields a discrete approximation $E_{\textrm{dis}}[\varphi]$ to the static energy functional $E[\varphi]$, which we can regard as a function $E_{\textrm{dis}}:\mathcal{C} \rightarrow \mathbb{R}$, where the discretised configuration space is the manifold $\mathcal{C}=(S^2)^{N_1\,N_2} \subset \mathbb{R}^{3\,N_1\,N_2}$ \cite{Winyard_2020,Speight_2021}.

To compute the minima of the discretised static energy, initially a gradient descent method was chosen.
However, gradient descent can be particularly slow method when the Hessian is of poor condition.
A more efficient way to is to simulate the time development using an accelerated gradient descent algorithm known as arrested Newton flow \cite{Gudnason_2020}.
The essence of the algorithm is as follows: we solve Newton's equations of motion for a particle on the discretised configuration space $\mathcal{C}$ with potential energy $E_{\textrm{dis}}$.
Explicitly, we are solving the system of 2nd order ODEs
\begin{equation}
    \ddot{\varphi} = -\frac{\delta E_{\textrm{dis}}}{\delta \varphi}[\varphi], \quad \varphi(0)=\varphi_0,
\label{eq: Numerical minimisation procedure - 2nd order ODEs}
\end{equation}
with initial velocity $\dot{\varphi}(0)=0$.
Setting $\psi:=\dot{\varphi}$ as the velocity with $\psi(0)=\dot{\varphi}(0)=0$ reduces the problem to a coupled system of 1st order ODEs.
We implement a 4th order Runge--Kutta method to solve this coupled system.

The main advantage in implementing the arrested Newton flow algorithm is that the field will naturally relax to a local minimum.
After each time step $t \mapsto t + \delta t$, we check to see if the energy is increasing.
If $E_{\textrm{dis}}(t + \delta t) > E_{\textrm{dis}}(t)$, we take out all the kinetic energy in the system by setting $\psi(t + \delta t)=0$ and restart the flow.
The flow then terminates when every component of the energy gradient $\frac{\delta E_{\textrm{dis}}}{\delta \varphi}$ is zero to within a given tolerance (we have used $10^{-4}$).
Unless stated otherwise, the plots shown throughout were simulated on a grid with $0.05$ lattice spacings and grid sizes $1000\times 1000$.

It is essential that we enforce the constraint $\varphi\cdot\varphi=1$.
This is normally done by including a Lagrange multiplier term into the Lagrangian, and the form for the Lagrange multiplier can be found taking the dot product of the field with the resulting Euler-Lagrange equations.
However, to do this numerically we have to pull our target space back onto $S^2$.
This is done by normalizing the Skyrme field $\varphi$ each loop,
\begin{equation}
    \varphi^a \rightarrow \frac{\varphi^a}{\sqrt{\varphi\cdot\varphi}}.
\label{eq: Numerical minimisation procedure - Field normalisation}
\end{equation}
We also need to project out the component of the energy gradient, and velocity, in the direction of Skyrme field, that is
\begin{equation}
    \frac{\delta \mathcal{E}}{\delta \varphi^a} \rightarrow \frac{\delta \mathcal{E}}{\delta \varphi^a} - \left( \frac{\delta \mathcal{E}}{\delta \varphi} \cdot \varphi \right) \frac{\varphi^a}{\sqrt{\varphi\cdot\varphi}}
\label{eq: Numerical minimisation procedure - Energy normalisation}
\end{equation}
and
\begin{equation}
    \psi^a \rightarrow \psi^a - \left( \psi\cdot \varphi \right) \frac{\varphi^a}{\sqrt{\varphi\cdot\varphi}}.
\label{eq: Numerical minimisation procedure - Velocity normalisation}
\end{equation}


\section{Baby Skyrmions on $\mathbb{R}^2$}
\label{sec: Baby Skyrmions on R2}

Consider the plane $\mathbb{R}^2$ with the usual flat Euclidean metric $g_{ij}=\delta_{ij}$.
The static energy functional of the baby Skyrme model on $\mathbb{R}^2$ takes the familiar form
\begin{equation}
    E[\varphi] = \int_{\mathbb{R}^2} \left\{ \frac{1}{2} (\partial_i \varphi)^2 + \frac{\kappa^2}{4} \left( \partial_i \varphi \times \partial_j \varphi \right)^2+ V(\varphi) \right\} \textrm{d}^2x.
\label{eq: Baby Skyrmions on R2 - Energy functional}
\end{equation}
For numerical analysis, it proves convenient to express the static energy functional using Einstein's summation notation, that is
\begin{eqnarray}
    E[\varphi] = \int_{\mathbb{R}^2} \left\{ \frac{1}{2}\left( \partial_i \varphi^a \right)^2 + \frac{\kappa^2}{4} \left( (\partial_i\varphi^a \, \partial_j\varphi^b)^2 \right. \right. \nonumber \\ \left. \left. - \partial_i\varphi^a \, \partial_j\varphi^a \, \partial_j\varphi^b \, \partial_i\varphi^b \right) + V(\varphi)\right\} \, \textrm{d}^2x,
\label{eq: Baby Skyrmions on R2 - Index energy functional}
\end{eqnarray}
where $i,j\in\{1,2\}$ and $a,b\in\{1,2,3\}$.
The energy functional has a continuous $O(3)$ symmetry before the symmetry is broken by the choice of potential term $V(\varphi)$.
To carry out arrested Newton flow, or a numerical relaxation method using the gradient of the energy, we need to calculate the energy gradient explicitly.
We will do this in index notation for numerical convenience.
The variation of the energy density with respect to field $\varphi^a$ is
\begin{eqnarray*}
    \frac{\delta \mathcal{E}}{\delta \varphi^a} = \frac{\delta V}{\delta \varphi^a} -\left\{ \partial_{ii}\varphi^a + \kappa^2 \left[ \partial_{ii}\varphi^a \left( \partial_j \varphi^b \right)^2 \right. \right. \\
        \left. \left. + \partial_i\varphi^a \left( \partial_{ij}\varphi^b\,\partial_j\varphi^b - \partial_{jj}\varphi^b\,\partial_i\varphi^b \right)  - \partial_{ij}\varphi^a \left( \partial_i\varphi^b\,\partial_j\varphi^b \right) \right] \right\}.
\label{eq: Baby Skyrmions on R2 - Index energy gradient}
\end{eqnarray*}


\subsection{Standard baby Skyrmions}
\label{subsec: Standard baby Skyrmions}

\begin{figure}[t]
	\centering
	\begin{subfigure}[b]{0.20\textwidth}
	\includegraphics[width=\textwidth]{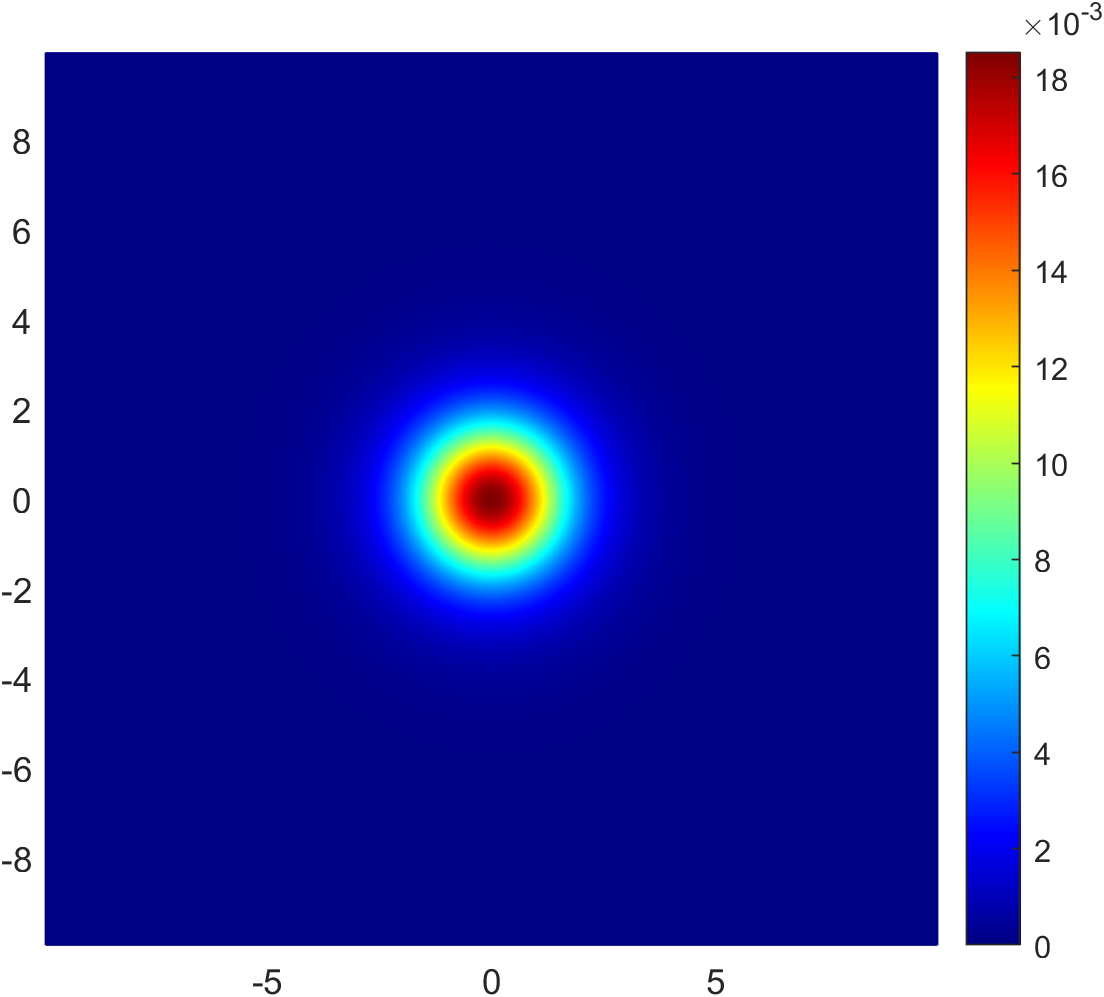}
	\caption{Standard charge-$1$ energy density.}
	\label{fig: Baby Skyrmions on R2 - EA Charge-1 (a)}
	\end{subfigure}
    ~
	\begin{subfigure}[b]{0.20\textwidth}
	\includegraphics[width=\textwidth]{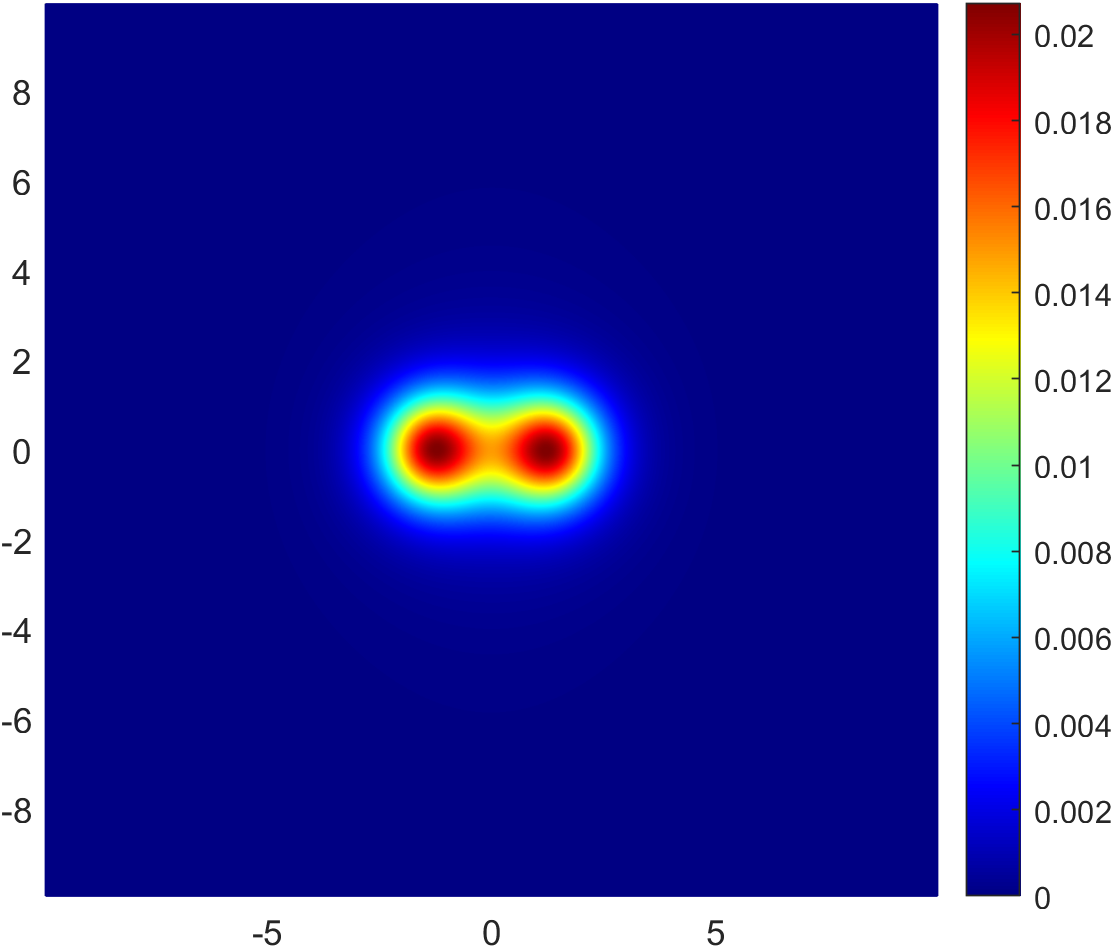}
	\caption{Easy plane charge-$1$ energy density.}
	\label{fig: Baby Skyrmions on R2 - EP Charge-1 (b)}
	\end{subfigure}
	\caption{Plots of the energy density of (a) the axially symmetric charge-$1$ baby Skyrmion, for the standard potential $V(\varphi)=m^2(1-\varphi^3)$, and (b) the charge-$1$ baby Skyrmion for the easy plane potential $V(\varphi)=\frac{1}{2}m^2 (\varphi^1)^2$.}
	\label{fig: Baby Skyrmions on R2 - EA & EP Charge-1}
\end{figure}
Numerous potentials have been proposed \cite{Leese_1990,Jaykka_2010,Jaykka_2012,Winyard_2013,Piette_1995,Salmi_2014,Ward_2004,Weidig_1999} and studied extensively in the literature.
However, there are two choices of potential that we are particularly interested in: the standard potential and the easy plane potential.
These two theories are quite distinct and, as we will describe below, we should expect different phenomena.
In the standard baby Skyrme model \cite{Piette_1995}, the standard potential is an analogue of the pion mass term in the Skyrme model, and takes the form
\begin{equation}
    V=m^2(1-\varphi^3).
\label{eq: Baby Skyrmions on R2 - Standard potential}
\end{equation}
If we consider excitations around our unique choice of vacuum $\varphi_\infty=(0,0,1)$, then the fields $\varphi^1$ and $\varphi^2$ acquire a mass $m$.
The standard potential~\eqref{eq: Baby Skyrmions on R2 - Standard potential} spontaneously breaks the $O(3)$ symmetry to an $O(2)$ symmetry that acts on the field components $\varphi^1, \varphi^2$.
For this potential, the charge-$1$ baby Skyrmion is axially symmetric and exponentially localised, see Fig.~\ref{fig: Baby Skyrmions on R2 - EA Charge-1 (a)}.
Piette \textit{et al.} \cite{Piette_1995} studied the asymptotic interactions of standard baby Skyrmions and found that two well-separated charge-$1$ solitons have an interaction energy that can be calculated using a dipole approximation, such that
\begin{equation}
    E_{\textrm{standard}} \propto \cos(\chi_1-\chi_2),
\label{eq: Baby Skyrmions on R2 - EA Interaction}
\end{equation}
where $\chi_1-\chi_2$ is the relative phase.
The attraction between these two well-separated baby Skyrmions is greatest when $\chi_1-\chi_2=\pi$.
This is known as the attractive channel.

\begin{figure}[t]
	\centering
	\begin{subfigure}[b]{0.20\textwidth}
	\includegraphics[width=\textwidth]{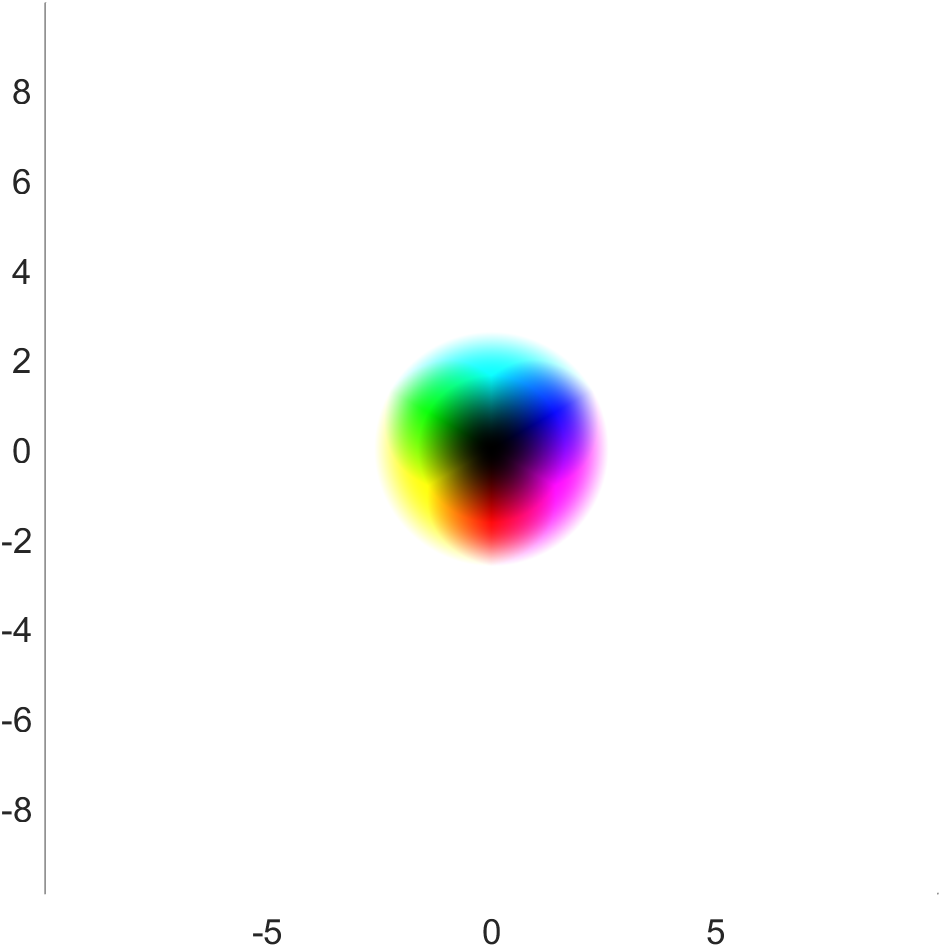}
	\caption{Standard charge-$1$ coloring.}
	\label{fig: Baby Skyrmions on R2 - Coloring scheme (a)}
	\end{subfigure}
    ~
	\begin{subfigure}[b]{0.20\textwidth}
	\includegraphics[width=\textwidth]{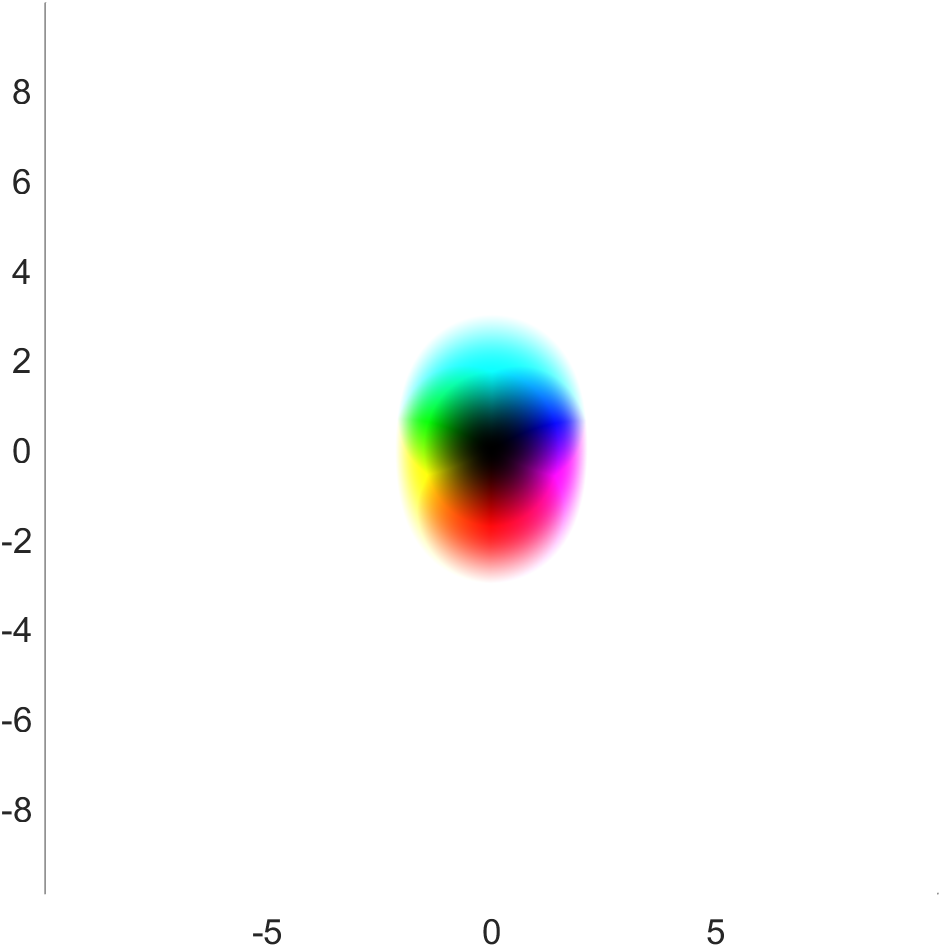}
	\caption{Easy plane charge-$1$ coloring.}
	\label{fig: Baby Skyrmions on R2 - Coloring scheme (b)}
	\end{subfigure}
	\caption{Plots of the coloring scheme detailed in the text for (a) the axially symmetric charge-$1$ baby Skyrmion, for the standard potential, and (b) the charge-$1$ baby Skyrmion for the easy plane potential.}
	\label{fig: Baby Skyrmions on R2 - Coloring scheme}
\end{figure}

There is a rather nice way to graphically represent the phase of a baby Skyrmion \cite{Halcrow_2020} using a HSV color model, which is almost analogous to the Runge color sphere coloring in the Skyrme model.
We begin by plotting the energy density and color it using the stereographic coordinate $W$, given in~\eqref{eq: Initial configurations - Riemann sphere coordinate}.
The phase of $W$, $\arg(\phi^1 + i \phi^2)$, gives the hue of the color and is defined such that $\arg(\phi^1 + i \phi^2)=0$ is red, $\arg(\phi^1 + i \phi^2)=2\pi/3$ is green and $\arg(\phi^1 + i \phi^2)=4\pi/3$ is blue.
We use the value of $\phi^3$ to determine the lightness, such that $\phi^3=+1$ is white and $\phi^3=-1$ is black \cite{Barsanti_2020}.
The coloring scheme detailed above is shown in Fig.~\ref{fig: Baby Skyrmions on R2 - Coloring scheme}.

\begin{figure}[t]
	\centering
	\begin{subfigure}[b]{0.2\textwidth}
	\includegraphics[width=\textwidth]{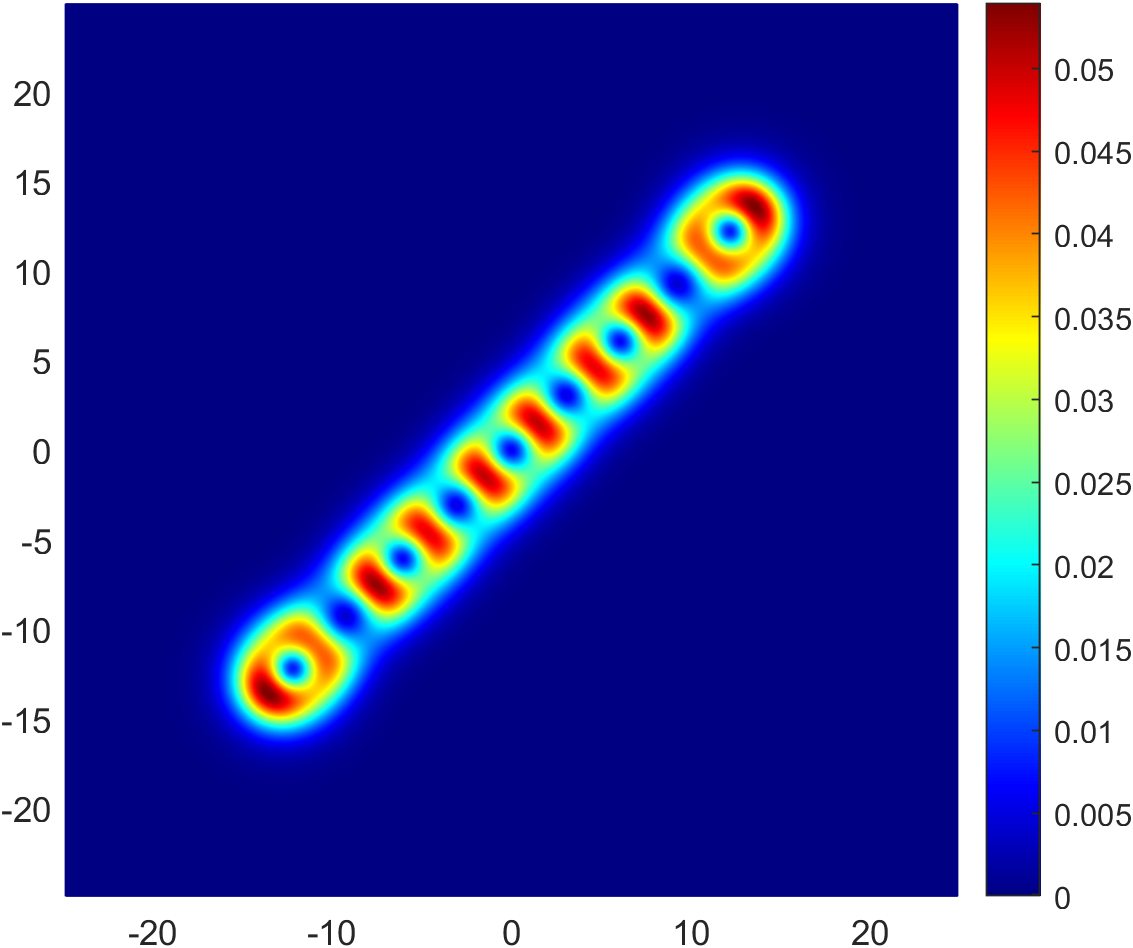}
	\caption{}
	\label{fig: Baby Skyrmions on R2 - Chain9 (a)}
	\end{subfigure}
    ~
    \begin{subfigure}[b]{0.17\textwidth}
	\includegraphics[width=\textwidth]{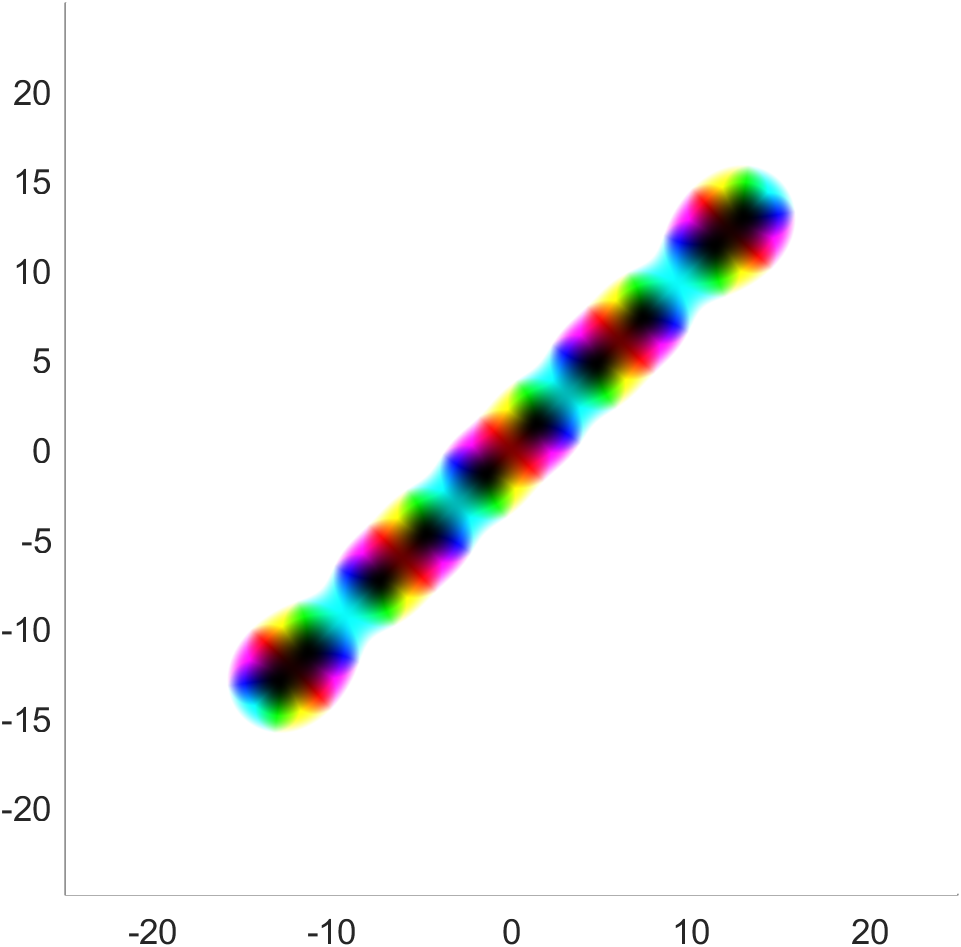}
	\caption{}
	\label{fig: Baby Skyrmions on R2 - Chain9 (d)}
	\end{subfigure} \\
	\begin{subfigure}[b]{0.2\textwidth}
	\includegraphics[width=\textwidth]{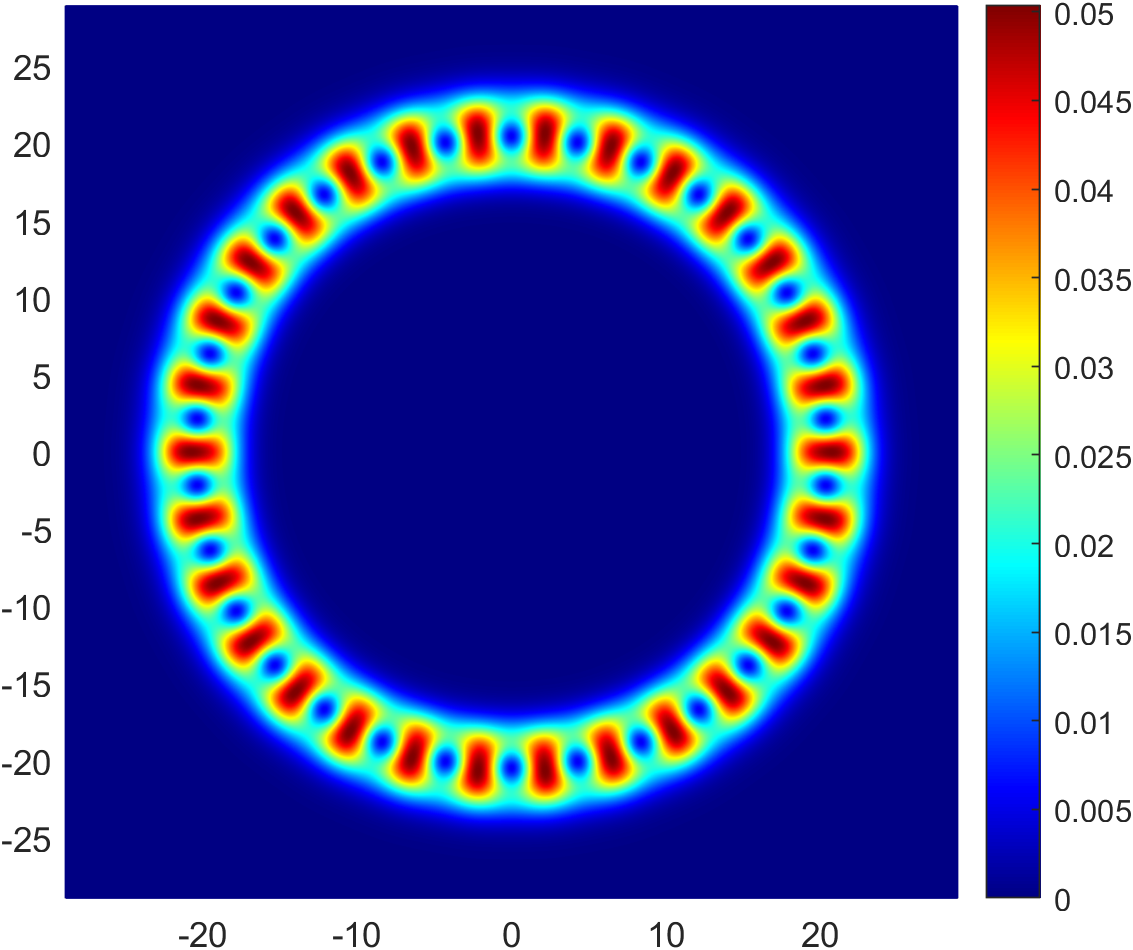}
	\caption{}
	\label{fig: Baby Skyrmions on R2 - Ring30 (b)}
	\end{subfigure}
    ~
	\begin{subfigure}[b]{0.17\textwidth}
	\includegraphics[width=\textwidth]{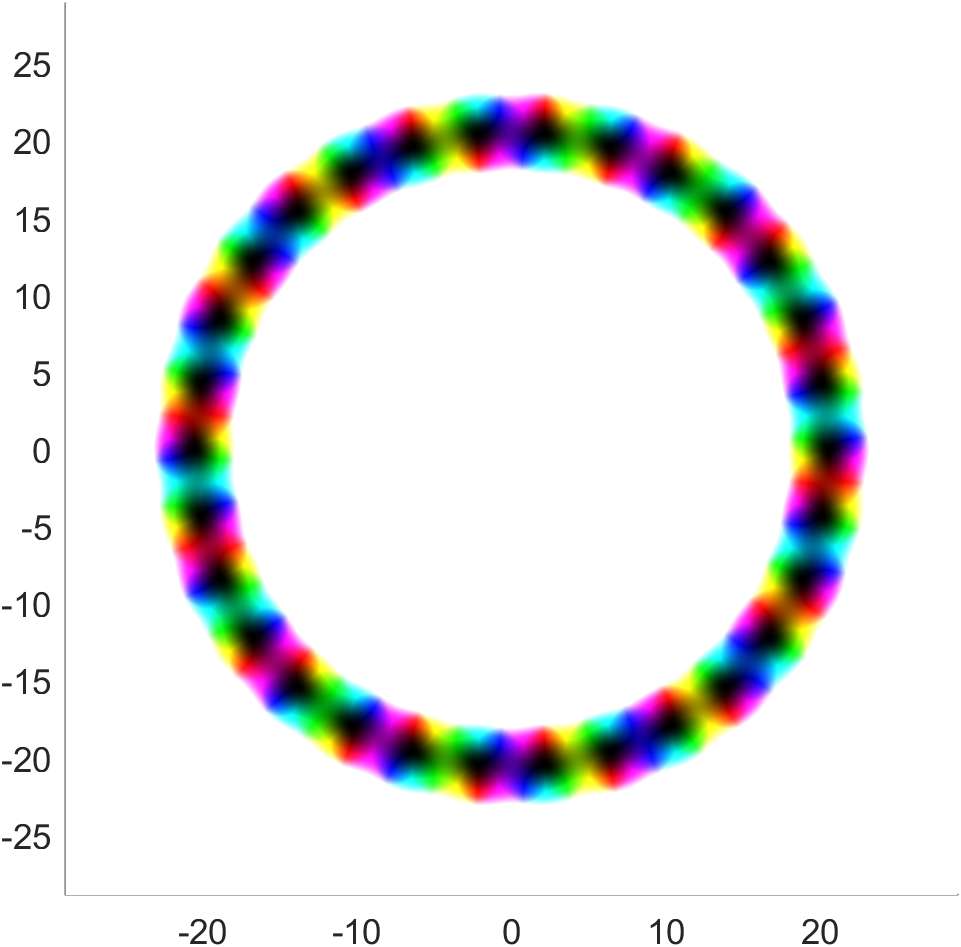}
	\caption{}
	\label{fig: Baby Skyrmions on R2 - Ring30 (d)}
	\end{subfigure} \\
	\begin{subfigure}[b]{0.1\textwidth}
	\includegraphics[width=\textwidth]{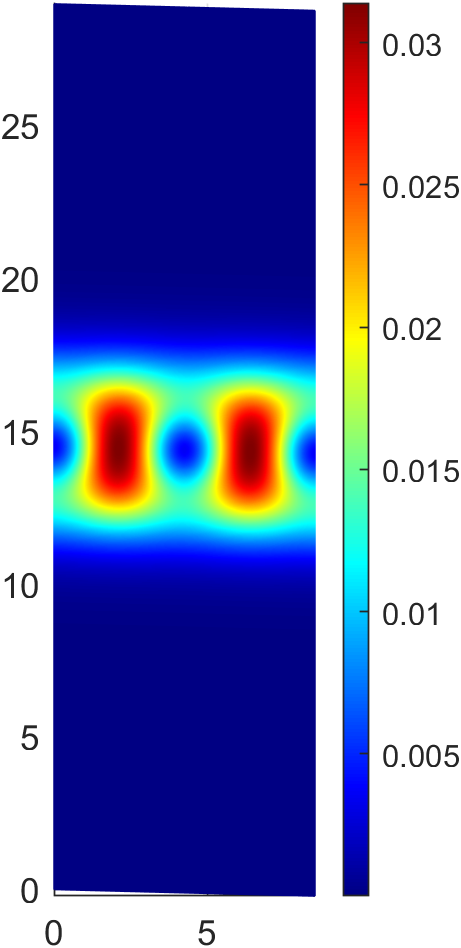}
	\caption{}
	\label{fig: Baby Skyrmions on R2 - Infinite chain (e)}
	\end{subfigure}
	~
	\begin{subfigure}[b]{0.07\textwidth}
	\includegraphics[width=\textwidth]{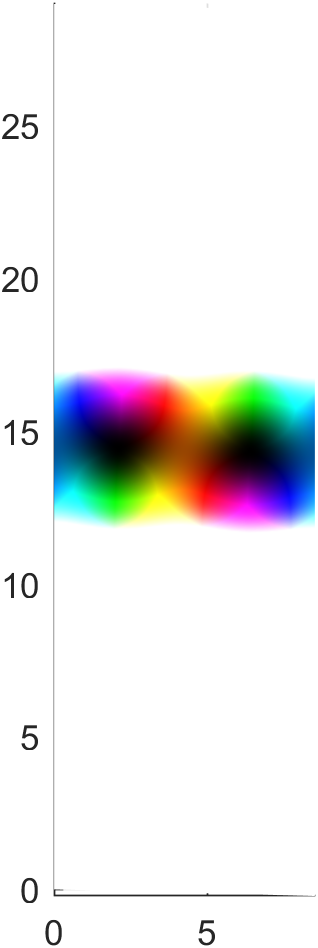}
	\caption{}
	\label{fig: Baby Skyrmions on R2 - Infinite chain (f)}
	\end{subfigure}
	\caption{Energy density plots of (a) the charge-$10$ chain solution, (c) the charge-$30$ ring solution and (e) the infinite chain.
	On the right hand side are the corresponding plots using the color scheme detailed in the text.}
	\label{fig: Baby Skyrmions on R2 - Chain and ring}
\end{figure}

For both potentials, multi-charged baby Skyrmion solutions have an underlying modular structure.
One such structure of interest for the standard potential~\eqref{eq: Baby Skyrmions on R2 - Standard potential} is that of chains of solitons.
This was first investigated  by Harland \cite{Harland_2008}, in the context of baby Skyrmions, and then later by Foster \cite{Foster_2010} and also Shnir \cite{Shnir_2021}.
Each chain has its ends capped by charge-$2$ solutions and the chain links are built from either charge-$1$ baby Skyrmions, with a relative phase of $\pi$ with each neighbour, or charge-$2$ solitons.
Shnir \cite{Shnir_2021} showed that a chain with charge-$1$ links has a lower energy than a chain with charge-$2$ links within each homotopy class.
For low charge solutions, chains appear to be good candidates for the global minima.
A typical chain configuration for the standard potential~\eqref{eq: Baby Skyrmions on R2 - Standard potential} is displayed in Fig.~\ref{fig: Baby Skyrmions on R2 - Chain9 (a)}.

Foster \cite{Foster_2010} also investigated baby Skyrmions on a cylinder $\mathbb{R} \times S^1$, and calculated the minimum energy per charge of an infinite chain to be $\mathcal{E}_{\textrm{chain}}=1.4549$.
We have carried out the same calculation using the lattice variation method detailed in Sec~\ref{sec: Baby Skyrmions on a lattice}.
This is done by imposing a Dirichlet boundary condition in the $x^2$-direction, $\lim_{|x^2| \rightarrow \infty}=\varphi_\infty$, and a periodic boundary condition in the $x^1$-direction, $\varphi(x^1,x^2)=\varphi(x^1+n_1 L,x^2)$, where $n_1\in\mathbb{Z}$.
The periodic cell length $L$ is then varied to minimise the energy, and a minimum energy of $\mathcal{E}_{\textrm{chain}}=1.4548$ was found for a periodic cell length of $L=8.53$.
Thus, our results provide excellent fidelity and are displayed in Fig.~\ref{fig: Baby Skyrmions on R2 - Infinite chain (e)}.

Later, it was realised by Winyard \cite{winyard_2016} that the energy density peaks at the ends of the chains could be reduced by joining the two ends into a ring-like solution with an added energy correction for the curvature of the ring.
Using a least squares fitting, they were able to obtain values for the energy contributions from the chain caps and the ring curvature.
They showed that ring solutions are a better candidate for the global minima for $B>B_{\textrm{ring}}\in\mathbb{Z}$.
For the mass $m^2=0.1$, this transition from chains to rings is numerically found to occur at $B_{\textrm{ring}}=15$.
A typical ring configuration is displayed in Fig.~\ref{fig: Baby Skyrmions on R2 - Ring30 (b)}.

Soliton crystals in the standard baby Skyrme model, with the standard potential~\eqref{eq: Baby Skyrmions on R2 - Standard potential}, were studied by Hen and Karliner \cite{Hen_2008,Hen_2009}.
Through their work they observed that the minimal energy soliton crystal was almost hexagonal by use of simulated annealing.
So one would expect chunks of the infinite hexagonal crystal to be global minima for some $B>B_{\textrm{crystal}}\in\mathbb{Z}$.
This prompts the basis of this paper: at what charge do chunks of the infinite soliton crystal become the global minima?


\subsection{Easy plane baby Skyrmions}
\label{subsec: Easy plane baby Skyrmions}

The second potential of particular interest is the easy plane potential,
\begin{equation}
    V(\varphi)=\frac{1}{2}m^2 (\varphi^1)^2,
\label{eq: Baby Skyrmions on R2 - Easy plane potential}
\end{equation}
proposed by J\"aykk\"a and Speight \cite{Jaykka_2010}.
As with the standard potential, the easy plane potential leaves an unbroken $O(2)$ symmetry.
However, the canonical choice of vacuum $\varphi_\infty=(0,0,1)$ distinguishes a point on the $O(2)$ orbit and breaks the symmetry further to a discrete $D_2$ symmetry.
Unlike the standard model, the charge-$1$ baby Skyrmion is not axially symmetric but rather is composed of two charge-$1/2$ baby Skyrmions.
This is shown in Fig.~\ref{fig: Baby Skyrmions on R2 - EP Charge-1 (b)}.
As we did before, let us consider elementary excitations around our canonical choice of vacuum $\varphi_\infty=(0,0,1)$, then the field $\varphi^1$ acquires a mass $m$ and the $\varphi^2$ field is massless.
Adapting the dipole approximation proposed by Piette \textit{et al.} \cite{Piette_1995}, and assuming that $\varphi^2$ mediates the dominant interaction asymptotically \cite{Jaykka_2010}, gives us an interaction energy
\begin{equation}
    E_{\textrm{easy-plane}} \propto \cos(\chi_1+\chi_2).
\label{eq: Baby Skyrmions on R2 - EP Interaction}
\end{equation}
This shows that the interaction energy depends only on the average phase of the dipoles, which is exactly opposite of the situation in the standard model.

While the phase coloring is particularly useful for the standard potential, it is actually more instructive to use the field structure of the field component $\varphi^1$ for the easy plane model.
The red peaks and blues peaks are attracted to one another, but peaks of the same color are repelled by each other.
As each individual peak in the energy density resembles a $\mathbb{C}P^1$ model lump, then we refer to each lump as a half charge lump.
Each of these half lumps are located at the red and blue peaks of $\varphi^1$ and come in pairs.
So more information can be gained by studying plots of the $\varphi^1$ density than the energy density itself.

\begin{figure}[tbp]
	\centering
	\begin{subfigure}[b]{0.2\textwidth}
	\includegraphics[width=\textwidth]{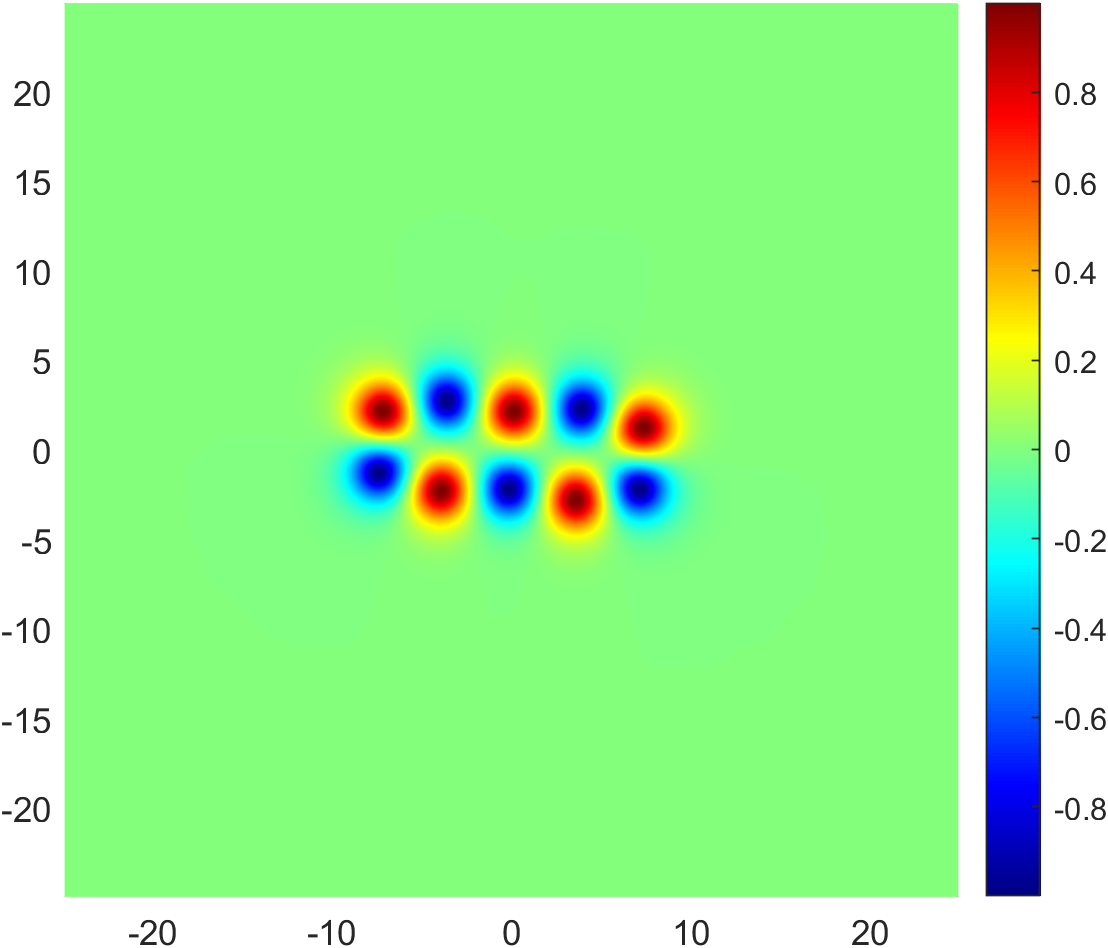}
	\caption{$B=5$.}
	\label{fig: Baby Skyrmions on R2 - EP B=5}
	\end{subfigure}
	~
	\begin{subfigure}[b]{0.2\textwidth}
	\includegraphics[width=\textwidth]{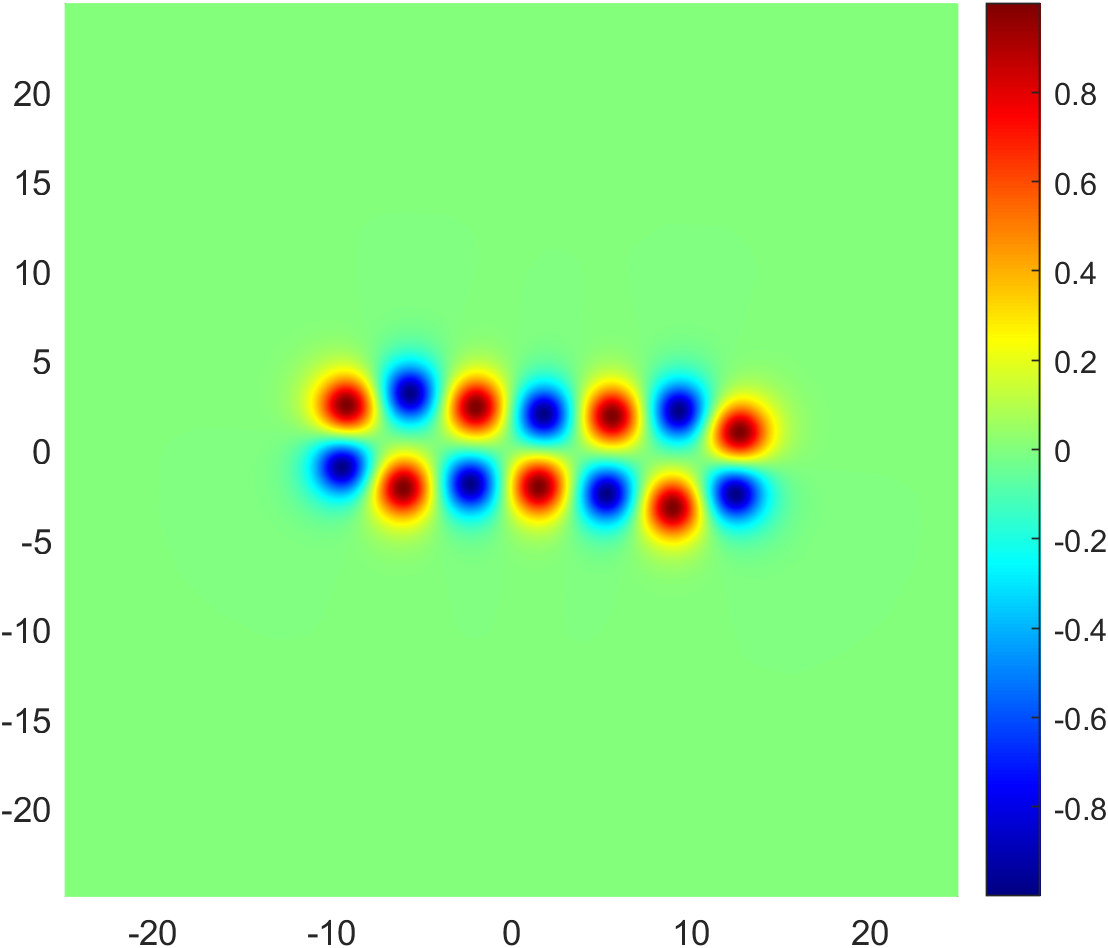}
	\caption{$B=7^*$.}
	\label{fig: Baby Skyrmions on R2 - EP B=7}
	\end{subfigure} \\
	\begin{subfigure}[b]{0.2\textwidth}
	\includegraphics[width=\textwidth]{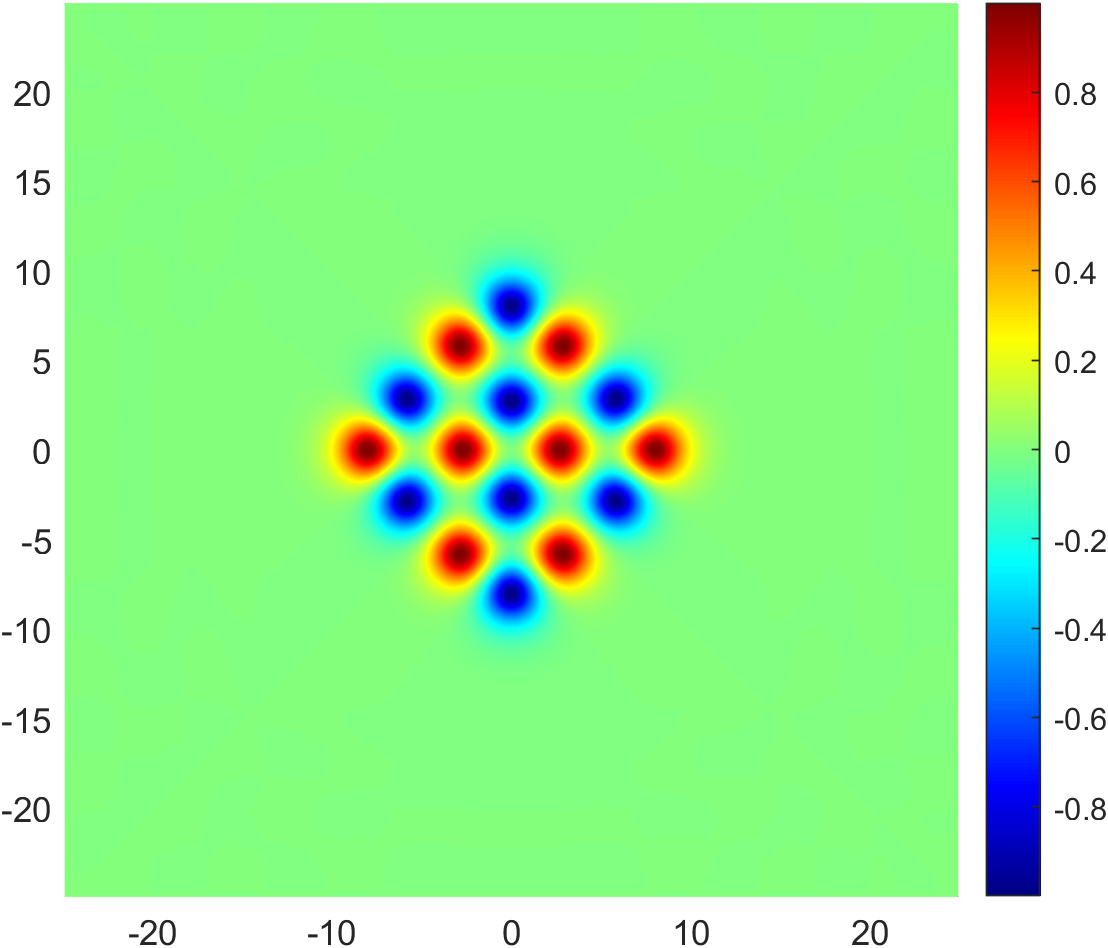}
	\caption{$B=8^*$.}
	\label{fig: Baby Skyrmions on R2 - EP B=8}
	\end{subfigure}
	~
	\begin{subfigure}[b]{0.2\textwidth}
	\includegraphics[width=\textwidth]{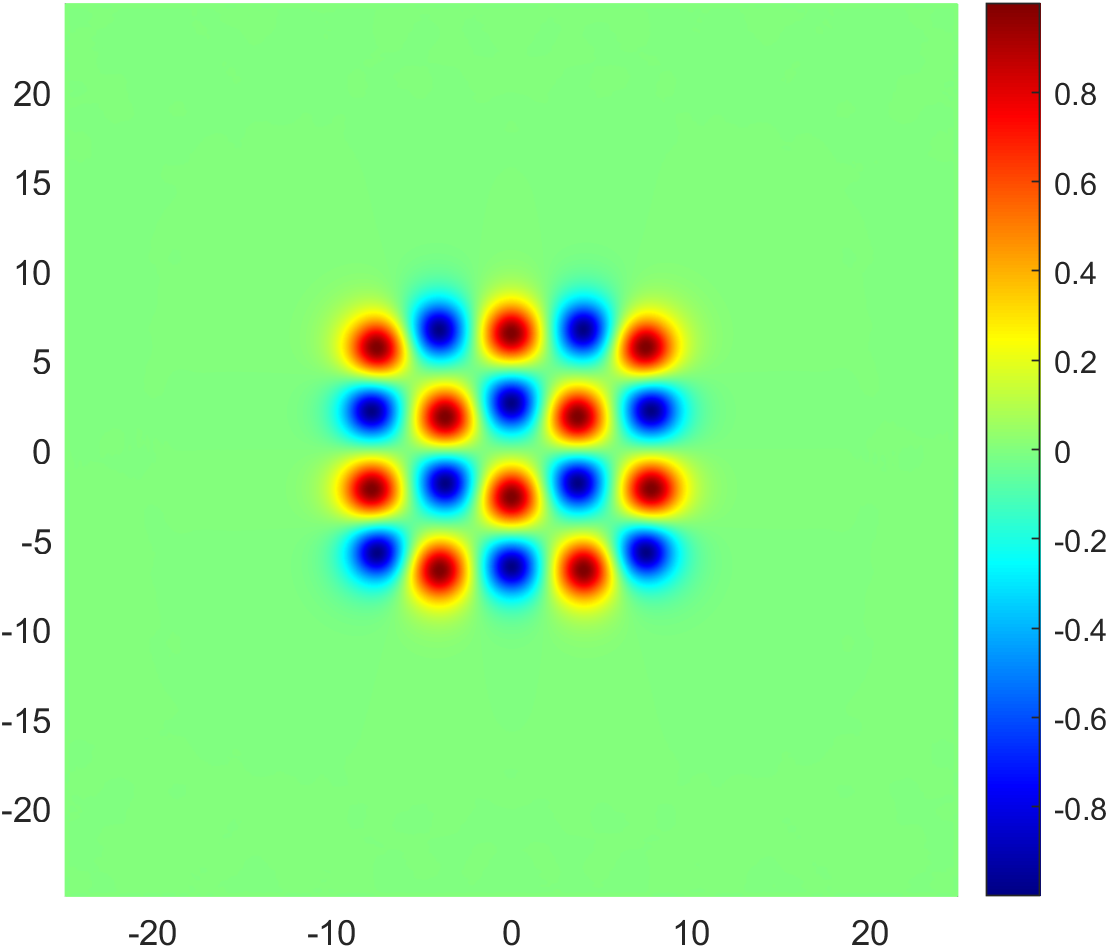}
	\caption{$B=10^*$.}
	\label{fig: Baby Skyrmions on R2 - EP B=10}
	\end{subfigure} \\
	\begin{subfigure}[b]{0.2\textwidth}
	\includegraphics[width=\textwidth]{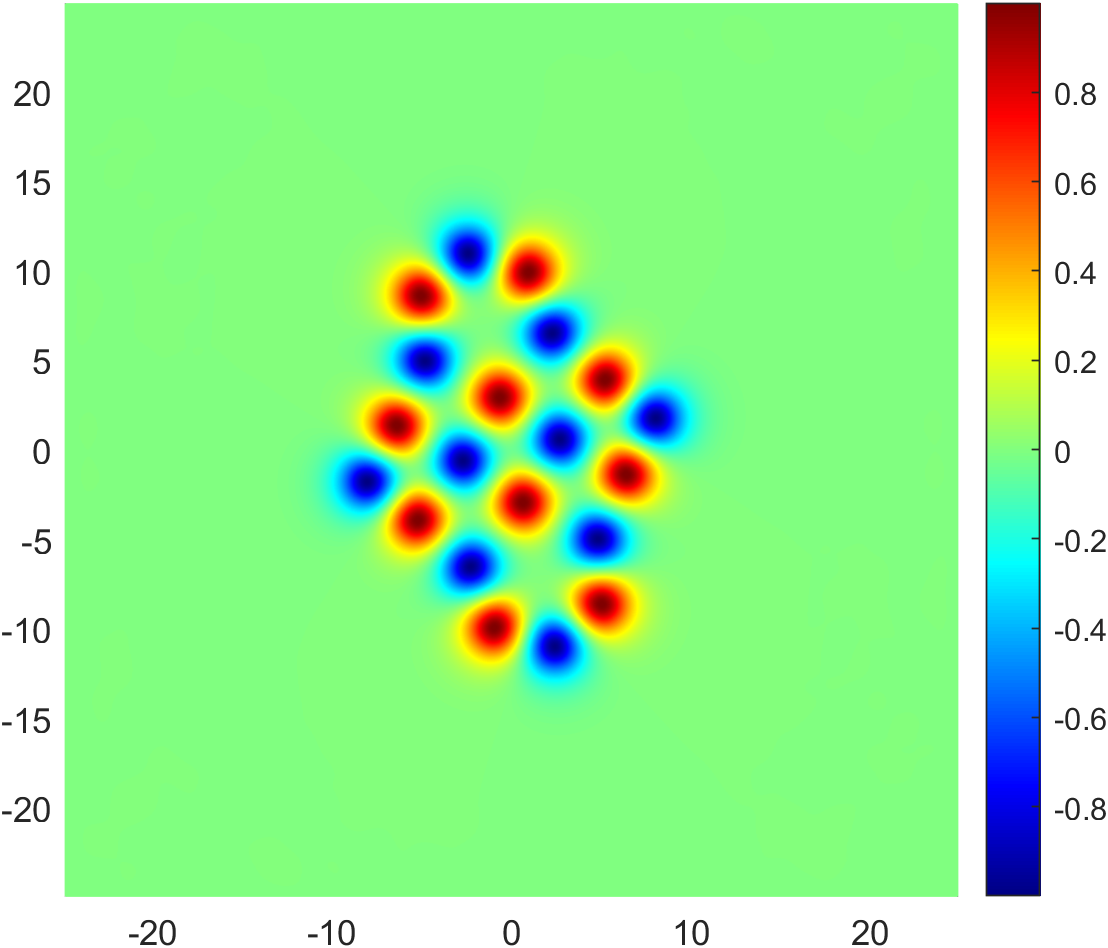}
	\caption{$B=10$.}
	\label{fig: Baby Skyrmions on R2 - EP B=10 (2)}
	\end{subfigure}
	~
	\begin{subfigure}[b]{0.2\textwidth}
	\includegraphics[width=\textwidth]{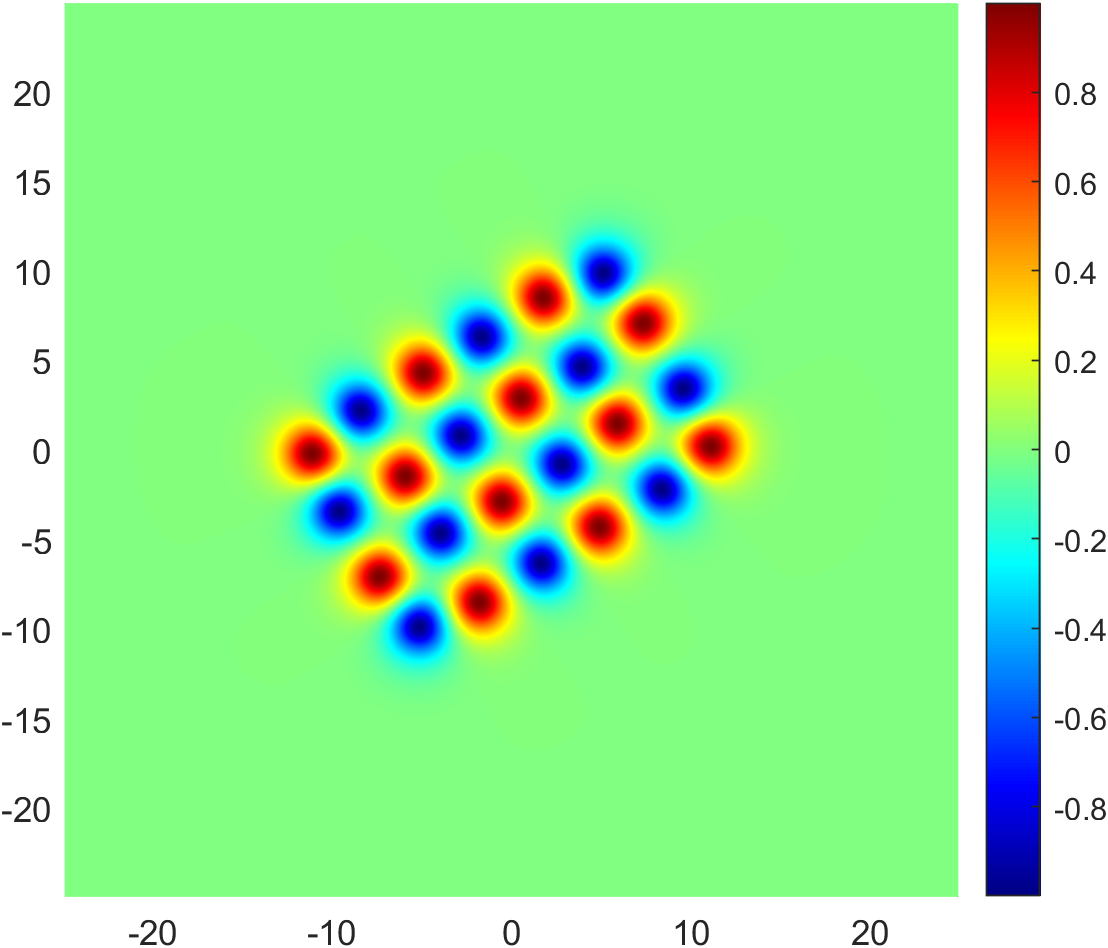}
	\caption{$B=12^*$.}
	\label{fig: Baby Skyrmions on R2 - EP B=12}
	\end{subfigure} \\
	\begin{subfigure}[b]{0.2\textwidth}
	\includegraphics[width=\textwidth]{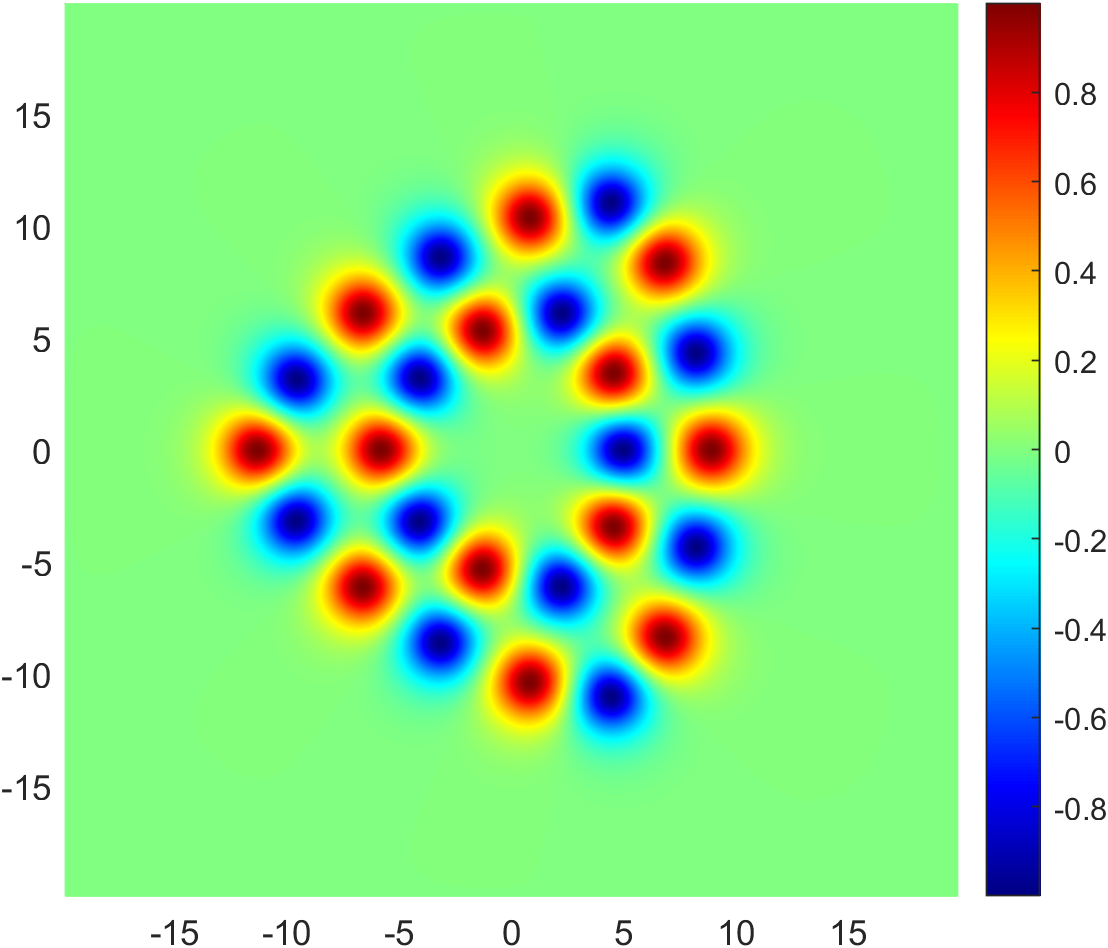}
	\caption{$B=13$.}
	\label{fig: Baby Skyrmions on R2 - EP B=13}
	\end{subfigure}
	~
	\begin{subfigure}[b]{0.2\textwidth}
	\includegraphics[width=\textwidth]{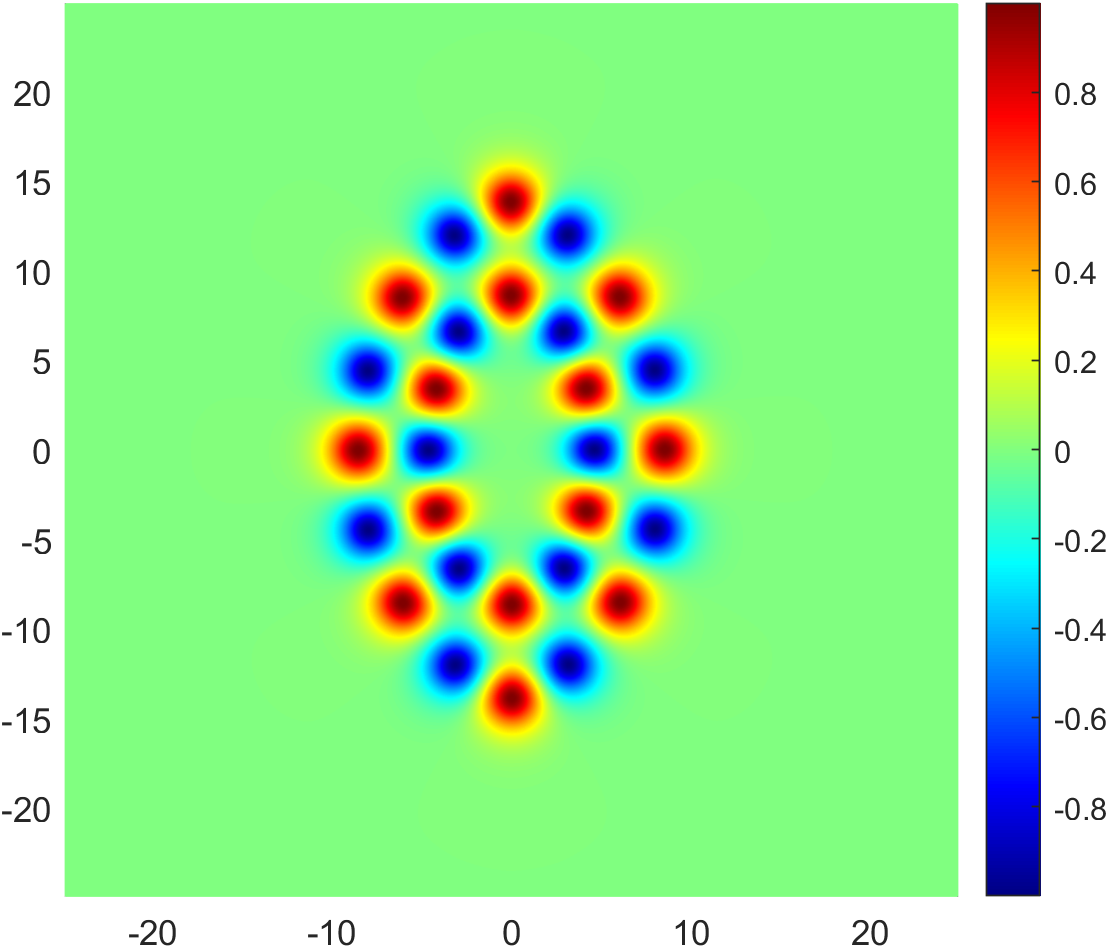}
	\caption{$B=14$.}
	\label{fig: Baby Skyrmions on R2 - EP B=14 (3)}
	\end{subfigure} \\
	\begin{subfigure}[b]{0.2\textwidth}
	\includegraphics[width=\textwidth]{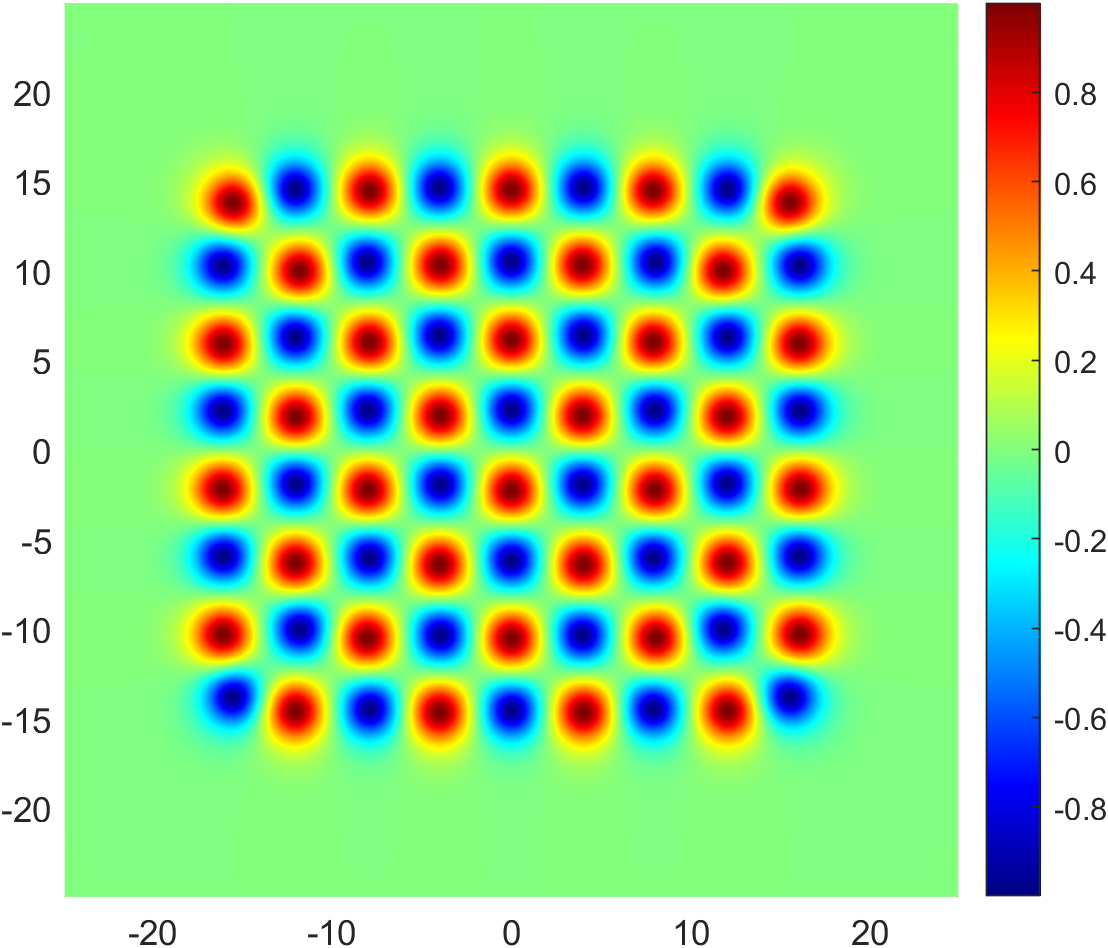}
	\caption{$B=36^*$.}
	\label{fig: Baby Skyrmions on R2 - EP B=36}
	\end{subfigure}
	~ 
	\begin{subfigure}[b]{0.2\textwidth}
	\includegraphics[width=\textwidth]{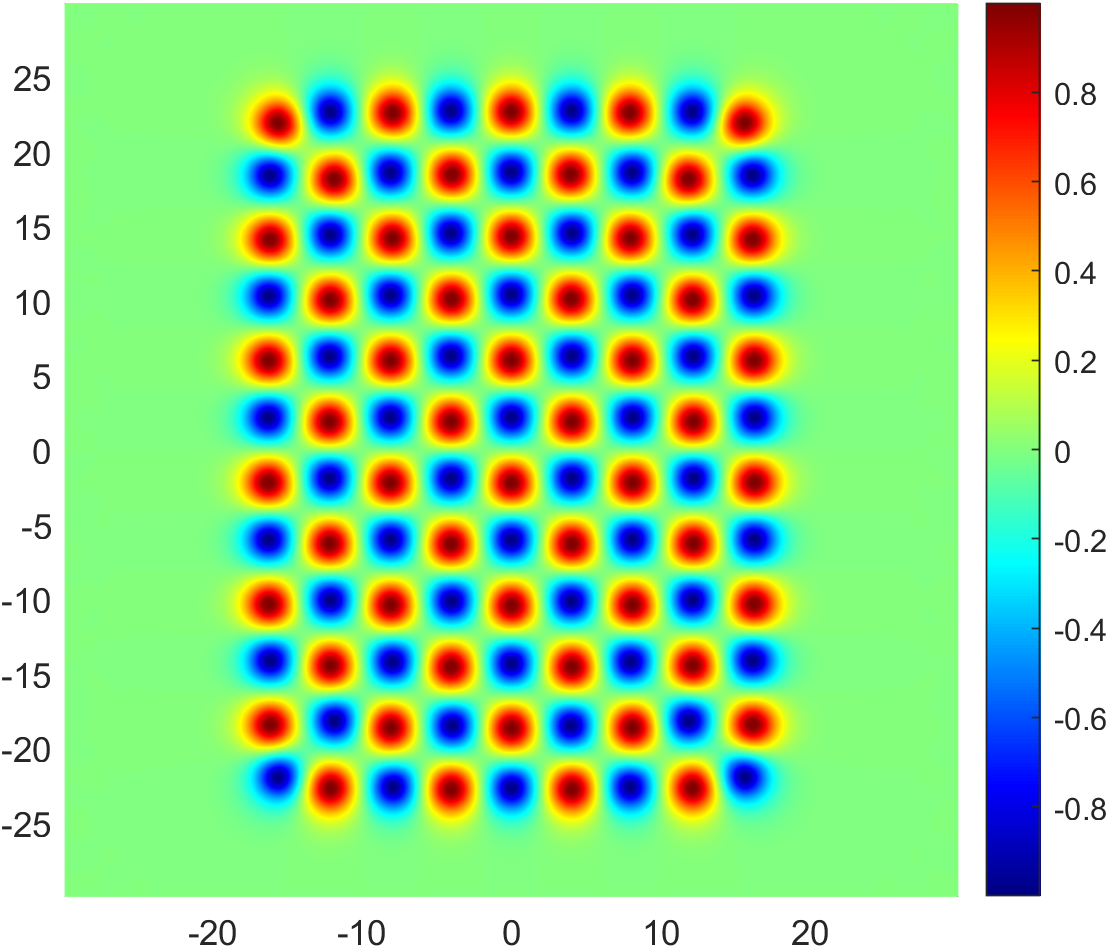}
	\caption{$B=54^*$.}
	\label{fig: Baby Skyrmions on R2 - EP B=54}
	\end{subfigure}
	\caption{$\varphi^1$ density plots of various local and global (*) energy minimisers.}
	\label{fig: Baby Skyrmions on R2 - EP solutions}
\end{figure}

In contrast to the standard model, chains do not appear to be the global minima for low charges in the easy plane model.
For charges $B\leq6$ with mass $m^2=1$, the global minima are $2B$-gons or ring-like solutions.
Chunks of an infinite crystal with a square/rectangular crystalline structure seem to be the global minima for \textit{almost} all charges $B>6$.
An example of such a global minimum for $B=8$ can be seen in Fig.~\ref{fig: Baby Skyrmions on R2 - EP B=8}.
The easy plane model also exhibits a modular structure with some more exotic local minima consisting of square and polygonal building blocks.
One such solution is the $B=10$ easy plane baby Skyrmion built from square and hexagonal units in Fig.~\ref{fig: Baby Skyrmions on R2 - EP B=10 (2)}.

Although chunks of the assumed infinite crystal are prevalent, ring-like solutions and chain solutions do exist as other local minima.
J\"aykk\"a and Speight \cite{Jaykka_2010} showed that $2B$-gon rings are the global minima for low charges, that is a single ring of $2B$ half lumps.
For higher charges, it is energetically favourable for the ring solutions to form a double ring structure with some discrete symmetry.
Example solutions for the easy plane model are shown in Fig.~\ref{fig: Baby Skyrmions on R2 - EP solutions}.
The charge-$5$ chain in Fig.~\ref{fig: Baby Skyrmions on R2 - EP B=5} is coincidentally a chunk of the assumed infinite soliton crystal and is only a local minimiser for $B=5$, rather a $10$-gon of half lumps is the global minimiser.
The particularly interesting aspect of the charge-$5$ chain is that its shape is closer to that of a square soliton crystal chain than that of a double hexagon.
This suggests that the square soliton crystal is a lower energy crystalline structure than the hexagonal soliton crystal.


\section{Lattice structure of baby Skyrmions}
\label{sec: Baby Skyrmions on a lattice}

In a series of papers by Hen and Karliner \cite{Hen_2008,Hen_2009}, they determine the minimal energy soliton crystal for the standard model to be hexagonal.
They scanned the parallelogram parameter space at a constant Skyrmion density to find the parallelogram that minimises the static energy.
Once they found the optimal parallelogram, they then varied the Skyrmion density to find the minimal energy Skyrmion structure.
In what follows, we refer to the shape of the lattice $\Lambda$ as the lattice structure and the energy minimising Skyrmion as the soliton/Skyrmion crystal.
Note that for an energy minimiser $\varphi$ to be a soliton crystal it has to satisfy the extended virial constraints detailed below.
We use a more robust method based on the work done by Speight \cite{Speight_2014} and propose an analytic method to determine the optimal lattice structure for an arbitrary potential.
We then apply this method to study Skyrmion crystals in the standard model and in the easy plane model.

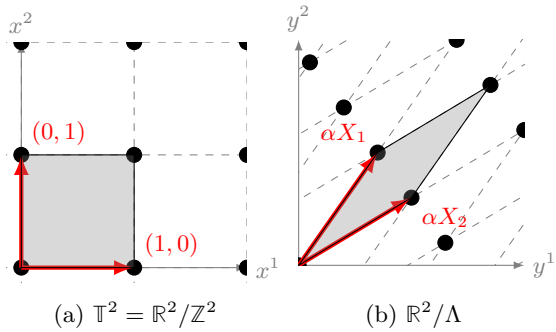
\begin{figure}[t]
  \centering
  \begin{subfigure}[b]{0.2\textwidth}
  \begin{tikzpicture}
    \coordinate (Origin)   at (0,0);
    \coordinate (XAxisMin) at (0,0);
    \coordinate (XAxisMax) at (3,0);
    \coordinate (YAxisMin) at (0,0);
    \coordinate (YAxisMax) at (0,3);
    \draw [thin, gray,-latex] (XAxisMin) -- (XAxisMax) node [right] {$x^1$};;
    \draw [thin, gray,-latex] (YAxisMin) -- (YAxisMax) node [above] {$x^2$};;

    \clip (-0.2,-0.2) rectangle (3cm,3cm); 
    \pgftransformcm{1}{0}{0}{1}{\pgfpoint{0cm}{0cm}}
    \coordinate (Xone) at (0,1.5);
    \coordinate (Xtwo) at (1.5,0);
    \draw[style=help lines,dashed] (-14,-14) grid[step=1.5cm] (14,14);
    \foreach \x in {-7,-6,...,7}{
      \foreach \y in {-7,-6,...,7}{
        \node[draw,circle,inner sep=2pt,fill] at (1.5*\x,1.5*\y) {};
      }
    }
    \draw [ultra thick,-latex,red] (Origin)
        -- (Xone) node [above right] {$(0,1)$};
    \draw [ultra thick,-latex,red] (Origin)
        -- (Xtwo) node [above right] {$(1,0)$};
    \filldraw[fill=gray, fill opacity=0.3, draw=black] (Origin)
        rectangle ($1*(Xone)+(Xtwo)$);
  \end{tikzpicture}
  \caption{$\mathbb{T}^2=\mathbb{R}^2/\mathbb{Z}^2$}
  \end{subfigure}
  \begin{subfigure}[b]{0.2\textwidth}
  \begin{tikzpicture}
    \coordinate (Origin)   at (0,0);
    \coordinate (XAxisMin) at (0,0);
    \coordinate (XAxisMax) at (3,0);
    \coordinate (YAxisMin) at (0,0);
    \coordinate (YAxisMax) at (0,3);
    \draw [thin, gray,-latex] (XAxisMin) -- (XAxisMax) node [right] {$y^1$};;
    \draw [thin, gray,-latex] (YAxisMin) -- (YAxisMax) node [above] {$y^2$};;

    \clip (0,0) rectangle (3cm,3cm); 
    \pgftransformcm{1}{0.6}{0.7}{1}{\pgfpoint{0cm}{0cm}}
    \coordinate (Xone) at (0,1.5);
    \coordinate (Xtwo) at (1.5,0);
    \draw[style=help lines,dashed] (-14,-14) grid[step=1.5cm] (14,14);
    \foreach \x in {-7,-6,...,7}{
      \foreach \y in {-7,-6,...,7}{
        \node[draw,circle,inner sep=2pt,fill] at (1.5*\x,1.5*\y) {};
      }
    }
    \draw [ultra thick,-latex,red] (Origin)
        -- (Xone) node [above left] {$\alpha X_1$};
    \draw [ultra thick,-latex,red] (Origin)
        -- (Xtwo) node [below right] {$\alpha X_2$};
    \filldraw[fill=gray, fill opacity=0.3, draw=black] (Origin)
        rectangle ($1*(Xone)+(Xtwo)$);
  \end{tikzpicture}
  \caption{$\mathbb{R}^2/\Lambda$}
  \end{subfigure}
  \caption{The (a) domain $2$-torus $\mathbb{T}^2$ and (b) target $2$-torus $\mathbb{R}^2/\Lambda$ for the diffeomorphism $F: \mathbb{T}^2 \rightarrow \mathbb{R}^2/\Lambda$.}
  \label{fig: Baby Skyrmions on a lattice - Period lattice}
\end{figure}

The physical space of interest is the $2$-torus $\mathbb{R}^2/\Lambda$, where $\Lambda$ is the set of all $2$-dimensional period lattices
\begin{equation}
    \Lambda = \left\{ \sum_{i=1}^2 n_i (\alpha X_i) \,\left.\right|\, n_i \in \mathbb{Z}, \alpha \in \mathbb{R}^* \right\},
\label{eq: Baby Skyrmions on a lattice - Period Lattice}
\end{equation}
$\alpha$ is a scaling parameter and $\{X_1,X_2\}$ is a basis for $\mathbb{R}^2$.
We have written the fundamental pair of periods in the form $Y_i=\alpha X_i\in\mathbb{R}^2$ for later convenience, where we will introduce a constraint such that the area of the period lattice is $\alpha^2$.
The crystallographic restriction theorem states that there are $5$ Bravais lattice types in $2$-dimensions \cite{Bamberg_2003}.
In each of these lattice types the fundamental unit cell is a certain type of a parallelogram.
To find the Skyrmion crystal we minimise the static energy functional over all period lattices.
Equivalently, we fix our domain of $\varphi$ to be the unit $2$-torus $\mathbb{R}^2/\mathbb{Z}^2$ and identify every other torus $\mathbb{R}^2/\Lambda$ with $\mathbb{R}^2/\mathbb{Z}^2$, but with a non-standard Riemannian metric $g$.
This metric $g$ on $\mathbb{R}^2/\mathbb{Z}^2$ is the pullback of the flat Euclidean metric $\bar{g}$ on $\mathbb{R}^2/\Lambda$ via the diffeomorphism $\mathbb{R}^2/\mathbb{Z}^2 \rightarrow \mathbb{R}^2/\Lambda$.
As we vary the period lattice $\Lambda$ then the metric $g$ varies \cite{Speight_2014}.

Now, let $F: \mathbb{T}^2 \rightarrow \mathbb{R}^2/\Lambda$ be a diffeomorphism with $F \in\GL^+(2,\mathbb{R})$ and $\mathbb{T}^2=\mathbb{R}^2/\mathbb{Z}^2$, as shown in Fig.~\ref{fig: Baby Skyrmions on a lattice - Period lattice}.
Using the identification $\GL^+(2,\mathbb{R})=\SL(2,\mathbb{R}) \times \mathbb{R}^* /\mathbb{Z}_2$, let $\mathcal{A} = [X_1 \, X_2] \in \SL(2,\mathbb{R})$ and $\alpha \in \mathbb{R}^*$ such that $F=\alpha \mathcal{A}\in\GL^+(2,\mathbb{R})$.
We will now identify the domain of $\varphi$ as $\Sigma=\mathbb{T}^2$, so that the Skyrme field is a map $\varphi: \mathbb{T}^2 \rightarrow S^2$.
The metric on $\mathbb{T}^2$ is the pullback $g=F^*\bar{g}$ of the flat Euclidean metric $\bar{g}$ on $\mathbb{R}^2/\Lambda$.
Explicitly, this is
\begin{equation}
    g = \alpha^2
    \begin{bmatrix}
        X_1 \cdot X_1 & X_1 \cdot X_2 \\
        X_1 \cdot X_2 & X_2 \cdot X_2
    \end{bmatrix},
\label{eq: Baby Skyrmions on a lattice - Torus metric}
\end{equation}
with inverse
\begin{equation}
    g^{-1} = \frac{1}{\alpha^2}
    \begin{bmatrix}
        X_2 \cdot X_2 & -X_1 \cdot X_2 \\
        -X_1 \cdot X_2 & X_1 \cdot X_1
    \end{bmatrix}.
\label{eq: Baby Skyrmions on a lattice - Inverse torus metric}
\end{equation}
The Riemannian volume form is simply $\textrm{vol}_g=\sqrt{\det g}\,\textrm{d}x^1\wedge\textrm{d}x^2=\alpha^2 \,\textrm{d}x^1\wedge\textrm{d}x^2$.
Then, using the local form for the Dirichlet term~\eqref{eq: Baby Skyrme model - Dirichlet energy} and the inverse metric~\eqref{eq: Baby Skyrmions on a lattice - Inverse torus metric}, we can compute the Dirichlet energy on $\mathbb{T}^2$ to be given by
\begin{eqnarray}
    E_2 & = \frac{1}{2}\int_{\mathbb{T}^2} \left\{ (X_2 \cdot X_2)(\partial_1 \varphi)^2 - 2(X_2\cdot X_1)(\partial_1\varphi\cdot\partial_2\varphi) \right. \nonumber \\ & \left. + (X_1 \cdot X_1)(\partial_2\varphi)^2 \right\} \textrm{d}x^1\,\textrm{d}x^2.
\label{eq: Baby Skyrmions on a lattice - Dirichlet energy}
\end{eqnarray}
Since the Dirichlet energy is conformally invariant, it does not have a dependence on the scaling parameter $\alpha$.
Likewise, using the local form for the Skyrme term~\eqref{eq: Baby Skyrme model - Skyrme energy}, the inverse metric~\eqref{eq: Baby Skyrmions on a lattice - Inverse torus metric} and the pullback of the area $2$-form~\eqref{eq: Baby Skyrme model - Area 2-form}, the Skyrme energy on $\mathbb{T}^2$ is 
\begin{equation}
    E_4 = \frac{\kappa^2}{2\alpha^2}\int_{\mathbb{T}^2} \left( \partial_1 \varphi \times \partial_2 \varphi \right)\cdot\left( \partial_1 \varphi \times \partial_2 \varphi \right) \textrm{d}x^1\,\textrm{d}x^2,
\label{eq: Baby Skyrmions on a lattice - Skyrme energy}
\end{equation}
and the potential energy is simply
\begin{equation}
    E_0 = \alpha^2 \int_{\mathbb{T}^2} V(\varphi)\, \textrm{d}x^1\,\textrm{d}x^2.
\label{eq: Baby Skyrmions on a lattice - Potential energy}
\end{equation}
Putting this together, we see that the static energy functional for baby Skyrmions on the unit area $2$-torus $\mathbb{T}^2$ with the non-standard Riemannian metric $g$ is
\begin{widetext}
\begin{equation}
    \begin{split}
        E = \frac{1}{2}\int_{\mathbb{T}^2} \left\{ X_2^2(\partial_1 \varphi)^2 - 2(X_2\cdot X_1)(\partial_1\varphi\cdot\partial_2\varphi) + X_1^2)(\partial_2\varphi)^2 \right\} \textrm{d}x^1\,\textrm{d}x^2 \\
        + \frac{\kappa^2}{2\alpha^2}\int_{\mathbb{T}^2} \left( \partial_1 \varphi \times \partial_2 \varphi \right)^2 \textrm{d}x^1\,\textrm{d}x^2 + \alpha^2 \int_{\mathbb{T}^2} V(\varphi)\, \textrm{d}x^1\,\textrm{d}x^2.
    \end{split}
\label{eq: Baby Skyrmions on a lattice - Static energy}
\end{equation}
\end{widetext}
As before, we need an explicit description of the energy gradient for our numerical analysis.
The variation of the energy density with respect to field $\varphi^a$ can be obtained from the Euler--Lagrange field equations, that is
\begin{equation*}
    \begin{split}
        \frac{\delta \mathcal{E}}{\delta \varphi^a} = \alpha^2 \frac{\delta V}{\delta \varphi^a} -\left\{ \alpha^2 g^{ij}\partial_{ij}\varphi^a + \frac{\kappa^2}{\alpha^2} \left[ \partial_{ii}\varphi^a \left( \partial_j \varphi^b \right)^2 \right. \right. \\
        \left. \left. + \partial_i\varphi^a \left( \partial_{ij}\varphi^b\,\partial_j\varphi^b - \partial_{jj}\varphi^b\,\partial_i\varphi^b \right)  - \partial_{ij}\varphi^a \left( \partial_i\varphi^b\,\partial_j\varphi^b \right) \right] \right\},
    \end{split}
\end{equation*}
where $i,j\in\{1,2\}$ and $a,b\in\{1,2,3\}$.

To find the optimal lattice structure, we must vary the static energy functional~\eqref{eq: Baby Skyrmions on a lattice - Static energy} with respect to the period lattice parameters $X_1,X_2$ and $\alpha$.
Firstly, taking the variation of the static energy functional~\eqref{eq: Baby Skyrmions on a lattice - Static energy} with respect to the scaling parameter $\alpha$,
\begin{equation*}
    \frac{\partial E}{\partial \alpha} = \int_{\mathbb{T}^2} \left\{ -\frac{\kappa^2}{\alpha^3}\left( \partial_1 \varphi \times \partial_2 \varphi \right)^2  + 2\alpha V(\varphi) \right\} \, \textrm{d}x^1\,\textrm{d}x^2 = 0,
\end{equation*}
yields the following relation for the scaling parameter:
\begin{equation}
    \alpha^2 = \sqrt{\frac{\frac{\kappa^2}{2}\int_{\mathbb{T}^2} \left( \partial_1 \varphi \times \partial_2 \varphi \right)^2 \textrm{d}x^1\,\textrm{d}x^2}{\int_{\mathbb{T}^2} V(\varphi)\, \textrm{d}x^1\,\textrm{d}x^2}}.
\label{eq: Baby Skyrmions on a lattice - Scaling parameter}
\end{equation}
Thus, the area of the period lattice is determined by the ratio of the flat Skyrme term to the flat potential term.
Determining the fundamental pair of periods $X_1, X_2$ which minimise the Dirichlet energy $E_2$ is a constrained quadratic optimization problem with the nonlinear constraint $\det([X_1\,X_2])=1$.
For notational convenience, let us write
\begin{equation}
    \mathcal{E}_{ij} = \int_{\mathbb{T}^2} \left( \partial_i \varphi \cdot \partial_j \varphi \right) \textrm{d}x^1\,\textrm{d}x^2.
\label{eq: Baby Skyrmions on a lattice - Compact Dirichlet}
\end{equation}
Then the Dirichlet energy~\eqref{eq: Baby Skyrmions on a lattice - Dirichlet energy} can be expressed in the quadratic form
\begin{equation}
    E_2 = \frac{1}{2}\bm{x}^T \mathcal{Q} \bm{x}, \quad
    \mathcal{Q} =
    \begin{bmatrix}
        \mathcal{E}_{22} & 0 & -\mathcal{E}_{12} & 0 \\
        0 & \mathcal{E}_{22} & 0 & -\mathcal{E}_{12} \\
        -\mathcal{E}_{12} & 0 & \mathcal{E}_{11} & 0 \\
        0 & -\mathcal{E}_{12} & 0 & \mathcal{E}_{11}
    \end{bmatrix},
\label{eq: Baby Skyrmions on a lattice - Quadratic Dirichlet}
\end{equation}
where $\bm{x}=\begin{bmatrix} X_1 \\ X_2 \end{bmatrix}$ is a $4$-vector and $\mathcal{Q}$ is a $4\times4$-symmetric matrix.
This constrained quadratic optimization problem can be solved by including the Lagrange term $\gamma(\det([X_1\,X_2])-1)$ into~\eqref{eq: Baby Skyrmions on a lattice - Quadratic Dirichlet}, where $\gamma\in\mathbb{R}^*$ is a Lagrange multiplier.
This reduces the problem to an eigenvalue problem
\begin{equation}
    \mathcal{B} \bm{x} = \gamma \bm{x}, \quad
    \mathcal{B} =
    \begin{bmatrix}
        0 & \mathcal{E}_{12} & 0 & -\mathcal{E}_{11} \\
        -\mathcal{E}_{12} & 0 & \mathcal{E}_{11} & 0 \\
        0 & \mathcal{E}_{22} & 0 & -\mathcal{E}_{12} \\
        -\mathcal{E}_{22} & 0 & \mathcal{E}_{12} & 0
    \end{bmatrix}.
\label{eq: Baby Skyrmions on a lattice - Eigenvalue problem}
\end{equation}

By definition, for an energy minimiser $\varphi: \mathbb{R}^2/\Lambda \rightarrow S^2$ to be a soliton lattice, its stress tensor $S[\varphi]$ must be $L^2$ orthogonal to the space of parallel symmetric bilinear forms $\mathbb{E}$ (a $3$-dimensional subspace of the space of sections of the rank $3$ vector bundle $T^*\mathbb{R}^2/\Lambda \odot T^*\mathbb{R}^2/\Lambda$).
Furthermore, if the Hessian of the soliton lattice is positive definite then it is a soliton crystal.
In fact, Speight \cite{Speight_2014} showed that \textit{every} baby Skyrme lattice is a soliton crystal.
The stress tensor of $\varphi$ is given by \cite{Jaykka_2012}
\begin{equation}
    S[\varphi] = \left( \frac{1}{2}|\textrm{d}\varphi|_{\bar{g}}^2 -\frac{1}{2}|\varphi^*\omega|_{\bar{g}}^2 + V(\varphi)\right) \bar{g} - \varphi^*h.
\label{eq: Baby Skyrmions on a lattice - Stress tensor}
\end{equation}
Let $\mathbb{E}_0$ be the $2$-dimensional space of traceless parallel symmetric bilinear forms.
Then the Skyrme field $\varphi$ is a soliton lattice if and only if $\varphi$ is $L^2$ orthogonal to $\bar{g}$ and $\mathbb{E}_0$, where we recall that $\bar{g}$ is the metric on $\mathbb{R}^2/\Lambda$ and not $\mathbb{T}^2$.
This gives us the familiar virial constraint
\begin{equation}
    \int_{\mathbb{R}^2/\Lambda} \left( -\frac{1}{2}|\varphi^*\omega|_{\bar{g}}^2 + V(\varphi) \right) \, \textrm{vol}_{\bar{g}} = E_0 - E_4 = 0.
\label{eq: Baby Skyrmions on a lattice - Virial constraint 1}
\end{equation}
Let $(y^1,y^2)$ be local orthonormal coordinates on $\mathbb{R}^2/\Lambda$ and $\varepsilon \in \mathbb{E}_0$.
Then for $S$ to be $L^2$ orthogonal to $\mathbb{E}_0$, we require
\begin{equation*}
    \braket{S,\varepsilon}_{L^2} = -\frac{1}{2}\braket{\varphi^*h,\varepsilon}_{L^2}=0.
\end{equation*}
As $\mathbb{E}_0$ is spanned by $\varepsilon_1=(\textrm{d}y^1)^2-(\textrm{d}y^2)^2$ and $\varepsilon_2=2\textrm{d}y^1\textrm{d}y^2$, we get the following additional virial constraints
\begin{equation}
    \int_{\mathbb{R}^2/\Lambda} \left( \left|\frac{\partial \varphi}{\partial y^1}\right|^2 - \left|\frac{\partial \varphi}{\partial y^2}\right|^2 \right) \textrm{d}y^1 \textrm{d}y^2 = 0
\label{eq: Baby Skyrmions on a lattice - Virial constraint 2}
\end{equation}
and
\begin{equation}
    \int_{\mathbb{R}^2/\Lambda} \frac{\partial \varphi}{\partial y^1} \cdot \frac{\partial \varphi}{\partial y^2} \, \textrm{d}y^1 \textrm{d}y^2 = 0.
\label{eq: Baby Skyrmions on a lattice - Virial constraint 3}
\end{equation}
These additional virial constraints state that the Skyrme map $\varphi$ must be conformal on average.
We have shown above that for the soliton lattice $\varphi$ to be critical with respect to variations of the period lattice $\Lambda$, it must satisfy the extended virial constraints in each lattice cell.

The extended Derrick virial constraints~\eqref{eq: Baby Skyrmions on a lattice - Virial constraint 1}--\eqref{eq: Baby Skyrmions on a lattice - Virial constraint 3} can be imposed as a consistency check when implementing the lattice optimisation method detailed above.
For numerics, the domain manifold of interest is the $2$-torus $\mathbb{T}^2=\mathbb{R}^2/\mathbb{Z}^2$ with metric $g=F^*\bar{g}$, where we previously introduced the diffeomorphism $F: \mathbb{T}^2 \rightarrow \mathbb{R}^2/\Lambda$.
Recall that we used the identification $F=\alpha \mathcal{A} \in \GL^+(2,\mathbb{R})$ for $\alpha\in\mathbb{R}^*$ and $\mathcal{A} = [X_1 \, X_2] \in \SL(2,\mathbb{R})$.
Thus, using the compact notation~\eqref{eq: Baby Skyrmions on a lattice - Compact Dirichlet}, the generalised virial constraints \eqref{eq: Baby Skyrmions on a lattice - Virial constraint 2} and \eqref{eq: Baby Skyrmions on a lattice - Virial constraint 3} on $\mathbb{T}^2$ are given, respectively, by
\begin{equation}
\begin{split}
    (\mathcal{A}_{22}^2-\mathcal{A}_{12}^2)\mathcal{E}_{11} + (\mathcal{A}_{21}^2-\mathcal{A}_{11}^2)\mathcal{E}_{22} \\ + 2(\mathcal{A}_{11}\mathcal{A}_{12}-\mathcal{A}_{21}\mathcal{A}_{22})\mathcal{E}_{12} = 0
\end{split}
\label{eq: Baby Skyrmions on a lattice - T2 Virial constraint 1}
\end{equation}
and
\begin{equation}
\begin{split}
    -\mathcal{A}_{12}\mathcal{A}_{22}\mathcal{E}_{11} - \mathcal{A}_{11}\mathcal{A}_{21}\mathcal{E}_{22} \\ + (\mathcal{A}_{11}\mathcal{A}_{22}+\mathcal{A}_{12}\mathcal{A}_{21})\mathcal{E}_{12} = 0,
\end{split}
\label{eq: Baby Skyrmions on a lattice - T2 Virial constraint 2}
\end{equation}
where $\mathcal{A}_{ij}=(X_j)_i$, the $i$\textsuperscript{th} component of $X_j$, and $(\mathcal{A}_{ij})=\mathcal{A}$.

In this section, we have shown that the problem of determining the optimal lattice structure that minimises the baby Skyrme energy~\eqref{eq: Baby Skyrmions on a lattice - Static energy} amounts to solving an eigenvalue problem~\eqref{eq: Baby Skyrmions on a lattice - Eigenvalue problem}.
During each iteration of our numerical minimisation algorithm, we perform an accelerated gradient descent then we compute the scaling parameter $\alpha$ via~\eqref{eq: Baby Skyrmions on a lattice - Scaling parameter} and solve the eigenvalue problem~\eqref{eq: Baby Skyrmions on a lattice - Eigenvalue problem} to give us the pair of periods $X_1,X_2$.
We also check that the generalised virial constraints~\eqref{eq: Baby Skyrmions on a lattice - Virial constraint 1}, \eqref{eq: Baby Skyrmions on a lattice - T2 Virial constraint 1} and \eqref{eq: Baby Skyrmions on a lattice - T2 Virial constraint 2} are satisfied each iteration, showing that the energy minimiser $\varphi: \mathbb{R}^2/\Lambda \rightarrow S^2$ is indeed a soliton lattice and thus a soliton crystal.
This determines the lattice $\Lambda$ and the algorithm in turn determines the Skyrmion crystal.
The numerics detailed throughout this section were carried out initially on a $200 \times 200$-grid with lattice spacing $\Delta x=0.005$, with a final higher accuracy simulation carried out on a $500 \times 500$-grid with lattice spacing $\Delta x=0.002$.
Finer meshes were tried but there were no considerable changes in the final energy.
Note that the coarser $200\times 200$-grid would be sufficient as this gives approximately the same accuracy as the numerics for the baby Skyrmions on $\mathbb{R}^2$.
It is also worth noting that the lattice spacings are fixed sizes on the discretised unit area $2$-torus $\mathbb{T}^2$, whereas the equivalent lattice spacings on the discretised $2$-torus $\mathbb{R}^2/\Lambda$ vary as the lattice $\Lambda$ varies.

\begin{figure}[t]
	\centering
	\begin{subfigure}[b]{0.20\textwidth}
	\includegraphics[width=\textwidth]{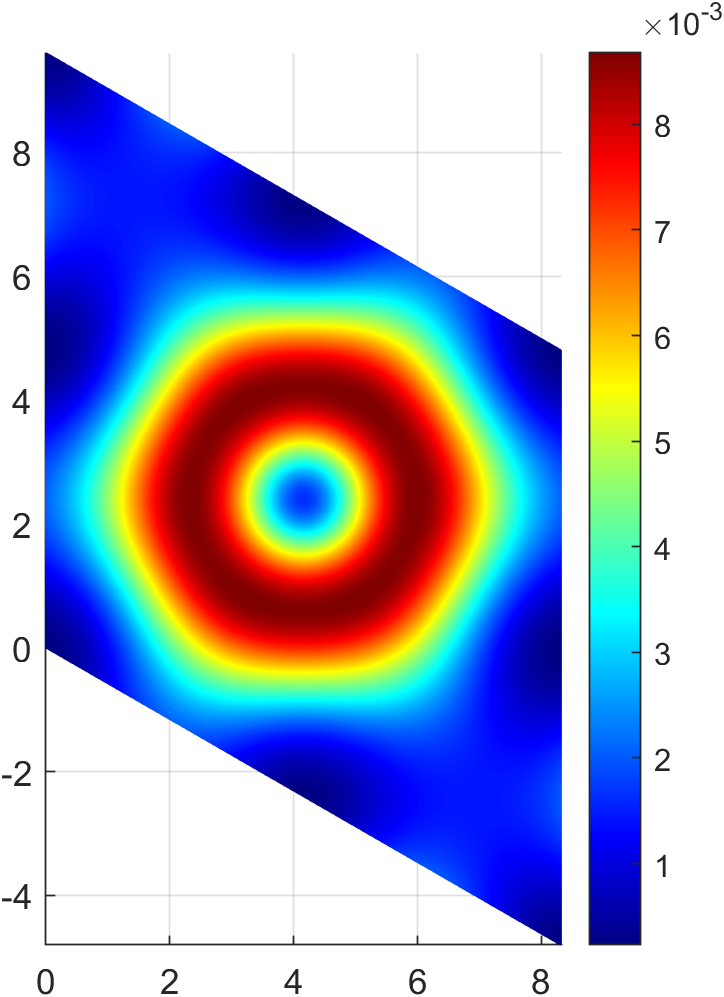}
	\caption{Energy density.}
	\label{fig: Standard lattice baby Skyrmions - Hexagonal lattice (a)}
	\end{subfigure}
    ~
	\begin{subfigure}[b]{0.16\textwidth}
	\includegraphics[width=\textwidth]{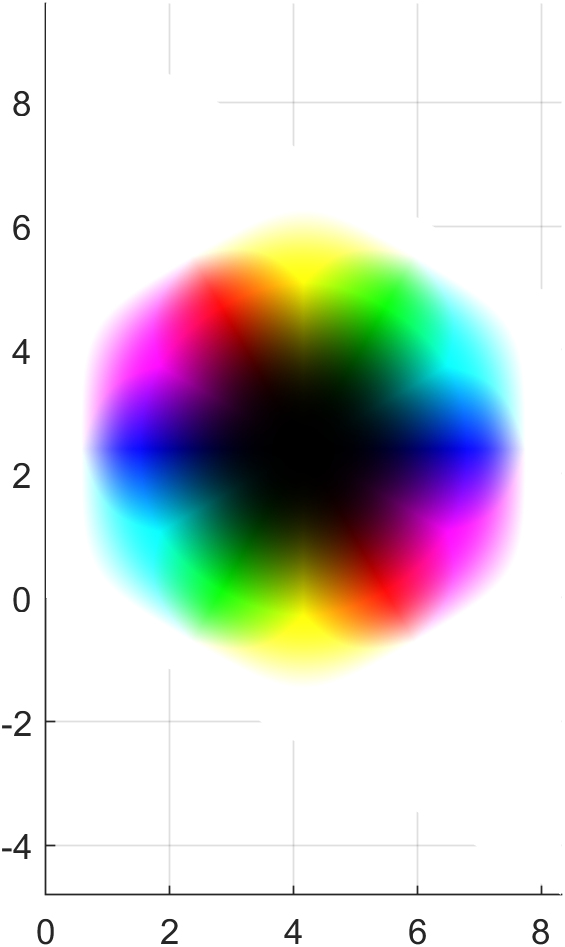}
	\caption{Phase coloring.}
	\label{fig: Standard lattice baby Skyrmions - Hexagonal lattice (b)}
	\end{subfigure}
	\caption{Hexagonal crystalline structure of the infinite crystal in the standard model.}
	\label{fig: Standard lattice baby Skyrmions - Hexagonal lattice}
\end{figure}


\subsection{Standard baby Skyrmion crystals}
\label{subsec: Standard lattice baby Skyrmions}

Employing the lattice optimisation method detailed in Sec~\ref{sec: Baby Skyrmions on a lattice} for the standard potential~\eqref{eq: Baby Skyrmions on R2 - Standard potential} with $m^2=0.1$ (and $\kappa^2$=1), the optimal lattice is found to be an equianharmonic lattice with the baby Skyrmions forming a hexagonal Skyrmion crystal with $D_6$ symmetry.
This is found for almost all $B=2$ initial configurations on random initial lattice geometry (with the exception of relaxing to the infinite chain solution occasionally).
Each unit cell contains a charge of $B=2$ and has sides of equal length $L_{\textrm{crystal}}=9.60$ with the angle between the two periods $X_1,X_2$ being $\theta=\frac{2\pi}{3}$, giving a unit cell area of $A=79.84$.
We find that the Skyrmion crystal has energy $\mathcal{E}_{\textrm{crystal}}=1.4543$, which is lower than the infinite chain energy $\mathcal{E}_{\textrm{chain}}=1.4548$.
Note that when we refer to energy values, we have normalised them by the Bogomolny bound, i.e. $\mathcal{E}:=E/(4\pi B)$.
The hexagonal Skyrmion crystal can be seen in Fig.~\ref{fig: Standard lattice baby Skyrmions - Hexagonal lattice}.
As a fidelity check with Hen and Karliner's work, we also determined the optimal lattice to be equianharmonic with a hexagonal Skyrmion crystal for $m^2=0.1$ and $\kappa^2=0.03$.
The energy for this crystal is found to be $\mathcal{E}_{\textrm{crystal}}=1.0799$, which is in excellent agreement with their numerically determined value of $\mathcal{E}_{\textrm{crystal}}=1.08$.

\begin{figure}[t]
	\centering
	\begin{subfigure}[b]{0.3\textwidth}
	\includegraphics[width=\textwidth]{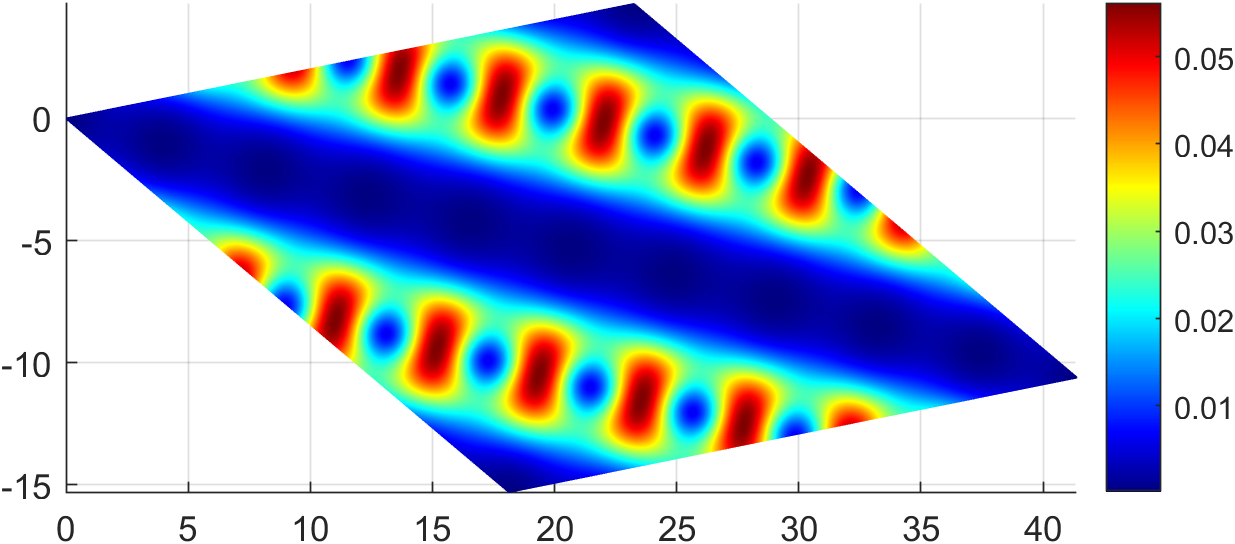}
	\caption{Energy density.}
	\label{fig: Standard lattice baby Skyrmions - B10 ED}
	\end{subfigure} \\
	\begin{subfigure}[b]{0.3\textwidth}
	\includegraphics[width=\textwidth]{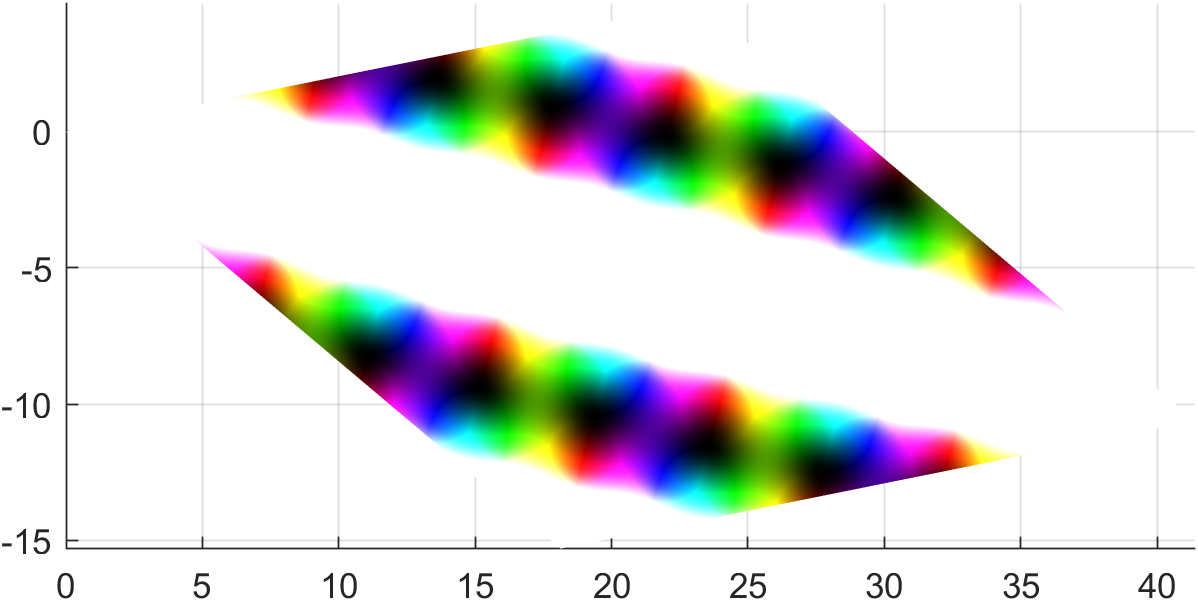}
	\caption{Phase coloring.}
	\label{fig: Easy plane lattice baby Skyrmions - B10 RC}
	\end{subfigure}
	\caption{Plots of the adjacent infinite chains in the attractive channel, yielding an energy lower than the infinite chain but higher than the hexagonal infinite crystal.}
	\label{fig: Easy plane lattice baby Skyrmions - B10}
\end{figure}

Other soliton crystals were searched for at numerous topological charges for various initial configurations and initial lattice geometry.
However, they all had a tendency to relax into a chain or rows of separate chains with the infinite chain energy $\mathcal{E}_{\textrm{chain}}=1.4548$.
A slightly lower energy configuration was found for rows of adjacent chains with all the charge-$1$ links rotated by $\pi$ in one chain relative to the other.
This attractive chains configuration has an energy of $\mathcal{E}_{\textrm{2-chains}}=1.4545$ and is shown in Fig.~\ref{fig: Easy plane lattice baby Skyrmions - B10}.


\subsection{Easy plane baby Skyrmion crystals}
\label{subsec: Easy plane lattice baby Skyrmions}

\begin{figure}[t]
	\centering
	\begin{subfigure}[b]{0.2\textwidth}
	\includegraphics[width=\textwidth]{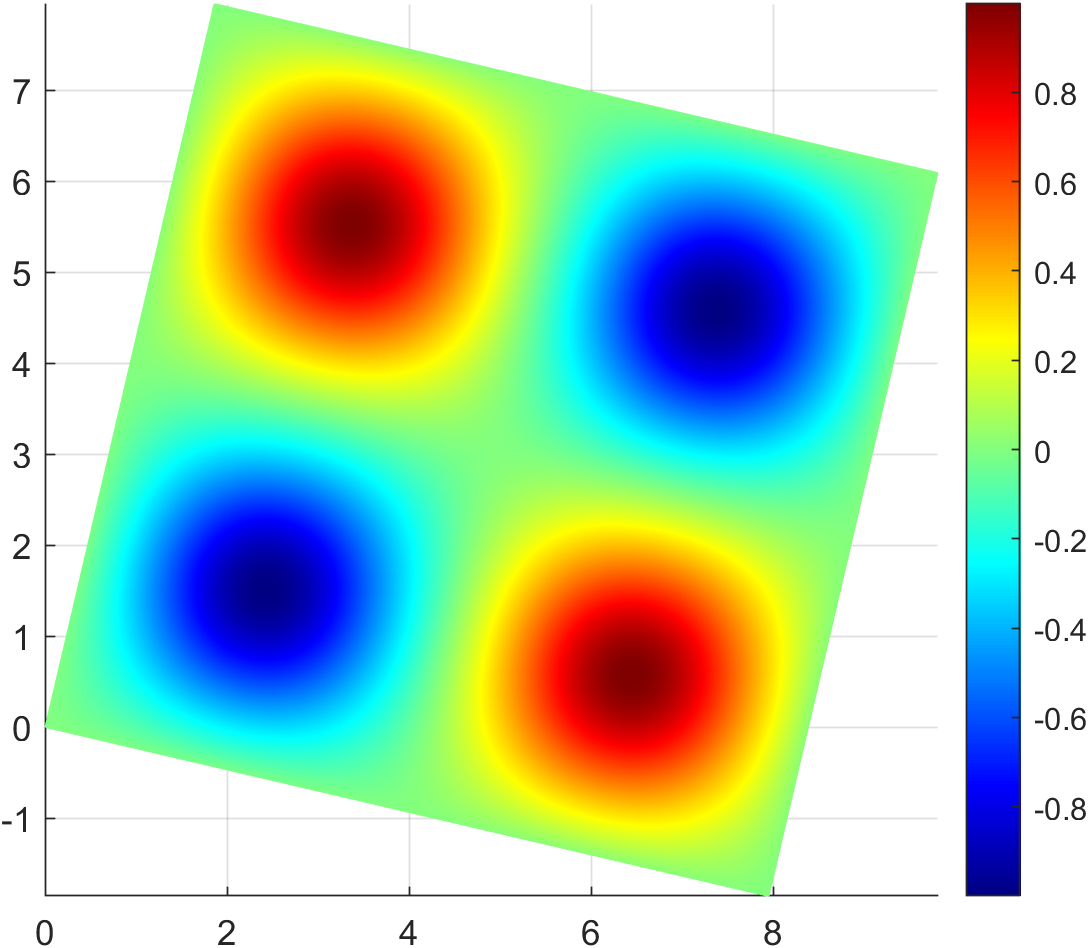}
	\caption{$B=2$.}
	\label{fig: Easy plane lattice baby Skyrmions - B2}
	\end{subfigure}
    ~ 
	\begin{subfigure}[b]{0.2\textwidth}
	\includegraphics[width=\textwidth]{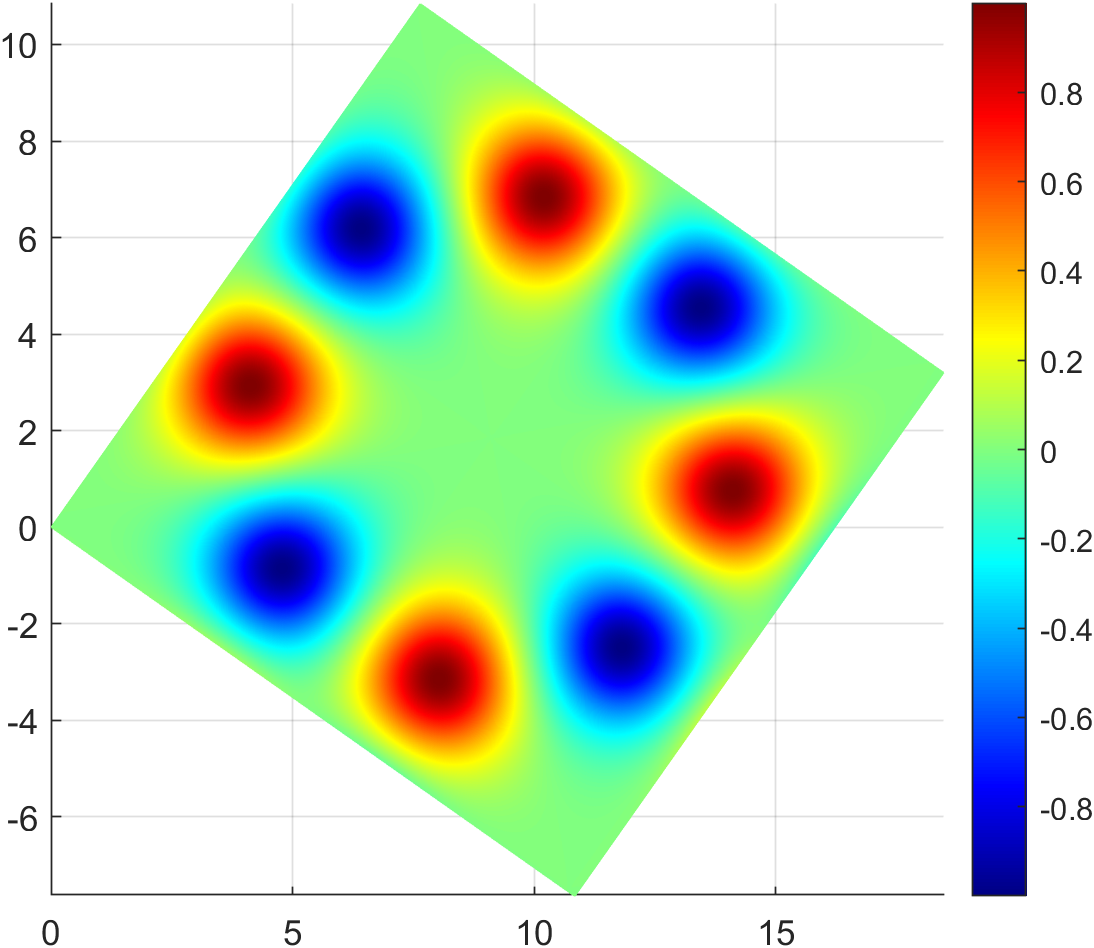}
	\caption{$B=4$.}
	\label{fig: Easy plane lattice baby Skyrmions - B4}
	\end{subfigure} \\
	\begin{subfigure}[b]{0.25\textwidth}
	\includegraphics[width=\textwidth]{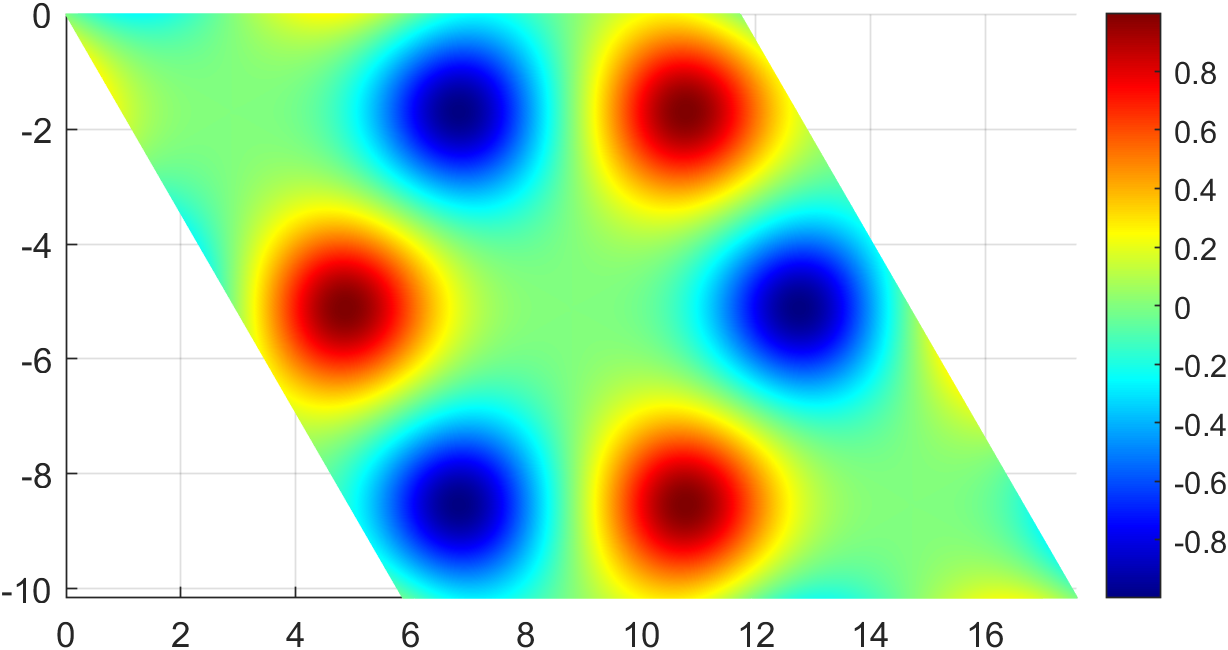}
	\caption{$B=3$.}
	\label{fig: Easy plane lattice baby Skyrmions - B3}
	\end{subfigure}
	\caption{$\varphi^1$ density plots of the optimal crystalline structures and their corresponding lattices.
	The lowest energy crystal structure is the $B=2$ and the highest is $B=4$.}
	\label{fig: Easy plane lattice baby Skyrmions - Crystal structures}
\end{figure}

As previously proposed in Sec~\ref{subsec: Easy plane baby Skyrmions}, it seems likely that there may possibly be a few soliton crystals for the easy plane model.
This prompts the search for Skyrmion crystals for a range of charges with various initial configurations.
The lowest energy Skyrmion crystal is a square of half lumps with $D_4$ symmetry in a square lattice for $B=2$, with energy $\mathcal{E}_{B=2}=1.5152$.
The square lattice has sides of equal length $L_{\textrm{crystal}}=8.20$, giving a unit cell area of $A=67.24$.
Two other Skyrmion crystals were found with slightly higher energies:
a hexagonal Skyrmion crystal in an equianharmonic lattice for $B=3$ with $D_6$ symmetry and energy $\mathcal{E}_{B=3}=1.5207$, and an octagonal Skyrmion crystal in a square lattice with $D_4$ symmetry and energy $\mathcal{E}_{B=4}=1.5228$.
These three Skyrmion crystals are shown in Fig.~\ref{fig: Easy plane lattice baby Skyrmions - Crystal structures}.

\section{Baby Skyrmion crystal chunks}
\label{sec: Baby Skyrme crystals}

The soliton crystal is the lowest energy solution, so one would expect chunks of the soliton crystal to be the global minima for charges past a critical charge $B_{\textrm{crystal}}$.
A starting point would be to split the crystal chunk energy into a bulk volume, or area, term and a surface term.
For a given charge $B$, we know the minimal energy soliton crystal, the corresponding lattice $\Lambda$ and the lattice area.
So, the bulk area term is easy to calculate.
However, the problem lies in minimising the surface energy contribution for a fixed area, which corresponds to minimising the crystal perimeter for a fixed area.
This is known as an isoperimetric problem.
Even once the minimal energy crystal chunk shape has been found, we still require an estimate of the surface energy (per unit length) to determine the surface energy of the crystal chunk.


\subsection{Surface energy of a baby Skyrmion crystal chunk}
\label{subsec: Surface energy of a baby Skyrme crystal}

\begin{figure}[t]
	\centering
	\begin{subfigure}[b]{0.12\textwidth}
	\includegraphics[width=\textwidth]{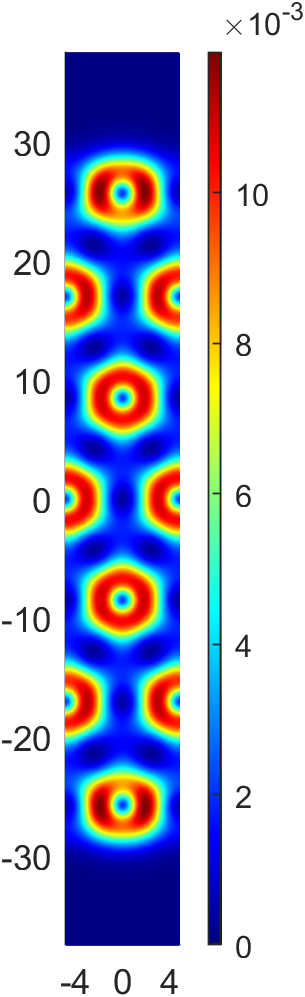}
	\caption{}
	\label{fig: Surface energy of a baby Skyrme crystal - Standard}
	\end{subfigure}
	~ 
	\begin{subfigure}[b]{0.1275\textwidth}
	\includegraphics[width=\textwidth]{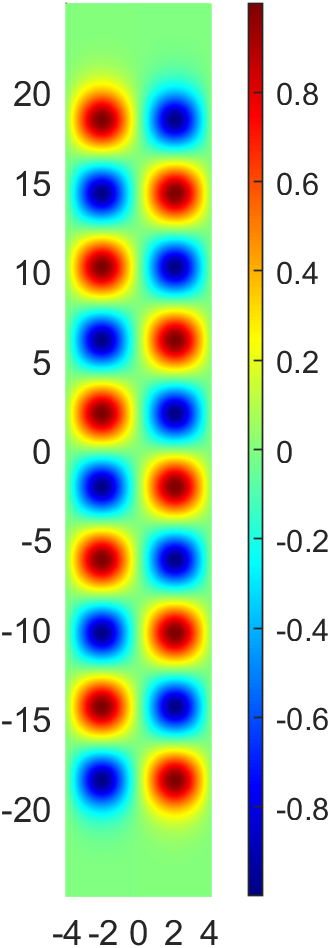}
	\caption{}
	\label{fig: Surface energy of a baby Skyrme crystal - EP}
	\end{subfigure}
	\caption{Energy density and $\varphi^1$ density plots showing a (a) $7$-layer standard hexagonal crystal slab and (b) a $5$-layer easy plane square crystal slab.}
	\label{fig: Surface energy of a baby Skyrme crystal - Crystal slabs}
\end{figure}
The surface energy per unit length of a Skyrmion crystal chunk can be predicted by using a crystal slab model.
Skyrmion crystals are layered on an infinite cylinder $\Sigma=\mathbb{R} \times S^1$ of width $L=L_{\textrm{crystal}}$, and the number of layers $n\in\mathbb{N}$ are increased to estimate the surface energy contribution.
As stated in Sec~\ref{sec: Baby Skyrme model}, this corresponds to a Dirichlet boundary condition in the $x^2$-direction, $\lim_{|x^2| \rightarrow \infty}=\varphi_\infty$, and a periodic boundary condition in the $x^1$-direction, $\varphi(x)=\varphi(x +n_1 X^1)$.
Each layer contributes a charge of $2$ in this model, giving an $n$-layer crystal slab a total charge of $B=2n$.
The crystal slab layering can be seen in Fig.~\ref{fig: Surface energy of a baby Skyrme crystal - Crystal slabs}.

For the standard potential, each hexagonal baby Skyrmion has $6$ \textit{bonding} sides (or nearest neighbours) and each crystal slab edge has $2$ unbonded sides.
Similarly, for easy plane crystal chunks, each baby Skyrmion has $4$ bonding sides and each crystal slab edge has $2$ unbonded sides.
We can express the crystal slab energy as
\begin{equation}
    E_{\textrm{slab}} = E_{\textrm{crystal}} + N_{\textrm{free}} E_{\textrm{bond}},
\label{eq: Surface energy of a baby Skyrme crystal - Crystal slab energy}
\end{equation}
where $N_{\textrm{free}}$ is the number of free bonds (or unbonded sides) and $E_{\textrm{bond}}$ is the energy of each free bond.
For the standard and easy plane crystal slabs in Fig.~\ref{fig: Surface energy of a baby Skyrme crystal - Crystal slabs} we have $N_{\textrm{free}}=4$.
We can approximate the surface energy contribution by applying a least squares fitting to the crystal slab energy normalised by the Bogomolny bound,
\begin{equation}
    \mathcal{E}_{\textrm{slab}} = \mathcal{E}_{\textrm{crystal}}+\frac{ N_{\textrm{free}}}{2n} \mathcal{E}_{\textrm{bond}},
\label{eq: Surface energy of a baby Skyrme crystal - Normalised crystal slab energy}
\end{equation}
where $\mathcal{E}_{\textrm{bond}}$ is the normalised free bond energy such that $E_{\textrm{bond}}=4\pi\mathcal{E}_{\textrm{bond}}$.
For approximating the surface energy, we computed the energies of various $n$-layer crystal slabs with $n\in\{3,\ldots,11\}$.
Using a trust region reflective algorithm, and the crystal slab approximation~\eqref{eq: Surface energy of a baby Skyrme crystal - Normalised crystal slab energy}, we find that for the standard potential $\mathcal{E}_{\textrm{bond}}=0.0031$ (with $m^2=0.1$) and for the easy plane potential $\mathcal{E}_{\textrm{bond}}=0.0103$ (with $m^2=1$).


\subsection{Standard crystal chunks}
\label{subsec: Standard crystal chunks}

To model a Skyrmion crystal chunk, we  split the crystal chunk energy into a bulk volume term and a surface term, $E_{\textrm{chunk}} = E_{\textrm{bulk}} + E_{\textrm{surface}}$.
The surface energy contribution of a baby Skyrme crystal is determined by the number of unbonded sides.
As stated before, for the standard potential, each hexagonal baby Skyrmion has $6$ \textit{bonding} sides (or nearest neighbours) which means there are many possible arrangements for crystal chunks for a given charge $B$.
For easy plane crystal chunks, the square lattice is the minimal energy crystal configuration so we only consider each easy plane baby Skyrmion to have $4$ bonding sides.
Our aim is to determine the shape of an equilibrium crystal by minimising the total surface energy associated to the crystal-vacuum interface.
In crystallography one normally employs the Wulff construction method to determine the equilibrium shape of a crystal chunk.
However, we take a simpler approach and only consider the perimeter of the crystal chunk boundary, not its shape.
Equivalently, we are considering the number of free bonds in a given crystal chunk.
This enables us to write the energy in the form
\begin{equation}
    E_{\textrm{chunk}} = E_{\textrm{bulk}} + N_{\textrm{free}} E_{\textrm{bond}}.
\label{eq: Crystal chunk approximation - General chunk}
\end{equation}
Therefore, we want to find the crystal chunk that minimises the number of free bonds, and hence its surface energy contribution, for a fixed charge $B$ and crystal area $A$.
\begin{figure}[tbp]
\centering
\includegraphics[width=0.45\textwidth]{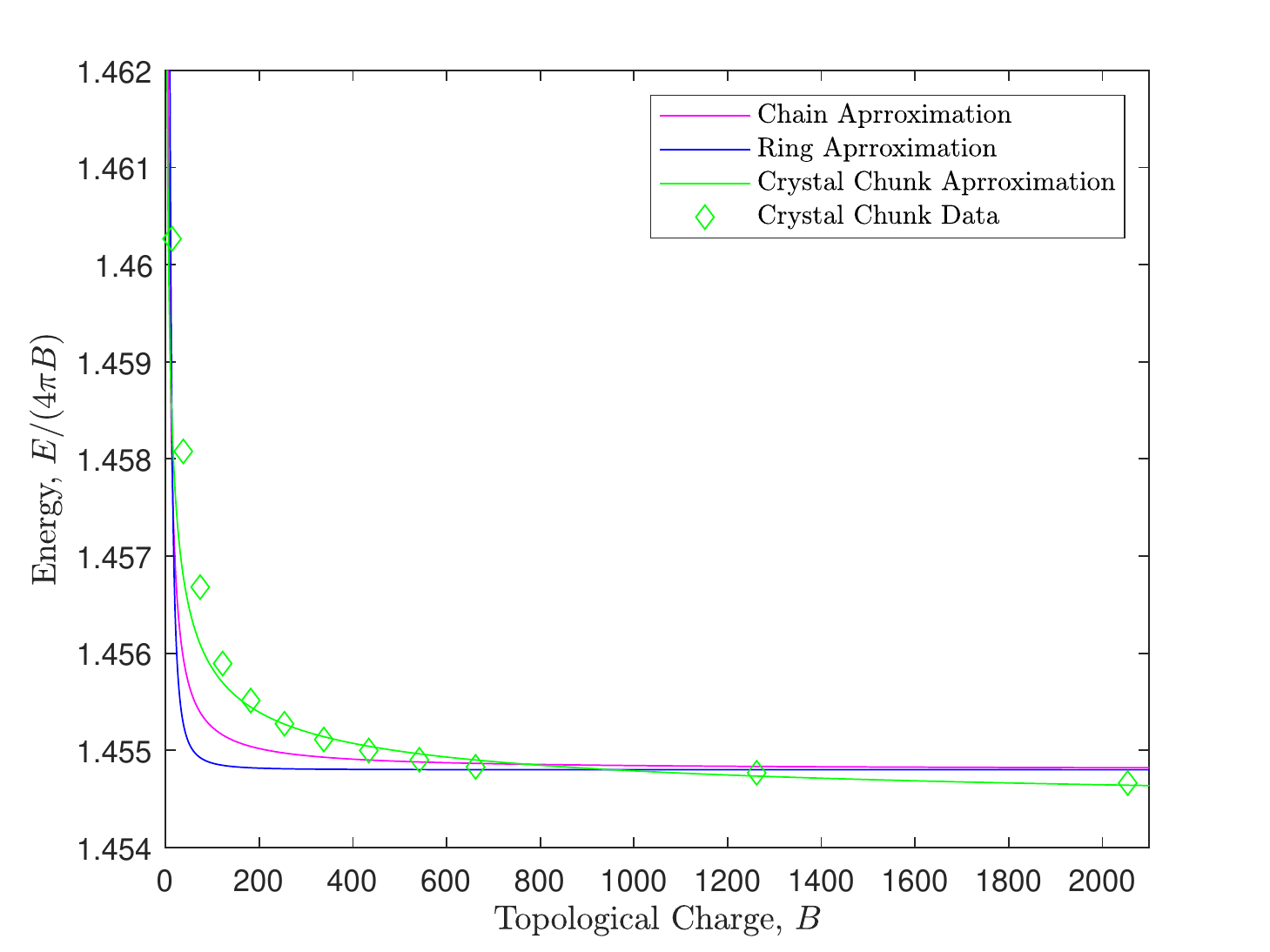}
\caption{Comparison of ring, chain and crystal chunk approximations in the standard model.}
\label{fig: EA Crystal chunk approximation - Comparison}
\end{figure}

\begin{figure}[t]
	\centering
	\begin{subfigure}[b]{0.2\textwidth}
	\includegraphics[width=\textwidth]{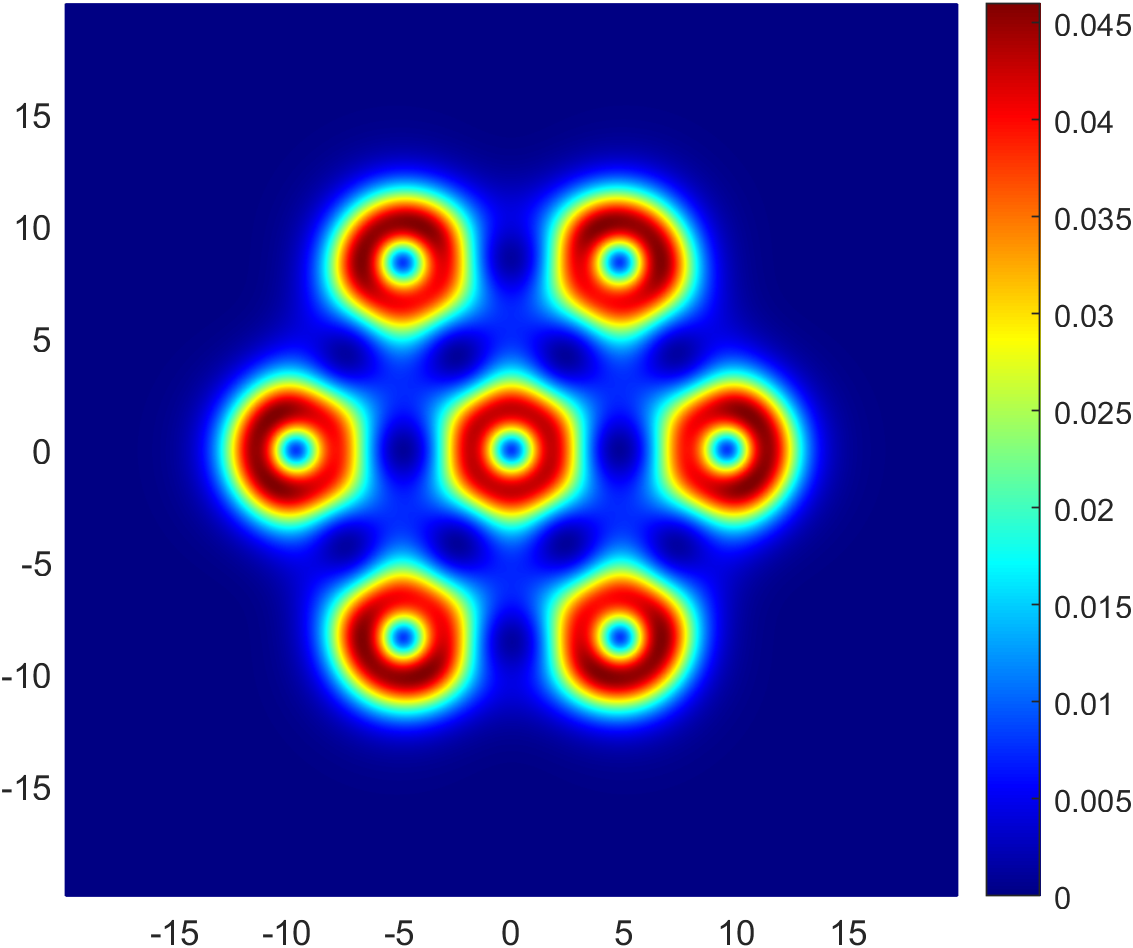}
	\caption{$n=1, B=14$.}
	\label{fig: Crystal chunk approximation - n1 (a)}
	\end{subfigure}
    ~ 
	\begin{subfigure}[b]{0.17\textwidth}
	\includegraphics[width=\textwidth]{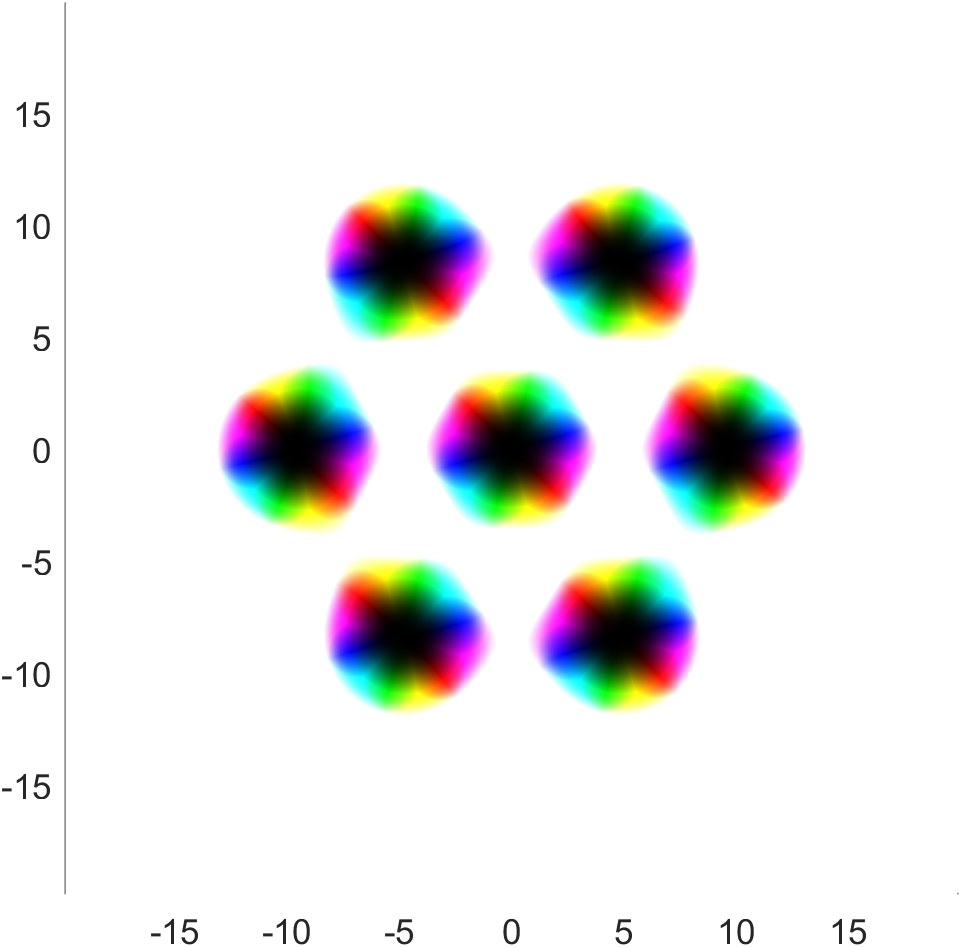}
	\caption{$B=14$ coloring.}
	\label{fig: Crystal chunk approximation - n1 (b)}
	\end{subfigure} \\
	\begin{subfigure}[b]{0.2\textwidth}
	\includegraphics[width=\textwidth]{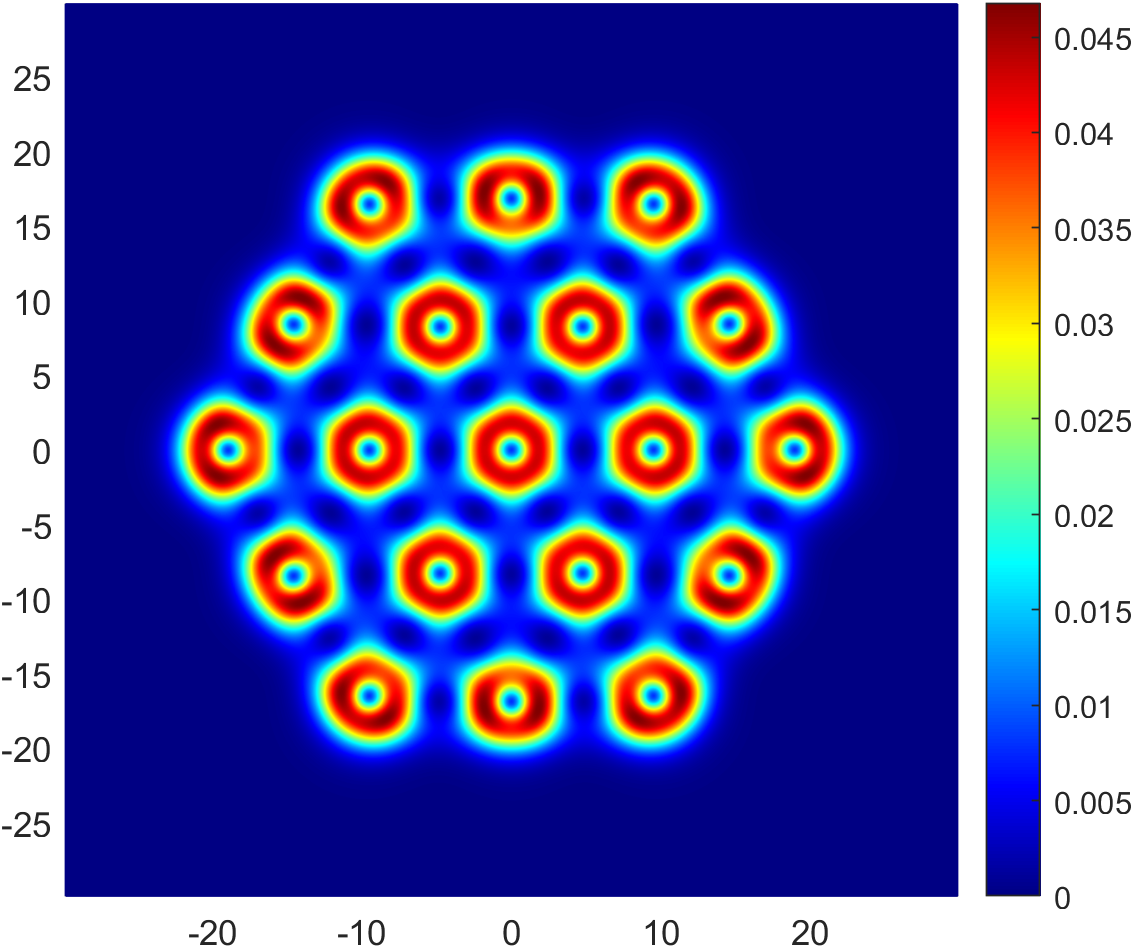}
	\caption{$n=2, B=38$.}
	\label{fig: Crystal chunk approximation - n2 (a)}
	\end{subfigure}
    ~ 
	\begin{subfigure}[b]{0.17\textwidth}
	\includegraphics[width=\textwidth]{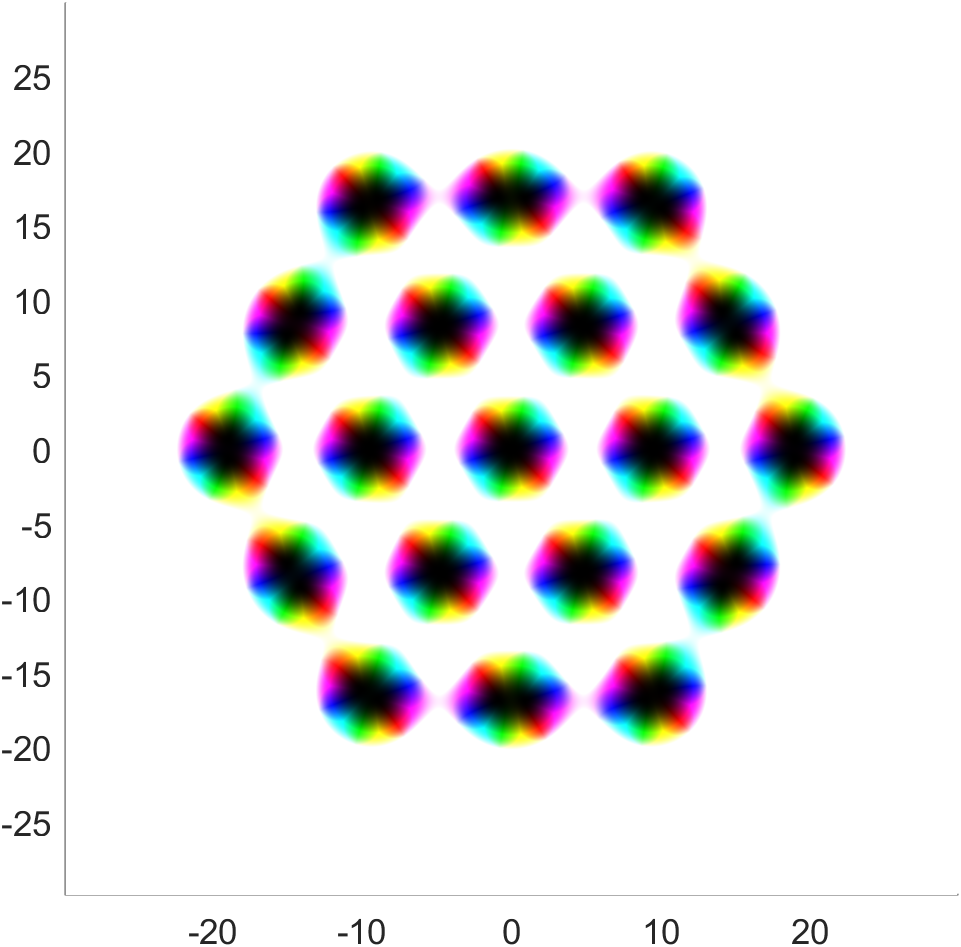}
	\caption{$B=38$ coloring.}
	\label{fig: Crystal chunk approximation - n2 (b)}
	\end{subfigure} \\
	\begin{subfigure}[b]{0.2\textwidth}
	\includegraphics[width=\textwidth]{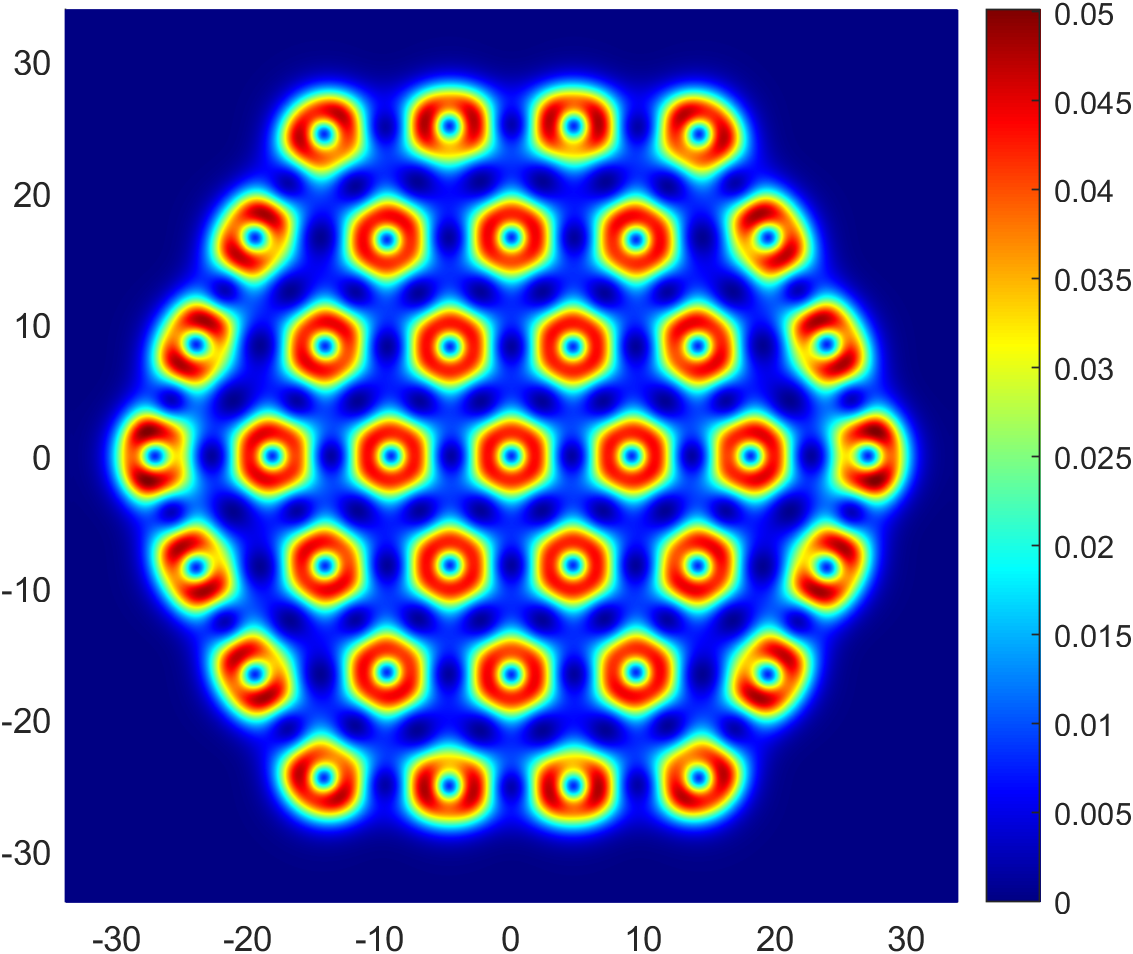}
	\caption{$n=3, B=74$.}
	\label{fig: Crystal chunk approximation - n3 (a)}
	\end{subfigure}
    ~ 
	\begin{subfigure}[b]{0.17\textwidth}
	\includegraphics[width=\textwidth]{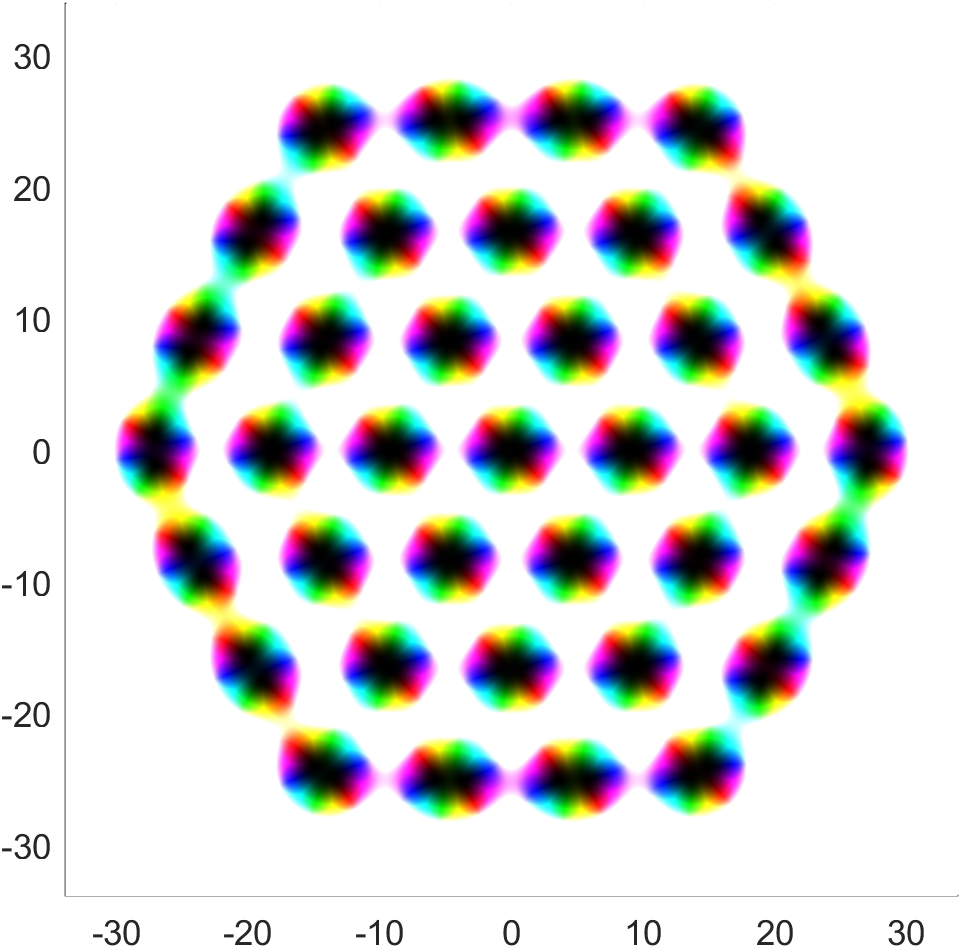}
	\caption{$B=74$ coloring.}
	\label{fig: Crystal chunk approximation - n3 (b)}
	\end{subfigure} \\
	\begin{subfigure}[b]{0.2\textwidth}
	\includegraphics[width=\textwidth]{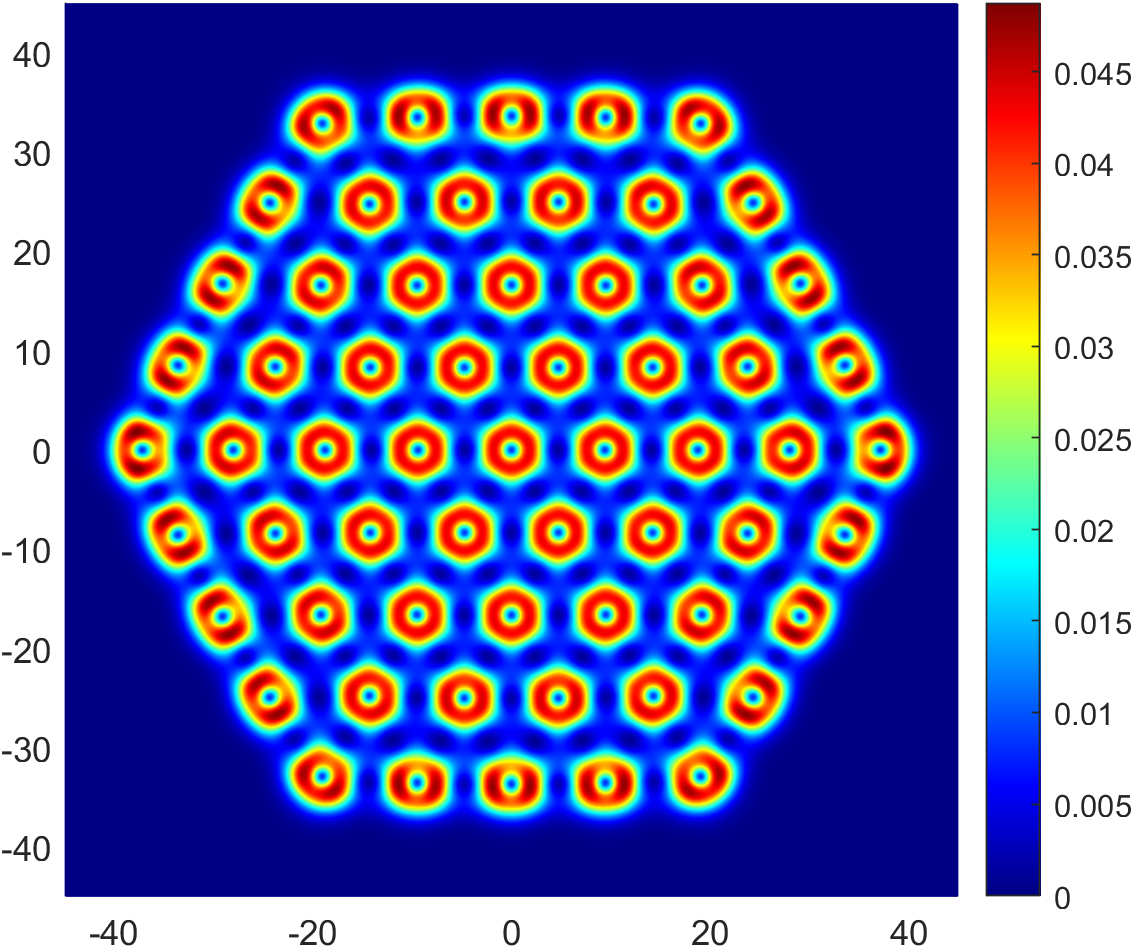}
	\caption{$n=4, B=122$.}
	\label{fig: Crystal chunk approximation - n4 (a)}
	\end{subfigure}
    ~ 
	\begin{subfigure}[b]{0.17\textwidth}
	\includegraphics[width=\textwidth]{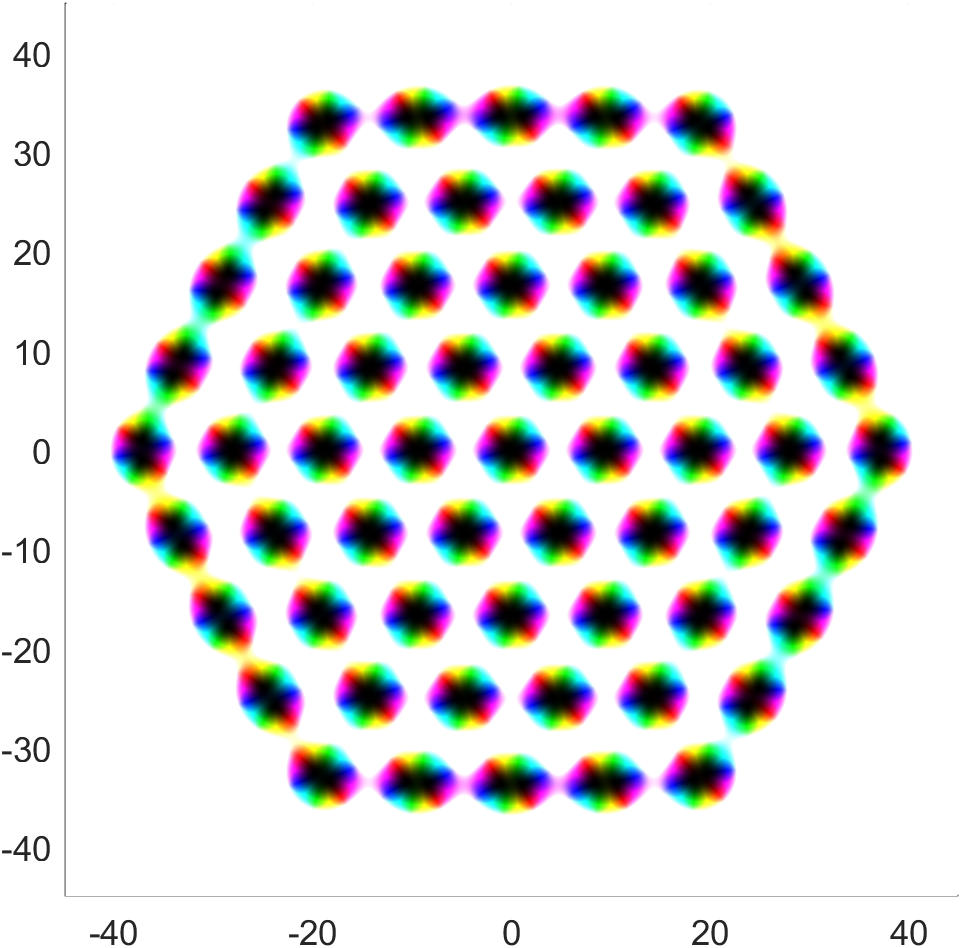}
	\caption{$B=122$ coloring.}
	\label{fig: Crystal chunk approximation - n4 (b)}
	\end{subfigure} \\
	\begin{subfigure}[b]{0.2\textwidth}
	\includegraphics[width=\textwidth]{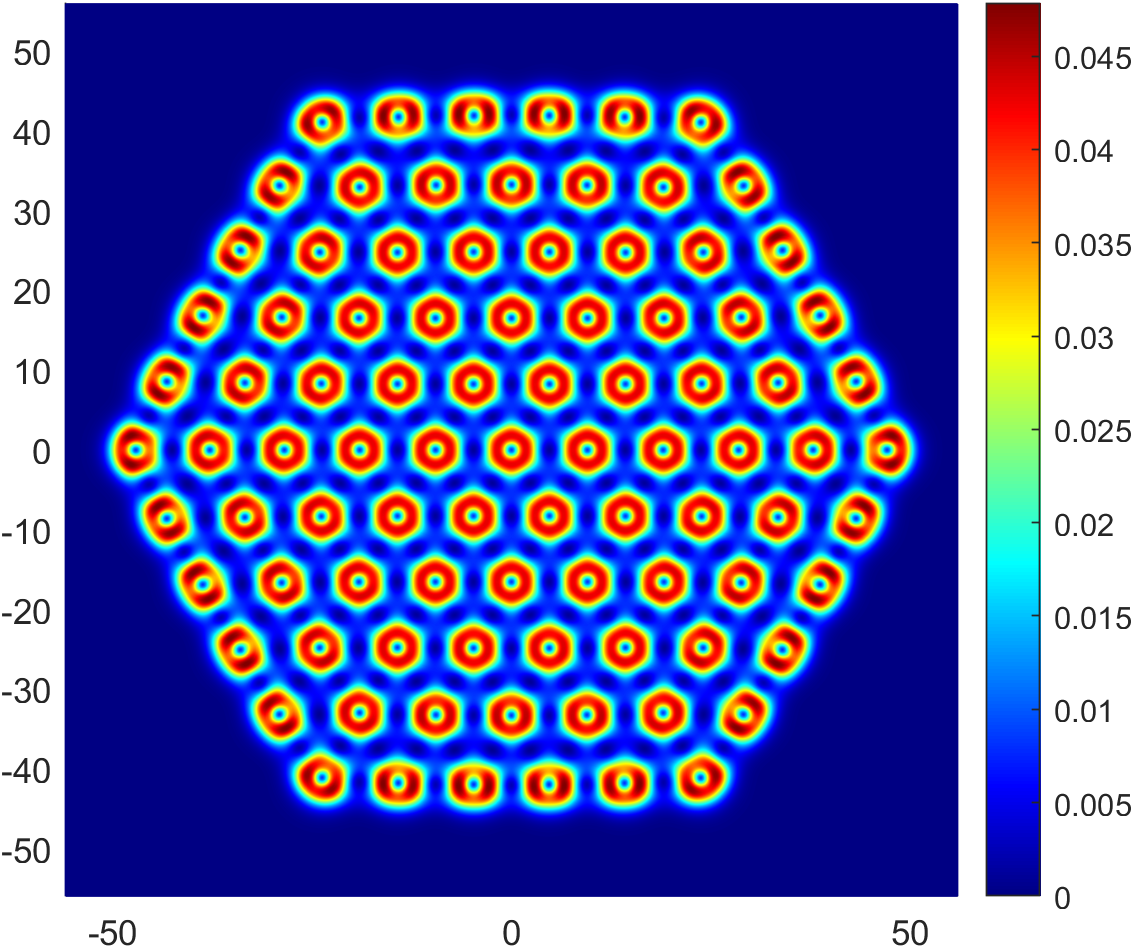}
	\caption{$n=5, B=182$.}
	\label{fig: Crystal chunk approximation - n5 (a)}
	\end{subfigure}
    ~ 
	\begin{subfigure}[b]{0.17\textwidth}
	\includegraphics[width=\textwidth]{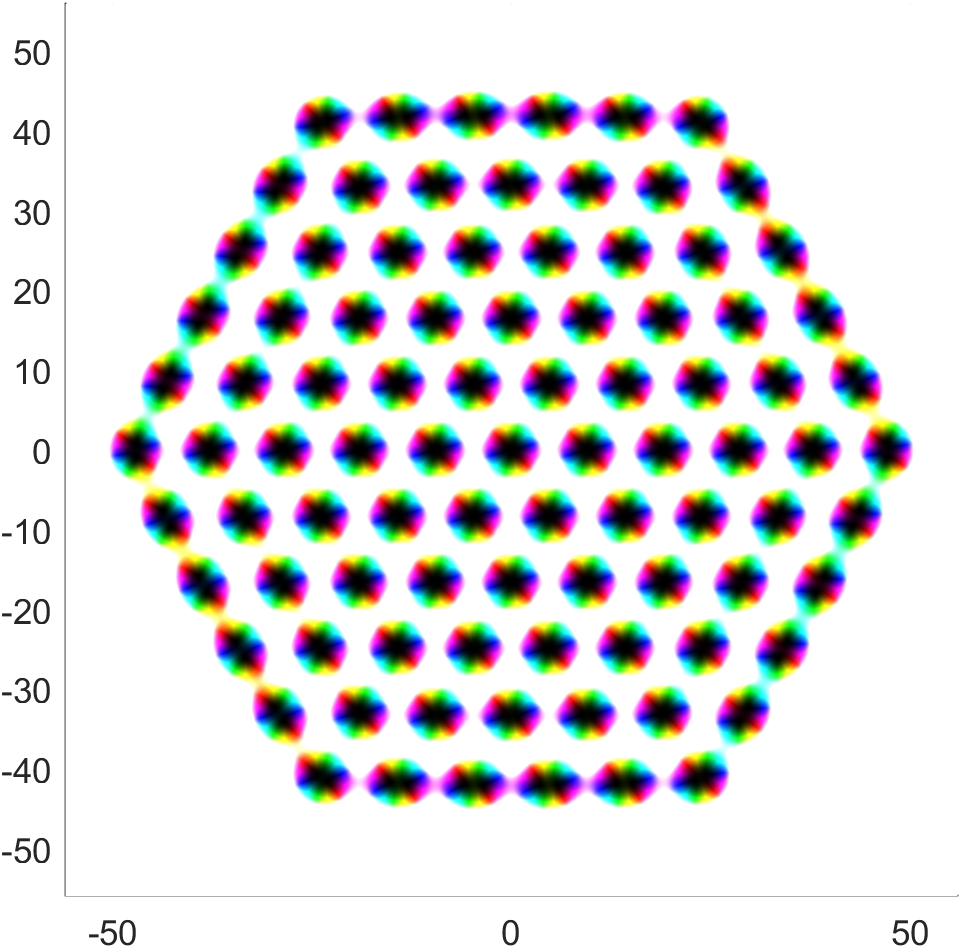}
	\caption{$B=182$ coloring.}
	\label{fig: Crystal chunk approximation - n5 (b)}
	\end{subfigure}
	\caption{Energy density plots of crystal chunk solutions in the standard model and their corresponding coloring on the right hand side.}
	\label{fig: Crystal chunk approximation - Crystals}
\end{figure}

The infinite standard crystal has a discrete $D_6$ symmetry, so we propose that minimal energy chunks of the infinite crystal take the form of layered hexagonal solitons, as can be seen in Fig.~\ref{fig: Crystal chunk approximation - Crystals}.
The number of charge-$2$ units in each layer is precisely $6n$.
As we consider each charge-$2$ baby Skyrme unit to have 6 bonding sides, we can determine the number of free bonds in an $n$-layer crystal chunk to be $N_{\textrm{free}}=(12n+6)$. 
This accounts for the $2$ free bonds on each outer charge-$2$ soliton plus the additional free bond at each vertex of the crystal chunk.
The total charge of a crystal chunk is $B = 2(3(n+1)n+1)$, such that
\begin{equation}
    n = \frac{1}{6}\left( \sqrt{6B-3}-3 \right).
\end{equation}
Thus, we can approximate the normalised energy of a hexagonal standard baby Skyrmion crystal chunk to be given by
\begin{align}
    \mathcal{E}_{\textrm{chunk}} & = \mathcal{E}_{\textrm{crystal}} + \frac{N_{\textrm{free}}}{B} \mathcal{E}_{\textrm{bond}} \nonumber \\
    & = \mathcal{E}_{\textrm{crystal}} + \frac{2\sqrt{6B-3}}{B} \mathcal{E}_{\textrm{bond}}.
\label{eq: Crystal chunk approximation - Empirical chunk}
\end{align}

To determine the transition charge $B_{\textrm{crystal}}$ where chunks of the infinite soliton crystal become the global minima, we need to compare the crystal chunk model~\eqref{eq: Crystal chunk approximation - Empirical chunk} to chain and ring models.
Using the models proposed by Winyard \cite{winyard_2016}, and our numerically determined value for the infinite chain, we are able to approximate ring and chain solutions.
The results are plotted in Fig.~\ref{fig: EA Crystal chunk approximation - Comparison}, which includes data points from crystal chunk solutions for numerous charges $B$ up to $B=2054$.
It can be observed that the crystal chunk approximation \eqref{eq: Crystal chunk approximation - Empirical chunk} fits the data very well.
We find that crystal chunk solutions become global minima for charges $B > B_{\textrm{crystal}} = 954$.
Energy density plots of the Skyrmion crystal chunk solutions are shown in Fig.~\ref{fig: Crystal chunk approximation - Crystals}.
All of these crystal chunks were found numerically on grids with lattice spacing $0.05$, with grid sizes ranging from $800\times 800$ to $2400 \times 2400$.
Crystal chunk solutions for $B=1262$ and $B=2054$ are not shown but were obtained on grids with lattice spacing $0.1$ and grid sizes $3000 \times 3000$ and $3800 \times 3800$, respectively.


\subsection{Easy plane crystal chunks}
\label{subsec: Easy plane crystal chunks}

To predict the energy of an easy plane crystal chunk is somewhat more challenging.
There exists three soliton crystals, all relatively close in energy with one other.
For charges $B$ such that $\sqrt{2B}\in\mathbb{Z}$, the minimal energy crystal chunk is the minimal perimeter $\sqrt{2B} \times \sqrt{2B}$-square of half lumps.
For non-square $2B$, it becomes exceedingly difficult to predict the global minima.
This is because there exists a smorgasbord of local minima for a given charge, which increases with the charge number.

Nevertheless, some progress can be made if we consider rectangular crystal chunks built from the square Skyrmion crystal.
In the first instance, if $2B$ has factors other than $2$ and $B$, say $a\in\mathbb{Z}$ and $2B/a\in\mathbb{Z}$, then the (local) minimal energy crystal chunk will be an $a \times 2B/a$-rectangle of half lumps such that the sum $a+2B/a$ is minimal with respect to the other pairs of possible factors.
We ignore the trivial factors $2B$ and $1$ as the $2B \times 1$-chain is most likely \textit{not} a local minimiser for the easy plane model.
Random initial configurations do not relax to a single linear chain, unlike the standard model.
Even starting with a single chain initial configuration in attractive channel orientations does not result in a relaxed final state of a single chain, it normally relaxes to the double chain.
To find the pair of factors with minimum sum, one would find the factor $a\in\mathbb{Z}$ of $2B$ that minimises the perimeter function $f(a)=2(a+2B/a)$.

\begin{figure}[t]
\centering
\includegraphics[width=0.45\textwidth]{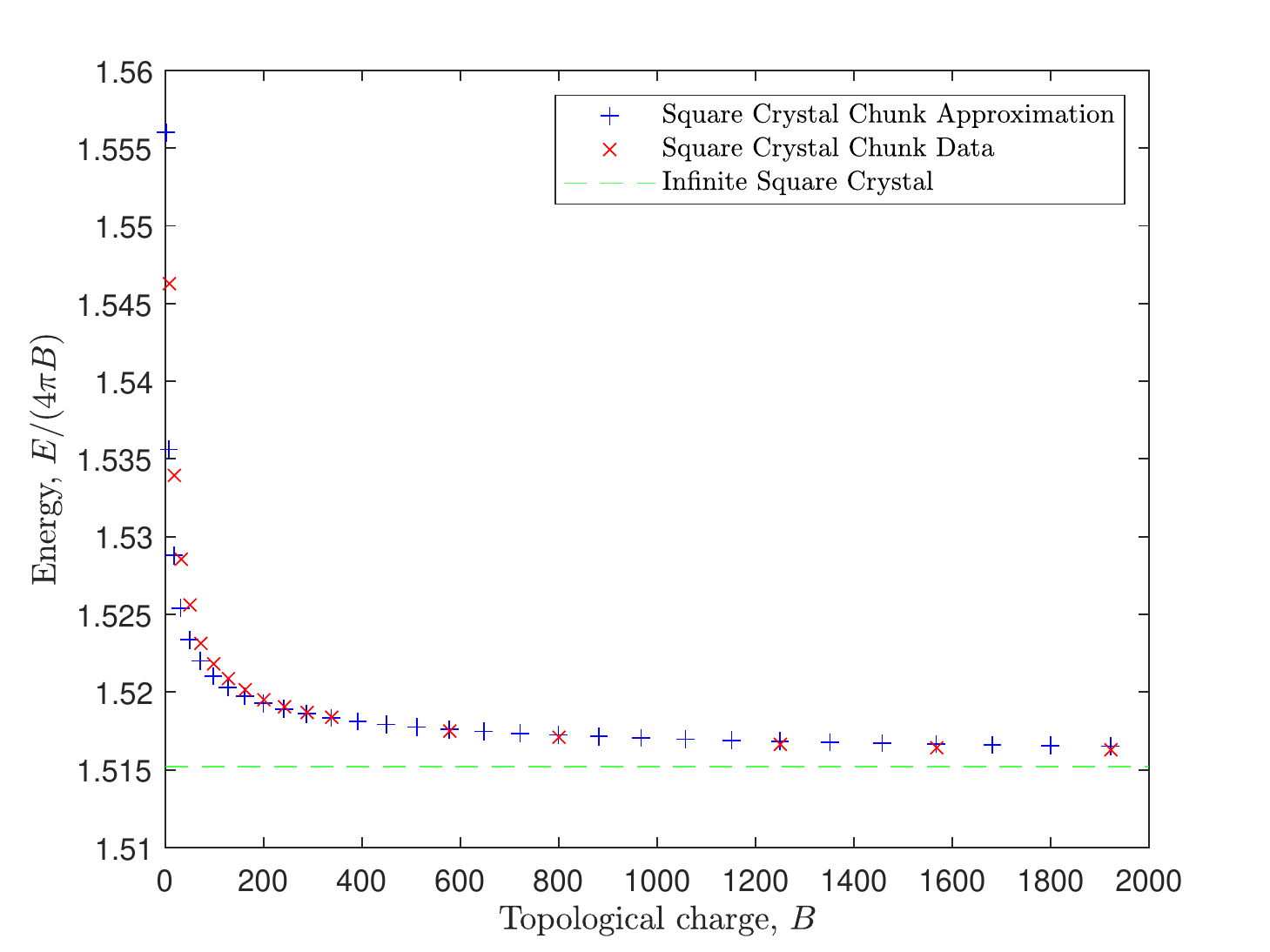}
\caption{Comparison of approximate and true square crystal chunks in the easy plane model.}
\label{fig: EP Crystal chunk approximation - Comparison}
\end{figure}

Using the above information, we are able estimate the energy of a crystal chunk for a given charge $B$.
Similar to the standard model, we split the crystal chunk energy into a bulk term and a surface term.
For charges $B$, we need to determine the pair $(a,2B/a)$ of minimal sum factors of $2B$.
Then we can calculate the surface energy and determining the bulk energy is straightforward.
Explicitly, the normalised energy for a charge-$B$ crystal chunk is given by
\begin{align}
    \mathcal{E}_{\textrm{chunk}} & = \mathcal{E}_{B=2} + \frac{N_{\textrm{free}}}{B} \mathcal{E}_{\textrm{bond}} \nonumber \\
    & = \mathcal{E}_{B=2} + 2 \left(a+\frac{2B}{a}\right) \frac{\mathcal{E}_{\textrm{bond}}}{B}.
\label{eq: Easy plane crystal chunks - Crystal chunk}
\end{align}
This approximation for square crystal chunks ($2B=a^2$) is plotted in Fig.~\ref{fig: EP Crystal chunk approximation - Comparison}, alongside the corresponding true numerically determined (normalised) energies.

Clearly, the further $a$ deviates from $\sqrt{2B}$ the higher the surface energy contribution.
So one would expect there to be normalised energy bands at high charges for this rectangular approximation.
These bands can be determined in the limit $B \rightarrow \infty$, and for such highly rectangular pairs $(a,2B/a)$ the bands in the limit $B \rightarrow \infty$ are given by
\begin{equation}
    \mathcal{E}_{\textrm{bands}}=\mathcal{E}_{B=2}+\frac{4}{a}\mathcal{E}_{\textrm{bond}}.
\label{eq: Easy plane crystal chunks - Energy bands}
\end{equation}

\begin{figure}[t]
	\centering
	\begin{subfigure}[b]{0.2\textwidth}
	\includegraphics[width=\textwidth]{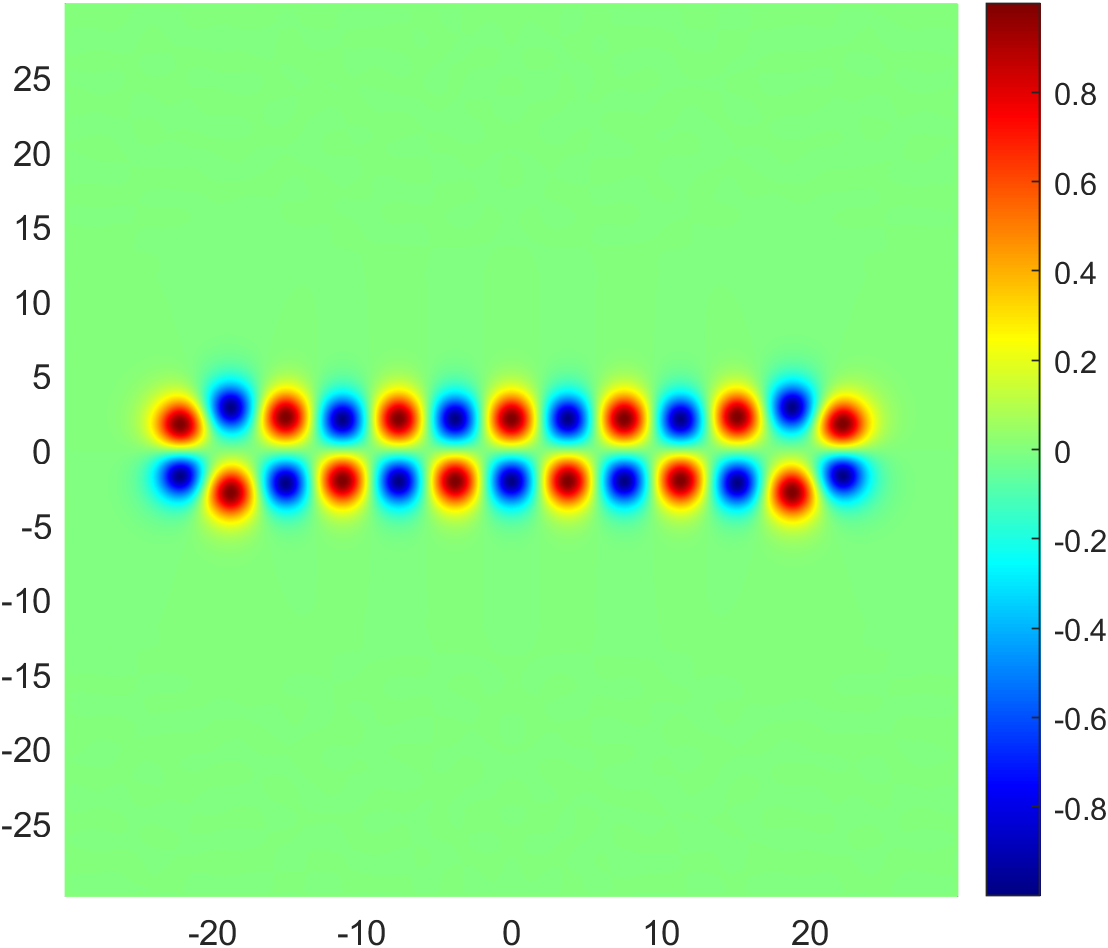}
	\caption{$\mathcal{E}=1.5411$.}
	\label{fig: Easy plane crystal chunks - Chain}
	\end{subfigure}
    ~ 
	\begin{subfigure}[b]{0.2\textwidth}
	\includegraphics[width=\textwidth]{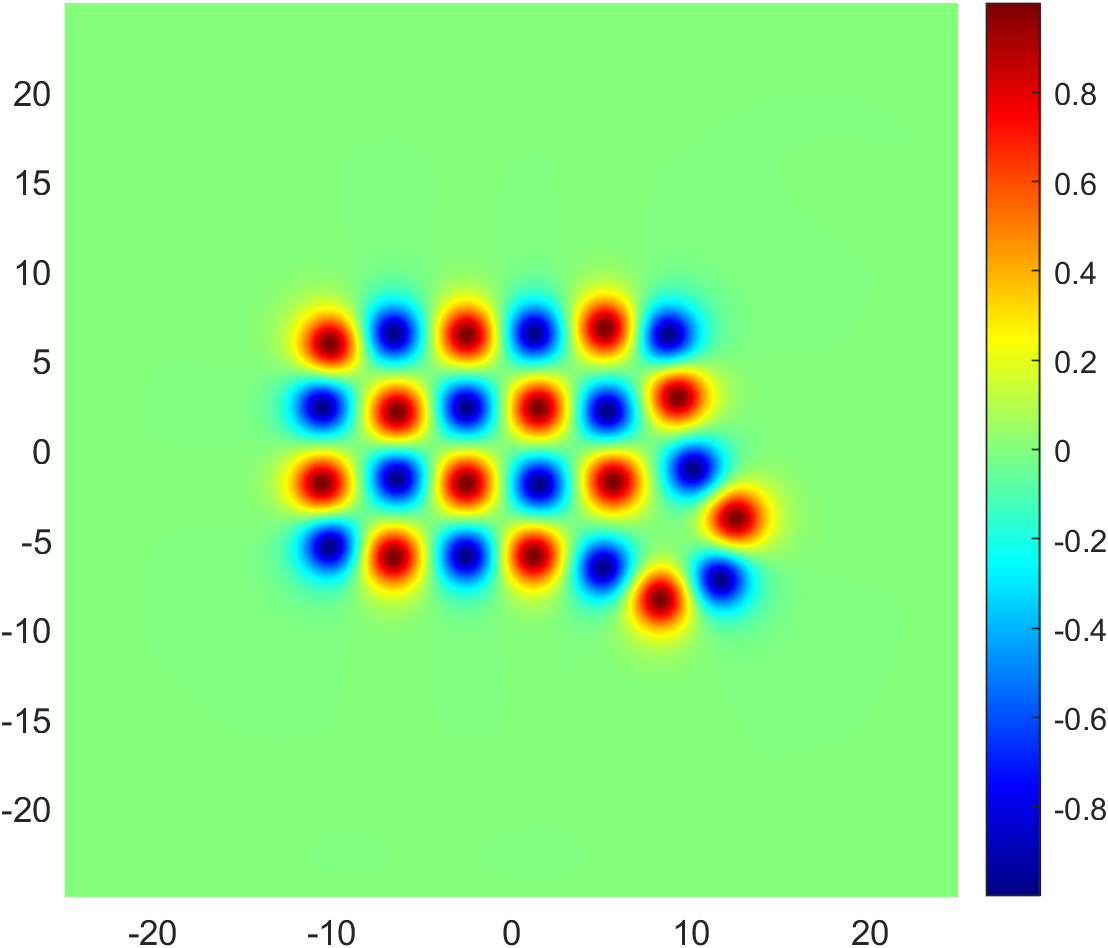}
	\caption{$\mathcal{E}=1.5395^*$.}
	\label{fig: Easy plane crystal chunks - Hole}
	\end{subfigure} \\
	\begin{subfigure}[b]{0.2\textwidth}
	\includegraphics[width=\textwidth]{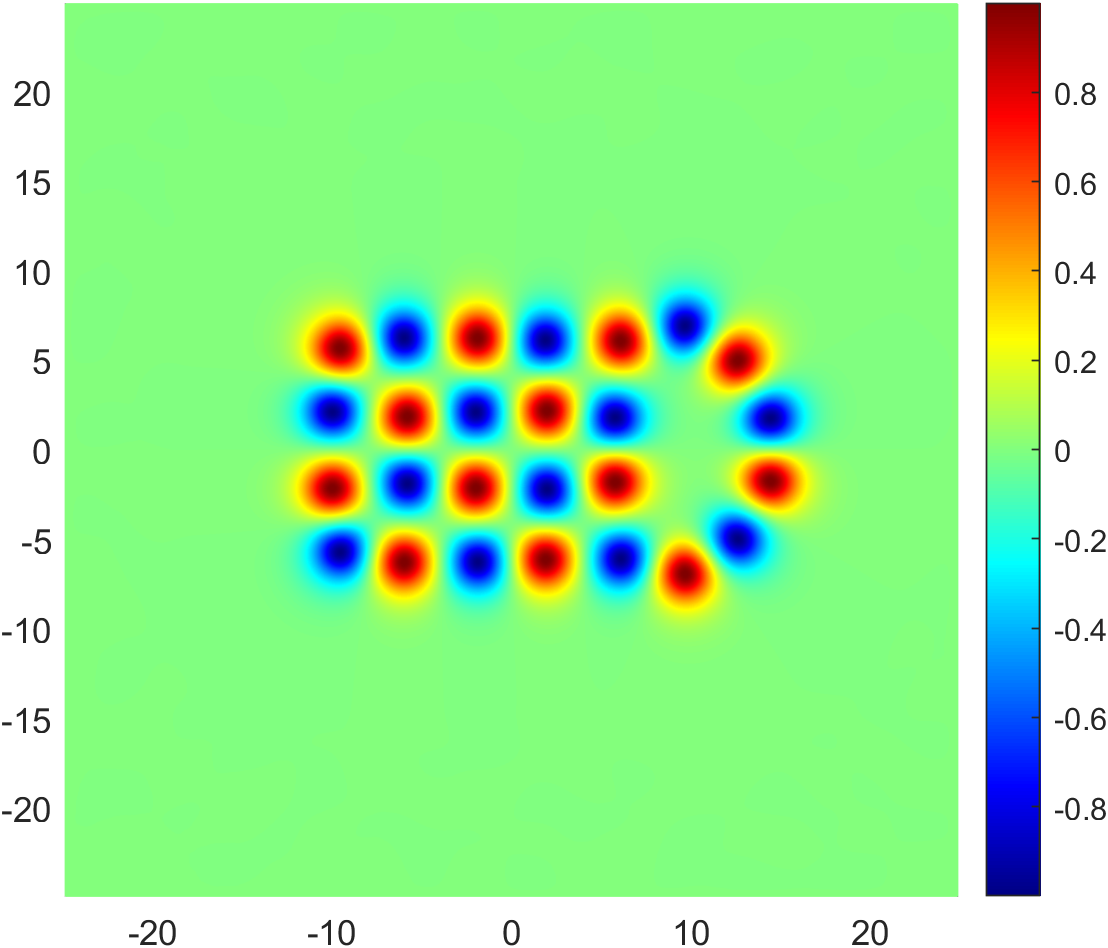}
	\caption{$\mathcal{E}=1.5407$.}
	\label{fig: Easy plane crystal chunks - 10-gon}
	\end{subfigure} 
	\caption{$\varphi^1$ density plots of the three candidates for the crystal chunk solution for a $B=13$ easy plane baby Skyrmion.
	The asterisk (*) indicates the global minimum.}
	\label{fig: Easy plane crystal chunks - Crystal structures}
\end{figure}

Since there are three soliton crystals, there are obviously better crystal chunk solutions for highly rectangular factors $(a,b)$.
As an example, lets consider local minima for the charge-$13$ easy plane soliton.
There are three solutions that one might expect to be contenders for the crystal chunk for this charge.
Firstly, the natural choice is the minimal perimeter rectangle, which will be a double chain, or simply a $2\times 13$-rectangle of half lumps.
This is depicted in Fig.~\ref{fig: Easy plane crystal chunks - Chain}.
The next idea is to consider the minimal perimeter rectangle of half lumps for a $B-1=12$ crystal chunk, then add a half lump pair to one of its corners to create a defect.
This results in a $6 \times 4$-rectangular crystal chunk with one distorted hexagonal corner, as shown in Fig.~\ref{fig: Easy plane crystal chunks - Hole}.
The third contender is akin to the corner cutting method used in the Skyrme model \cite{Manton_2012,Feist_2013}, in which we try to remove half lumps (one blue and one red) from two corners of the minimal perimeter $7 \times 4$-rectangular $B=14$ crystal chunk.
This does not have the intended effect of missing half lumps on corners, rather it pulls the rest of the row, and the adjacent row, away from the chunk to form an arc with two half lumps more than the half lump height of the rectangular chunk.
This can be seen in Fig.~\ref{fig: Easy plane crystal chunks - 10-gon}.
Out of these three most likely crystal chunk candidates for a charge-$13$ soliton, the rectangular crystal chunk with one distorted hexagonal corner is the minimal energy solution.

So for a charge-$13$ baby Skyrmion the minimal perimeter rectangle model fails as a candidate for the global minimal energy crystal chunk in the easy plane model.
So even at charge-$13$ we have already found a lower energy crystal arrangement than the rectangular crystal chunk.
One would expect that adding hexagonal/octagonal defects to nearly square crystal chunks would result in lower energy solutions than rectangular crystal chunks.
However, for square crystal chunks, such that $B\in\sqrt{2B}$, the rectangular crystal chunk model~\eqref{eq: Easy plane crystal chunks - Crystal chunk} is an excellent approximation.

In brief, we conjecture that the prevalent minimal energy crystal chunks for the easy plane model are squares of half lumps if $\sqrt{2B} \in \mathbb{Z}$, otherwise they are minimal perimeter rectangles of half lumps, with some crystal chunks having distorted hexagonal defects.
We have also proposed an empirical model to determine the energy for a given square/rectangular crystal chunk.


\section{Concluding remarks}
\label{sec: Concluding remarks}

In this paper, we have presented a method to determine soliton crystals on an optimised lattice for arbitrary potentials in the baby Skyrme model.
Once the minimal energy soliton crystal is known, the solitons can be layered by use of a crystal slab model and the surface energy per unit length obtained numerically.
Using insight obtained from the soliton crystal and the surface energy, chunks of the Skyrmion crystal can be constructed and their corresponding energies determined.

For the standard potential, we have demonstrated that the global minimum energy Skyrmion crystal exhibits a clear hexagonal $D_6$ symmetry.
This hexagonal soliton crystal has a lower normalised energy than the infinite chain solution proposed by Foster \cite{Foster_2010}.
We propose that the global minima are layered hexagonal crystals for $B>954$ with $m^2=0.1$.

We determined that rows of adjacent infinite chains in attractive orientations have normalised energies close to that of the the infinite hexagonal crystal.
So it is quite possible that concentric ring solutions in attractive orientations could be the global minima for charges $B$ with $B_{\textrm{ring}}<B<B_{\textrm{crystal}}$, this would need to be investigated.
Likewise, Winyward~\cite{winyard_2016} showed that chain solutions could also intersect to form junctions, and proposed that networks of standard baby Skyrmions could be the global minima between rings and crystal chunks.
This too would need to be studied.

Solitons in the easy plane model take the form of configurations of half lumps.
This model has three soliton crystals all relatively close in energy: square, hexagonal and octagonal.
Of these, the square Skyrmion crystal of half lumps is the global minimum.
This is more reminiscent of the three dimensional Skyrme system and, in a manner of respect, a better analogue.
The easy plane model exhibits a plethora of local minima with various different types of symmetries.
We conjecture that, when $2B$ is a perfect square, square crystal chunks are the global minima.
For rectangular $2B$, the minimal energy crystal chunks are, as close to square as possible, rectangular crystal chunks of half lumps with some chunks having hexagonal surface defects.
The study of internal anomalies has not been carried out, so it is possible that the inclusion of a defect into the bulk is more energetically favourable over a surface defect.

\begin{figure}[t]
	\centering
	\includegraphics[width=0.30\textwidth]{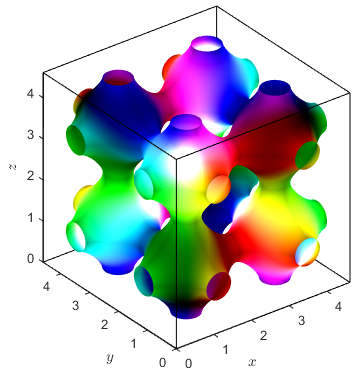}
	\caption{Baryon density plot of the cubic lattice of half Skyrmions in the $3$D Skyrme model.
	The $3$D crystal is colored using the Runge coloring scheme detailed in \cite{Feist_2013}.}
	\label{fig: Concluding remarks - Standard crystal}
\end{figure}
The obvious extension of the work detailed in this paper is to the three dimensional Skyrme model.
The crystal structure has been studied extensively in the literature \cite{Feist_2013,Kugler_1988,Kugler_1989,Manton_1995,Battye_1998,Castillejo_1989,Goldhaber_1987,Jackson_1988}.
All the lattice variations that have been studied were for a cubic lattice of side length $L$, in which only $L$ is varied.
In this case, the ``optimal'' soliton crystal is the cubic arrangement of half Skyrmions \cite{Kugler_1988} as shown in Fig.~\ref{fig: Concluding remarks - Standard crystal}.
It is believed that the simple cubic arrangement of half Skyrmions is the lowest energy Skyrmion crystal.
However, if the Skyrme model and the baby Skyrme model truly are analogous, it is possible that a hexagonal Skyrmion configuration could prevail (or even an entirely different crystalline structure).


\begin{acknowledgments}
The author would like to thank Derek Harland, Martin Speight, Thomas Winyard, Chris Halcrow and Sven Bjarke Gudnason for many useful discussions.
This work was supported by a PhD studentship from UKRI, Grant No. EP/V520081/1.
\end{acknowledgments}

\bibliography{Baby_Skyrme_Crystals}

\begin{thebibliography}{50}%
\makeatletter
\providecommand \@ifxundefined [1]{%
 \@ifx{#1\undefined}
}%
\providecommand \@ifnum [1]{%
 \ifnum #1\expandafter \@firstoftwo
 \else \expandafter \@secondoftwo
 \fi
}%
\providecommand \@ifx [1]{%
 \ifx #1\expandafter \@firstoftwo
 \else \expandafter \@secondoftwo
 \fi
}%
\providecommand \natexlab [1]{#1}%
\providecommand \enquote  [1]{``#1''}%
\providecommand \bibnamefont  [1]{#1}%
\providecommand \bibfnamefont [1]{#1}%
\providecommand \citenamefont [1]{#1}%
\providecommand \href@noop [0]{\@secondoftwo}%
\providecommand \href [0]{\begingroup \@sanitize@url \@href}%
\providecommand \@href[1]{\@@startlink{#1}\@@href}%
\providecommand \@@href[1]{\endgroup#1\@@endlink}%
\providecommand \@sanitize@url [0]{\catcode `\\12\catcode `\$12\catcode
  `\&12\catcode `\#12\catcode `\^12\catcode `\_12\catcode `\%12\relax}%
\providecommand \@@startlink[1]{}%
\providecommand \@@endlink[0]{}%
\providecommand \url  [0]{\begingroup\@sanitize@url \@url }%
\providecommand \@url [1]{\endgroup\@href {#1}{\urlprefix }}%
\providecommand \urlprefix  [0]{URL }%
\providecommand \Eprint [0]{\href }%
\providecommand \doibase [0]{https://doi.org/}%
\providecommand \selectlanguage [0]{\@gobble}%
\providecommand \bibinfo  [0]{\@secondoftwo}%
\providecommand \bibfield  [0]{\@secondoftwo}%
\providecommand \translation [1]{[#1]}%
\providecommand \BibitemOpen [0]{}%
\providecommand \bibitemStop [0]{}%
\providecommand \bibitemNoStop [0]{.\EOS\space}%
\providecommand \EOS [0]{\spacefactor3000\relax}%
\providecommand \BibitemShut  [1]{\csname bibitem#1\endcsname}%
\let\auto@bib@innerbib\@empty
\bibitem [{\citenamefont {Skyrme}(1961)}]{Skyrme_1961}%
  \BibitemOpen
  \bibfield  {author} {\bibinfo {author} {\bibfnamefont {T.~H.~R.}\
  \bibnamefont {Skyrme}},\ }\bibfield  {title} {\bibinfo {title} {A non-linear
  field theory},\ }\href {https://doi.org/10.1098/rspa.1961.0018} {\bibfield
  {journal} {\bibinfo  {journal} {Proc. R. Soc. Lond. A}\ }\textbf {\bibinfo
  {volume} {260}},\ \bibinfo {pages} {127} (\bibinfo {year}
  {1961})}\BibitemShut {NoStop}%
\bibitem [{\citenamefont {Witten}(1983{\natexlab{a}})}]{Witten_1983_1}%
  \BibitemOpen
  \bibfield  {author} {\bibinfo {author} {\bibfnamefont {E.}~\bibnamefont
  {Witten}},\ }\bibfield  {title} {\bibinfo {title} {Global aspects of current
  algebra},\ }\href {https://doi.org/10.1016/0550-3213(83)90063-9} {\bibfield
  {journal} {\bibinfo  {journal} {Nucl. Phys. B}\ }\textbf {\bibinfo {volume}
  {223}},\ \bibinfo {pages} {422} (\bibinfo {year}
  {1983}{\natexlab{a}})}\BibitemShut {NoStop}%
\bibitem [{\citenamefont {Witten}(1983{\natexlab{b}})}]{Witten_1983_2}%
  \BibitemOpen
  \bibfield  {author} {\bibinfo {author} {\bibfnamefont {E.}~\bibnamefont
  {Witten}},\ }\bibfield  {title} {\bibinfo {title} {Current algebra, baryons,
  and quark confinement},\ }\href
  {https://doi.org/10.1016/0550-3213(83)90064-0} {\bibfield  {journal}
  {\bibinfo  {journal} {Nucl. Phys. B}\ }\textbf {\bibinfo {volume} {223}},\
  \bibinfo {pages} {433} (\bibinfo {year} {1983}{\natexlab{b}})}\BibitemShut
  {NoStop}%
\bibitem [{\citenamefont {Sakai}\ and\ \citenamefont
  {Sugimoto}(2005)}]{Sugimoto_2005}%
  \BibitemOpen
  \bibfield  {author} {\bibinfo {author} {\bibfnamefont {T.}~\bibnamefont
  {Sakai}}\ and\ \bibinfo {author} {\bibfnamefont {S.}~\bibnamefont
  {Sugimoto}},\ }\bibfield  {title} {\bibinfo {title} {Low energy hadron
  physics in holographic {QCD}},\ }\href {https://doi.org/10.1143/PTP.113.843}
  {\bibfield  {journal} {\bibinfo  {journal} {Prog. Theor. Phys.}\ }\textbf
  {\bibinfo {volume} {113}},\ \bibinfo {pages} {843} (\bibinfo {year}
  {2005})}\BibitemShut {NoStop}%
\bibitem [{\citenamefont {Ma}\ \emph {et~al.}(2019)\citenamefont {Ma},
  \citenamefont {Halcrow},\ and\ \citenamefont {Zhang}}]{Ma_2019}%
  \BibitemOpen
  \bibfield  {author} {\bibinfo {author} {\bibfnamefont {N.}~\bibnamefont
  {Ma}}, \bibinfo {author} {\bibfnamefont {C.~J.}\ \bibnamefont {Halcrow}},\
  and\ \bibinfo {author} {\bibfnamefont {H.}~\bibnamefont {Zhang}},\ }\bibfield
   {title} {\bibinfo {title} {Effect of the {Coulomb} energy on {Skyrmions}},\
  }\href {https://doi.org/10.1103/PhysRevC.99.044312} {\bibfield  {journal}
  {\bibinfo  {journal} {Phys. Rev. C}\ }\textbf {\bibinfo {volume} {99}},\
  \bibinfo {pages} {044312} (\bibinfo {year} {2019})}\BibitemShut {NoStop}%
\bibitem [{\citenamefont {Kugler}\ and\ \citenamefont
  {Shtrikman}(1988)}]{Kugler_1988}%
  \BibitemOpen
  \bibfield  {author} {\bibinfo {author} {\bibfnamefont {M.}~\bibnamefont
  {Kugler}}\ and\ \bibinfo {author} {\bibfnamefont {S.}~\bibnamefont
  {Shtrikman}},\ }\bibfield  {title} {\bibinfo {title} {A new {Skyrmion}
  crystal},\ }\href
  {https://doi.org/https://doi.org/10.1016/0370-2693(88)90653-3} {\bibfield
  {journal} {\bibinfo  {journal} {Phys. Lett. B}\ }\textbf {\bibinfo {volume}
  {208}},\ \bibinfo {pages} {491} (\bibinfo {year} {1988})}\BibitemShut
  {NoStop}%
\bibitem [{\citenamefont {Piette}\ \emph {et~al.}(1995)\citenamefont {Piette},
  \citenamefont {Schroers},\ and\ \citenamefont {Zakrzewski}}]{Piette_1995}%
  \BibitemOpen
  \bibfield  {author} {\bibinfo {author} {\bibfnamefont {B.}~\bibnamefont
  {Piette}}, \bibinfo {author} {\bibfnamefont {B.}~\bibnamefont {Schroers}},\
  and\ \bibinfo {author} {\bibfnamefont {W.}~\bibnamefont {Zakrzewski}},\
  }\bibfield  {title} {\bibinfo {title} {Dynamics of baby {Skyrmions}},\ }\href
  {https://doi.org/10.1016/0550-3213(95)00011-G} {\bibfield  {journal}
  {\bibinfo  {journal} {Nucl. Phys. B}\ }\textbf {\bibinfo {volume} {439}},\
  \bibinfo {pages} {205} (\bibinfo {year} {1995})}\BibitemShut {NoStop}%
\bibitem [{\citenamefont {Yu}\ \emph {et~al.}(2010)\citenamefont {Yu},
  \citenamefont {Onose}, \citenamefont {Kanazawa}, \citenamefont {Park},
  \citenamefont {Han}, \citenamefont {Matsui}, \citenamefont {Nagaosa},\ and\
  \citenamefont {Tokura}}]{Yu_2010}%
  \BibitemOpen
  \bibfield  {author} {\bibinfo {author} {\bibfnamefont {X.~Z.}\ \bibnamefont
  {Yu}}, \bibinfo {author} {\bibfnamefont {Y.}~\bibnamefont {Onose}}, \bibinfo
  {author} {\bibfnamefont {N.}~\bibnamefont {Kanazawa}}, \bibinfo {author}
  {\bibfnamefont {J.~H.}\ \bibnamefont {Park}}, \bibinfo {author}
  {\bibfnamefont {J.~H.}\ \bibnamefont {Han}}, \bibinfo {author} {\bibfnamefont
  {Y.}~\bibnamefont {Matsui}}, \bibinfo {author} {\bibfnamefont
  {N.}~\bibnamefont {Nagaosa}},\ and\ \bibinfo {author} {\bibfnamefont
  {Y.}~\bibnamefont {Tokura}},\ }\bibfield  {title} {\bibinfo {title}
  {Real-space observation of a two-dimensional {Skyrmion} crystal},\ }\href
  {https://doi.org/10.1038/nature09124} {\bibfield  {journal} {\bibinfo
  {journal} {Nature}\ }\textbf {\bibinfo {volume} {465}},\ \bibinfo {pages}
  {901} (\bibinfo {year} {2010})}\BibitemShut {NoStop}%
\bibitem [{\citenamefont {Kovalev}\ and\ \citenamefont
  {Sandhoefner}(2018)}]{Kovalev_2018}%
  \BibitemOpen
  \bibfield  {author} {\bibinfo {author} {\bibfnamefont {A.~A.}\ \bibnamefont
  {Kovalev}}\ and\ \bibinfo {author} {\bibfnamefont {S.}~\bibnamefont
  {Sandhoefner}},\ }\bibfield  {title} {\bibinfo {title} {Skyrmions and
  anti-{Skyrmions} in quasi-two-dimensional magnets},\ }\href
  {https://doi.org/10.3389/fphy.2018.00098} {\bibfield  {journal} {\bibinfo
  {journal} {Frontiers in Physics}\ }\textbf {\bibinfo {volume} {6}},\ \bibinfo
  {pages} {98} (\bibinfo {year} {2018})}\BibitemShut {NoStop}%
\bibitem [{\citenamefont {Sondhi}\ \emph {et~al.}(1993)\citenamefont {Sondhi},
  \citenamefont {Karlhede}, \citenamefont {Kivelson},\ and\ \citenamefont
  {Rezayi}}]{Sondhi_1993}%
  \BibitemOpen
  \bibfield  {author} {\bibinfo {author} {\bibfnamefont {S.~L.}\ \bibnamefont
  {Sondhi}}, \bibinfo {author} {\bibfnamefont {A.}~\bibnamefont {Karlhede}},
  \bibinfo {author} {\bibfnamefont {S.~A.}\ \bibnamefont {Kivelson}},\ and\
  \bibinfo {author} {\bibfnamefont {E.~H.}\ \bibnamefont {Rezayi}},\ }\bibfield
   {title} {\bibinfo {title} {Skyrmions and the crossover from the integer to
  fractional quantum {Hall} effect at small {Zeeman} energies},\ }\href
  {https://doi.org/10.1103/PhysRevB.47.16419} {\bibfield  {journal} {\bibinfo
  {journal} {Phys. Rev. B}\ }\textbf {\bibinfo {volume} {47}},\ \bibinfo
  {pages} {16419} (\bibinfo {year} {1993})}\BibitemShut {NoStop}%
\bibitem [{\citenamefont {Kuchkin}\ \emph {et~al.}(2021)\citenamefont
  {Kuchkin}, \citenamefont {Chichay}, \citenamefont {Barton-Singer},
  \citenamefont {Rybakov}, \citenamefont {Bl\"ugel}, \citenamefont {Schroers},\
  and\ \citenamefont {Kiselev}}]{Schroers_2021}%
  \BibitemOpen
  \bibfield  {author} {\bibinfo {author} {\bibfnamefont {V.~M.}\ \bibnamefont
  {Kuchkin}}, \bibinfo {author} {\bibfnamefont {K.}~\bibnamefont {Chichay}},
  \bibinfo {author} {\bibfnamefont {B.}~\bibnamefont {Barton-Singer}}, \bibinfo
  {author} {\bibfnamefont {F.~N.}\ \bibnamefont {Rybakov}}, \bibinfo {author}
  {\bibfnamefont {S.}~\bibnamefont {Bl\"ugel}}, \bibinfo {author}
  {\bibfnamefont {B.~J.}\ \bibnamefont {Schroers}},\ and\ \bibinfo {author}
  {\bibfnamefont {N.~S.}\ \bibnamefont {Kiselev}},\ }\bibfield  {title}
  {\bibinfo {title} {Geometry and symmetry in skyrmion dynamics},\ }\href
  {https://doi.org/10.1103/PhysRevB.104.165116} {\bibfield  {journal} {\bibinfo
   {journal} {Phys. Rev. B}\ }\textbf {\bibinfo {volume} {104}},\ \bibinfo
  {pages} {165116} (\bibinfo {year} {2021})}\BibitemShut {NoStop}%
\bibitem [{\citenamefont {Ackerman}\ \emph {et~al.}(2017)\citenamefont
  {Ackerman}, \citenamefont {Boyle},\ and\ \citenamefont
  {Smalyukh}}]{Ackerman_2017}%
  \BibitemOpen
  \bibfield  {author} {\bibinfo {author} {\bibfnamefont {P.~J.}\ \bibnamefont
  {Ackerman}}, \bibinfo {author} {\bibfnamefont {T.}~\bibnamefont {Boyle}},\
  and\ \bibinfo {author} {\bibfnamefont {I.~I.}\ \bibnamefont {Smalyukh}},\
  }\bibfield  {title} {\bibinfo {title} {Squirming motion of baby {Skyrmions}
  in nematic fluids},\ }\href {https://doi.org/10.1038/s41467-017-00659-5}
  {\bibfield  {journal} {\bibinfo  {journal} {Nat Commun}\ }\textbf {\bibinfo
  {volume} {8}},\ \bibinfo {pages} {673} (\bibinfo {year} {2017})}\BibitemShut
  {NoStop}%
\bibitem [{\citenamefont {Hen}\ and\ \citenamefont
  {Karliner}(2008{\natexlab{a}})}]{Hen_2008}%
  \BibitemOpen
  \bibfield  {author} {\bibinfo {author} {\bibfnamefont {I.}~\bibnamefont
  {Hen}}\ and\ \bibinfo {author} {\bibfnamefont {M.}~\bibnamefont {Karliner}},\
  }\bibfield  {title} {\bibinfo {title} {Hexagonal structure of baby {Skyrmion}
  lattices},\ }\href {https://doi.org/10.1103/PhysRevD.77.054009} {\bibfield
  {journal} {\bibinfo  {journal} {Phys. Rev. D}\ }\textbf {\bibinfo {volume}
  {77}},\ \bibinfo {pages} {054009} (\bibinfo {year}
  {2008}{\natexlab{a}})}\BibitemShut {NoStop}%
\bibitem [{\citenamefont {Hen}\ and\ \citenamefont
  {Karliner}(2009)}]{Hen_2009}%
  \BibitemOpen
  \bibfield  {author} {\bibinfo {author} {\bibfnamefont {I.}~\bibnamefont
  {Hen}}\ and\ \bibinfo {author} {\bibfnamefont {M.}~\bibnamefont {Karliner}},\
  }\bibfield  {title} {\bibinfo {title} {Lattice structure of baby
  {Skyrmions}},\ }\href {https://doi.org/10.1007/s11232-009-0083-6} {\bibfield
  {journal} {\bibinfo  {journal} {Theor. Math. Phys.}\ }\textbf {\bibinfo
  {volume} {160}},\ \bibinfo {pages} {933} (\bibinfo {year}
  {2009})}\BibitemShut {NoStop}%
\bibitem [{\citenamefont {Rybakov}\ and\ \citenamefont
  {Kiselev}(2019)}]{Rybakov_2019}%
  \BibitemOpen
  \bibfield  {author} {\bibinfo {author} {\bibfnamefont {F.~N.}\ \bibnamefont
  {Rybakov}}\ and\ \bibinfo {author} {\bibfnamefont {N.~S.}\ \bibnamefont
  {Kiselev}},\ }\bibfield  {title} {\bibinfo {title} {Chiral magnetic
  {Skyrmions} with arbitrary topological charge},\ }\href
  {https://doi.org/10.1103/PhysRevB.99.064437} {\bibfield  {journal} {\bibinfo
  {journal} {Phys. Rev. B}\ }\textbf {\bibinfo {volume} {99}},\ \bibinfo
  {pages} {064437} (\bibinfo {year} {2019})}\BibitemShut {NoStop}%
\bibitem [{\citenamefont {Bogdanov}\ and\ \citenamefont
  {Hubert}(1994)}]{Bogdanov_1994}%
  \BibitemOpen
  \bibfield  {author} {\bibinfo {author} {\bibfnamefont {A.}~\bibnamefont
  {Bogdanov}}\ and\ \bibinfo {author} {\bibfnamefont {A.}~\bibnamefont
  {Hubert}},\ }\bibfield  {title} {\bibinfo {title} {Thermodynamically stable
  magnetic vortex states in magnetic crystals},\ }\href
  {https://doi.org/10.1016/0304-8853(94)90046-9} {\bibfield  {journal}
  {\bibinfo  {journal} {J. Magn. Magn. Mater.}\ }\textbf {\bibinfo {volume}
  {138}},\ \bibinfo {pages} {255} (\bibinfo {year} {1994})}\BibitemShut
  {NoStop}%
\bibitem [{\citenamefont {Kleiner}\ \emph {et~al.}(1964)\citenamefont
  {Kleiner}, \citenamefont {Roth},\ and\ \citenamefont
  {Autler}}]{Kleiner_1964}%
  \BibitemOpen
  \bibfield  {author} {\bibinfo {author} {\bibfnamefont {W.~H.}\ \bibnamefont
  {Kleiner}}, \bibinfo {author} {\bibfnamefont {L.~M.}\ \bibnamefont {Roth}},\
  and\ \bibinfo {author} {\bibfnamefont {S.~H.}\ \bibnamefont {Autler}},\
  }\bibfield  {title} {\bibinfo {title} {Bulk solution of {Ginzburg-Landau}
  equations for type {II} superconductors: upper critical field region},\
  }\href {https://doi.org/10.1103/PhysRev.133.A1226} {\bibfield  {journal}
  {\bibinfo  {journal} {Phys. Rev.}\ }\textbf {\bibinfo {volume} {133}},\
  \bibinfo {pages} {A1226} (\bibinfo {year} {1964})}\BibitemShut {NoStop}%
\bibitem [{\citenamefont {Battye}\ and\ \citenamefont
  {Sutcliffe}(1998)}]{Battye_1998}%
  \BibitemOpen
  \bibfield  {author} {\bibinfo {author} {\bibfnamefont {R.~A.}\ \bibnamefont
  {Battye}}\ and\ \bibinfo {author} {\bibfnamefont {P.}~\bibnamefont
  {Sutcliffe}},\ }\bibfield  {title} {\bibinfo {title} {A {Skyrme} lattice with
  hexagonal symmetry},\ }\href
  {https://doi.org/https://doi.org/10.1016/S0370-2693(97)01196-9} {\bibfield
  {journal} {\bibinfo  {journal} {Phys. Lett. B}\ }\textbf {\bibinfo {volume}
  {416}},\ \bibinfo {pages} {385} (\bibinfo {year} {1998})}\BibitemShut
  {NoStop}%
\bibitem [{\citenamefont {J\"aykk\"a}\ and\ \citenamefont
  {Speight}(2010)}]{Jaykka_2010}%
  \BibitemOpen
  \bibfield  {author} {\bibinfo {author} {\bibfnamefont {J.}~\bibnamefont
  {J\"aykk\"a}}\ and\ \bibinfo {author} {\bibfnamefont {M.}~\bibnamefont
  {Speight}},\ }\bibfield  {title} {\bibinfo {title} {Easy plane baby
  {Skyrmions}},\ }\href {https://doi.org/10.1103/PhysRevD.82.125030} {\bibfield
   {journal} {\bibinfo  {journal} {Phys. Rev. D}\ }\textbf {\bibinfo {volume}
  {82}},\ \bibinfo {pages} {125030} (\bibinfo {year} {2010})}\BibitemShut
  {NoStop}%
\bibitem [{\citenamefont {Kobayashi}\ and\ \citenamefont
  {Nitta}(2013)}]{Nitta_2013}%
  \BibitemOpen
  \bibfield  {author} {\bibinfo {author} {\bibfnamefont {M.}~\bibnamefont
  {Kobayashi}}\ and\ \bibinfo {author} {\bibfnamefont {M.}~\bibnamefont
  {Nitta}},\ }\bibfield  {title} {\bibinfo {title} {Fractional vortex molecules
  and vortex polygons in a baby {Skyrme} model},\ }\href
  {https://doi.org/10.1103/PhysRevD.87.125013} {\bibfield  {journal} {\bibinfo
  {journal} {Phys. Rev. D}\ }\textbf {\bibinfo {volume} {87}},\ \bibinfo
  {pages} {125013} (\bibinfo {year} {2013})}\BibitemShut {NoStop}%
\bibitem [{\citenamefont {Kobayashi}\ and\ \citenamefont
  {Nitta}(2014)}]{Nitta_2014}%
  \BibitemOpen
  \bibfield  {author} {\bibinfo {author} {\bibfnamefont {M.}~\bibnamefont
  {Kobayashi}}\ and\ \bibinfo {author} {\bibfnamefont {M.}~\bibnamefont
  {Nitta}},\ }\bibfield  {title} {\bibinfo {title} {Vortex polygons and their
  stabilities in {Bose-Einstein} condensates and field theory},\ }\href
  {https://doi.org/10.1007/s10909-013-0977-4} {\bibfield  {journal} {\bibinfo
  {journal} {J. Low Temp. Phys.}\ }\textbf {\bibinfo {volume} {175}},\ \bibinfo
  {pages} {208} (\bibinfo {year} {2014})}\BibitemShut {NoStop}%
\bibitem [{\citenamefont {Lin}\ \emph {et~al.}(2015)\citenamefont {Lin},
  \citenamefont {Saxena},\ and\ \citenamefont {Batista}}]{Batista_2015}%
  \BibitemOpen
  \bibfield  {author} {\bibinfo {author} {\bibfnamefont {S.-Z.}\ \bibnamefont
  {Lin}}, \bibinfo {author} {\bibfnamefont {A.}~\bibnamefont {Saxena}},\ and\
  \bibinfo {author} {\bibfnamefont {C.~D.}\ \bibnamefont {Batista}},\
  }\bibfield  {title} {\bibinfo {title} {Skyrmion fractionalization and merons
  in chiral magnets with easy-plane anisotropy},\ }\href
  {https://doi.org/10.1103/PhysRevB.91.224407} {\bibfield  {journal} {\bibinfo
  {journal} {Phys. Rev. B}\ }\textbf {\bibinfo {volume} {91}},\ \bibinfo
  {pages} {224407} (\bibinfo {year} {2015})}\BibitemShut {NoStop}%
\bibitem [{\citenamefont {Baird}\ and\ \citenamefont
  {Wood}(2003)}]{Baird_Wood_2003}%
  \BibitemOpen
  \bibfield  {author} {\bibinfo {author} {\bibfnamefont {P.}~\bibnamefont
  {Baird}}\ and\ \bibinfo {author} {\bibfnamefont {J.~C.}\ \bibnamefont
  {Wood}},\ }\href {https://doi.org/10.1093/acprof:oso/9780198503620.001.0001}
  {\emph {\bibinfo {title} {Harmonic Morphisms Between Riemannian
  Manifolds}}},\ London Mathematical Society monographs\ (\bibinfo  {publisher}
  {Clarendon Press},\ \bibinfo {year} {2003})\BibitemShut {NoStop}%
\bibitem [{\citenamefont {Speight}(2010)}]{Speight_2010}%
  \BibitemOpen
  \bibfield  {author} {\bibinfo {author} {\bibfnamefont {J.~M.}\ \bibnamefont
  {Speight}},\ }\bibfield  {title} {\bibinfo {title} {Compactons and
  semi-compactons in the extreme baby skyrme model},\ }\href
  {https://doi.org/10.1088/1751-8113/43/40/405201} {\bibfield  {journal}
  {\bibinfo  {journal} {J. Phys. A: Math. Theor.}\ }\textbf {\bibinfo {volume}
  {43}},\ \bibinfo {pages} {405201} (\bibinfo {year} {2010})}\BibitemShut
  {NoStop}%
\bibitem [{\citenamefont {Foster}\ and\ \citenamefont
  {Sutcliffe}(2009)}]{Foster_2009}%
  \BibitemOpen
  \bibfield  {author} {\bibinfo {author} {\bibfnamefont {D.}~\bibnamefont
  {Foster}}\ and\ \bibinfo {author} {\bibfnamefont {P.}~\bibnamefont
  {Sutcliffe}},\ }\bibfield  {title} {\bibinfo {title} {Baby {Skyrmions}
  stabilized by vector mesons},\ }\href
  {https://doi.org/10.1103/PhysRevD.79.125026} {\bibfield  {journal} {\bibinfo
  {journal} {Phys. Rev. D}\ }\textbf {\bibinfo {volume} {79}},\ \bibinfo
  {pages} {125026} (\bibinfo {year} {2009})}\BibitemShut {NoStop}%
\bibitem [{\citenamefont {Salmi}\ and\ \citenamefont
  {Sutcliffe}(2015)}]{Salmi_2014}%
  \BibitemOpen
  \bibfield  {author} {\bibinfo {author} {\bibfnamefont {P.}~\bibnamefont
  {Salmi}}\ and\ \bibinfo {author} {\bibfnamefont {P.}~\bibnamefont
  {Sutcliffe}},\ }\bibfield  {title} {\bibinfo {title} {Aloof baby
  {Skyrmions}},\ }\href {https://doi.org/10.1088/1751-8113/48/3/035401}
  {\bibfield  {journal} {\bibinfo  {journal} {J. Phys. A}\ }\textbf {\bibinfo
  {volume} {48}},\ \bibinfo {pages} {035401} (\bibinfo {year}
  {2015})}\BibitemShut {NoStop}%
\bibitem [{\citenamefont {Hen}\ and\ \citenamefont
  {Karliner}(2008{\natexlab{b}})}]{Karliner_2008}%
  \BibitemOpen
  \bibfield  {author} {\bibinfo {author} {\bibfnamefont {I.}~\bibnamefont
  {Hen}}\ and\ \bibinfo {author} {\bibfnamefont {M.}~\bibnamefont {Karliner}},\
  }\bibfield  {title} {\bibinfo {title} {Rotational symmetry breaking in baby
  {Skyrme} models},\ }\href {https://doi.org/10.1088/0951-7715/21/3/002}
  {\bibfield  {journal} {\bibinfo  {journal} {Nonlinearity}\ }\textbf {\bibinfo
  {volume} {21}},\ \bibinfo {pages} {399} (\bibinfo {year}
  {2008}{\natexlab{b}})}\BibitemShut {NoStop}%
\bibitem [{\citenamefont {Speight}\ and\ \citenamefont
  {Winyard}(2020)}]{Winyard_2020}%
  \BibitemOpen
  \bibfield  {author} {\bibinfo {author} {\bibfnamefont {J.~M.}\ \bibnamefont
  {Speight}}\ and\ \bibinfo {author} {\bibfnamefont {T.}~\bibnamefont
  {Winyard}},\ }\bibfield  {title} {\bibinfo {title} {Skyrmions and spin waves
  in frustrated ferromagnets at low applied magnetic field},\ }\href
  {https://doi.org/10.1103/PhysRevB.101.134420} {\bibfield  {journal} {\bibinfo
   {journal} {Phys. Rev. B}\ }\textbf {\bibinfo {volume} {101}},\ \bibinfo
  {pages} {134420} (\bibinfo {year} {2020})}\BibitemShut {NoStop}%
\bibitem [{\citenamefont {Speight}\ and\ \citenamefont
  {Winyard}(2021)}]{Speight_2021}%
  \BibitemOpen
  \bibfield  {author} {\bibinfo {author} {\bibfnamefont {M.}~\bibnamefont
  {Speight}}\ and\ \bibinfo {author} {\bibfnamefont {T.}~\bibnamefont
  {Winyard}},\ }\bibfield  {title} {\bibinfo {title} {Intervortex forces in
  competing-order superconductors},\ }\href
  {https://doi.org/10.1103/PhysRevB.103.014514} {\bibfield  {journal} {\bibinfo
   {journal} {Phys. Rev. B}\ }\textbf {\bibinfo {volume} {103}},\ \bibinfo
  {pages} {014514} (\bibinfo {year} {2021})}\BibitemShut {NoStop}%
\bibitem [{\citenamefont {Gudnason}\ and\ \citenamefont
  {Speight}(2020)}]{Gudnason_2020}%
  \BibitemOpen
  \bibfield  {author} {\bibinfo {author} {\bibfnamefont {S.~B.}\ \bibnamefont
  {Gudnason}}\ and\ \bibinfo {author} {\bibfnamefont {J.~M.}\ \bibnamefont
  {Speight}},\ }\bibfield  {title} {\bibinfo {title} {Realistic classical
  binding energies in the $\omega$-{Skyrme} model},\ }\href
  {https://doi.org/https://doi.org/10.1007/JHEP07(2020)184} {\bibfield
  {journal} {\bibinfo  {journal} {J. High Energ. Phys.}\ }\textbf {\bibinfo
  {volume} {2020}},\ \bibinfo {pages} {184}}\BibitemShut {NoStop}%
\bibitem [{\citenamefont {Leese}\ \emph {et~al.}(1990)\citenamefont {Leese},
  \citenamefont {Peyrard},\ and\ \citenamefont {Zakrzewski}}]{Leese_1990}%
  \BibitemOpen
  \bibfield  {author} {\bibinfo {author} {\bibfnamefont {R.~A.}\ \bibnamefont
  {Leese}}, \bibinfo {author} {\bibfnamefont {M.}~\bibnamefont {Peyrard}},\
  and\ \bibinfo {author} {\bibfnamefont {W.~J.}\ \bibnamefont {Zakrzewski}},\
  }\bibfield  {title} {\bibinfo {title} {Soliton scatterings in some
  relativistic models in (2+1) dimensions},\ }\href
  {https://doi.org/10.1088/0951-7715/3/3/011} {\bibfield  {journal} {\bibinfo
  {journal} {Nonlinearity}\ }\textbf {\bibinfo {volume} {3}},\ \bibinfo {pages}
  {773} (\bibinfo {year} {1990})}\BibitemShut {NoStop}%
\bibitem [{\citenamefont {J\"aykk\"a}\ \emph {et~al.}(2012)\citenamefont
  {J\"aykk\"a}, \citenamefont {Speight},\ and\ \citenamefont
  {Sutcliffe}}]{Jaykka_2012}%
  \BibitemOpen
  \bibfield  {author} {\bibinfo {author} {\bibfnamefont {J.}~\bibnamefont
  {J\"aykk\"a}}, \bibinfo {author} {\bibfnamefont {M.}~\bibnamefont
  {Speight}},\ and\ \bibinfo {author} {\bibfnamefont {P.}~\bibnamefont
  {Sutcliffe}},\ }\bibfield  {title} {\bibinfo {title} {Broken baby
  {Skyrmions}},\ }\href {https://doi.org/10.1098/rspa.2011.0543} {\bibfield
  {journal} {\bibinfo  {journal} {Proc. R. Soc. A.}\ }\textbf {\bibinfo
  {volume} {468}},\ \bibinfo {pages} {1085} (\bibinfo {year}
  {2012})}\BibitemShut {NoStop}%
\bibitem [{\citenamefont {Jennings}\ and\ \citenamefont
  {Winyard}(2014)}]{Winyard_2013}%
  \BibitemOpen
  \bibfield  {author} {\bibinfo {author} {\bibfnamefont {P.}~\bibnamefont
  {Jennings}}\ and\ \bibinfo {author} {\bibfnamefont {T.}~\bibnamefont
  {Winyard}},\ }\bibfield  {title} {\bibinfo {title} {Broken planar {Skyrmions}
  -- statics and dynamics},\ }\href {https://doi.org/10.1007/JHEP01(2014)122}
  {\bibfield  {journal} {\bibinfo  {journal} {J. High Energ. Phys.}\ }\textbf
  {\bibinfo {volume} {2014}},\ \bibinfo {pages} {122}}\BibitemShut {NoStop}%
\bibitem [{\citenamefont {Ward}(2004)}]{Ward_2004}%
  \BibitemOpen
  \bibfield  {author} {\bibinfo {author} {\bibfnamefont {R.}~\bibnamefont
  {Ward}},\ }\bibfield  {title} {\bibinfo {title} {Planar {Skyrmions} at high
  and low density},\ }\href {https://doi.org/10.1088/0951-7715/17/3/014}
  {\bibfield  {journal} {\bibinfo  {journal} {Nonlinearity}\ }\textbf {\bibinfo
  {volume} {17}},\ \bibinfo {pages} {1033} (\bibinfo {year}
  {2004})}\BibitemShut {NoStop}%
\bibitem [{\citenamefont {Weidig}(1999)}]{Weidig_1999}%
  \BibitemOpen
  \bibfield  {author} {\bibinfo {author} {\bibfnamefont {T.}~\bibnamefont
  {Weidig}},\ }\bibfield  {title} {\bibinfo {title} {The baby skyrme models and
  their multi-skyrmions},\ }\href {https://doi.org/10.1088/0951-7715/12/6/303}
  {\bibfield  {journal} {\bibinfo  {journal} {Nonlinearity}\ }\textbf {\bibinfo
  {volume} {12}},\ \bibinfo {pages} {1489} (\bibinfo {year}
  {1999})}\BibitemShut {NoStop}%
\bibitem [{\citenamefont {Halcrow}(2020)}]{Halcrow_2020}%
  \BibitemOpen
  \bibfield  {author} {\bibinfo {author} {\bibfnamefont {C.}~\bibnamefont
  {Halcrow}},\ }\bibfield  {title} {\bibinfo {title} {Quantum soliton
  scattering manifolds},\ }\href {https://doi.org/10.1007/JHEP07(2020)182}
  {\bibfield  {journal} {\bibinfo  {journal} {J. High Energ. Phys.}\ }\textbf
  {\bibinfo {volume} {2020}},\ \bibinfo {pages} {182}}\BibitemShut {NoStop}%
\bibitem [{\citenamefont {Gudnason}\ \emph {et~al.}(2020)\citenamefont
  {Gudnason}, \citenamefont {Barsanti},\ and\ \citenamefont
  {Bolognesi}}]{Barsanti_2020}%
  \BibitemOpen
  \bibfield  {author} {\bibinfo {author} {\bibfnamefont {S.~B.}\ \bibnamefont
  {Gudnason}}, \bibinfo {author} {\bibfnamefont {B.}~\bibnamefont {Barsanti}},\
  and\ \bibinfo {author} {\bibfnamefont {S.}~\bibnamefont {Bolognesi}},\
  }\bibfield  {title} {\bibinfo {title} {Near-{BPS} baby {Skyrmions}},\ }\href
  {https://doi.org/10.1007/JHEP11(2020)062} {\bibfield  {journal} {\bibinfo
  {journal} {J. High Energ. Phys.}\ }\textbf {\bibinfo {volume} {2020}},\
  \bibinfo {pages} {62}}\BibitemShut {NoStop}%
\bibitem [{\citenamefont {Harland}\ and\ \citenamefont
  {Ward}(2008)}]{Harland_2008}%
  \BibitemOpen
  \bibfield  {author} {\bibinfo {author} {\bibfnamefont {D.}~\bibnamefont
  {Harland}}\ and\ \bibinfo {author} {\bibfnamefont {R.~S.}\ \bibnamefont
  {Ward}},\ }\bibfield  {title} {\bibinfo {title} {Walls and chains of planar
  {Skyrmions}},\ }\href {https://doi.org/10.1103/PhysRevD.77.045009} {\bibfield
   {journal} {\bibinfo  {journal} {Phys. Rev. D}\ }\textbf {\bibinfo {volume}
  {77}},\ \bibinfo {pages} {045009} (\bibinfo {year} {2008})}\BibitemShut
  {NoStop}%
\bibitem [{\citenamefont {Foster}(2010)}]{Foster_2010}%
  \BibitemOpen
  \bibfield  {author} {\bibinfo {author} {\bibfnamefont {D.}~\bibnamefont
  {Foster}},\ }\bibfield  {title} {\bibinfo {title} {Baby {Skyrmion} chains},\
  }\href {https://doi.org/10.1088/0951-7715/23/3/001} {\bibfield  {journal}
  {\bibinfo  {journal} {Nonlinearity}\ }\textbf {\bibinfo {volume} {23}},\
  \bibinfo {pages} {465} (\bibinfo {year} {2010})}\BibitemShut {NoStop}%
\bibitem [{\citenamefont {Shnir}(2021)}]{Shnir_2021}%
  \BibitemOpen
  \bibfield  {author} {\bibinfo {author} {\bibfnamefont {Y.~M.}\ \bibnamefont
  {Shnir}},\ }\bibfield  {title} {\bibinfo {title} {Chains of interacting
  solitons},\ }\href {https://doi.org/10.3390/sym13020284} {\bibfield
  {journal} {\bibinfo  {journal} {Symmetry}\ }\textbf {\bibinfo {volume}
  {284}},\ \bibinfo {pages} {13} (\bibinfo {year} {2021})}\BibitemShut
  {NoStop}%
\bibitem [{\citenamefont {Winyard}(2016)}]{winyard_2016}%
  \BibitemOpen
  \bibfield  {author} {\bibinfo {author} {\bibfnamefont {T.~S.}\ \bibnamefont
  {Winyard}},\ }\emph {\bibinfo {title} {The Skyrme Model: Curved Space,
  Symmetries and Mass}},\ \href@noop {} {Ph.D. thesis},\ \bibinfo  {school}
  {Durham Theses} (\bibinfo {year} {2016})\BibitemShut {NoStop}%
\bibitem [{\citenamefont {Speight}(2014)}]{Speight_2014}%
  \BibitemOpen
  \bibfield  {author} {\bibinfo {author} {\bibfnamefont {J.~M.}\ \bibnamefont
  {Speight}},\ }\bibfield  {title} {\bibinfo {title} {Solitons on tori and
  soliton crystals},\ }\href {https://doi.org/10.1007/s00220-014-2104-z}
  {\bibfield  {journal} {\bibinfo  {journal} {Comm. Math. Phys.}\ }\textbf
  {\bibinfo {volume} {332}},\ \bibinfo {pages} {355} (\bibinfo {year}
  {2014})}\BibitemShut {NoStop}%
\bibitem [{\citenamefont {Bamberg}\ \emph {et~al.}(2003)\citenamefont
  {Bamberg}, \citenamefont {Cairns},\ and\ \citenamefont
  {Kilminster}}]{Bamberg_2003}%
  \BibitemOpen
  \bibfield  {author} {\bibinfo {author} {\bibfnamefont {J.}~\bibnamefont
  {Bamberg}}, \bibinfo {author} {\bibfnamefont {G.}~\bibnamefont {Cairns}},\
  and\ \bibinfo {author} {\bibfnamefont {D.}~\bibnamefont {Kilminster}},\
  }\bibfield  {title} {\bibinfo {title} {The crystallographic restriction,
  permutations, and {Goldbach's} conjecture},\ }\href
  {https://doi.org/10.2307/3647934} {\bibfield  {journal} {\bibinfo  {journal}
  {Am. Math. Mon.}\ }\textbf {\bibinfo {volume} {110}},\ \bibinfo {pages} {202}
  (\bibinfo {year} {2003})}\BibitemShut {NoStop}%
\bibitem [{\citenamefont {Manton}(2012)}]{Manton_2012}%
  \BibitemOpen
  \bibfield  {author} {\bibinfo {author} {\bibfnamefont {N.~S.}\ \bibnamefont
  {Manton}},\ }\bibfield  {title} {\bibinfo {title} {Classical {Skyrmions} -
  static solutions and dynamics},\ }\href {https://doi.org/10.1002/mma.2512}
  {\bibfield  {journal} {\bibinfo  {journal} {Math. Meth. Appl. Sci.}\ }\textbf
  {\bibinfo {volume} {35}},\ \bibinfo {pages} {1188} (\bibinfo {year}
  {2012})}\BibitemShut {NoStop}%
\bibitem [{\citenamefont {Feist}\ \emph {et~al.}(2013)\citenamefont {Feist},
  \citenamefont {Lau},\ and\ \citenamefont {Manton}}]{Feist_2013}%
  \BibitemOpen
  \bibfield  {author} {\bibinfo {author} {\bibfnamefont {D.~T.~J.}\
  \bibnamefont {Feist}}, \bibinfo {author} {\bibfnamefont {P.~H.~C.}\
  \bibnamefont {Lau}},\ and\ \bibinfo {author} {\bibfnamefont {N.~S.}\
  \bibnamefont {Manton}},\ }\bibfield  {title} {\bibinfo {title} {Skyrmions up
  to baryon number 108},\ }\href {https://doi.org/10.1103/PhysRevD.87.085034}
  {\bibfield  {journal} {\bibinfo  {journal} {Phys. Rev. D}\ }\textbf {\bibinfo
  {volume} {87}},\ \bibinfo {pages} {085034} (\bibinfo {year}
  {2013})}\BibitemShut {NoStop}%
\bibitem [{\citenamefont {Kugler}\ and\ \citenamefont
  {Shtrikman}(1989)}]{Kugler_1989}%
  \BibitemOpen
  \bibfield  {author} {\bibinfo {author} {\bibfnamefont {M.}~\bibnamefont
  {Kugler}}\ and\ \bibinfo {author} {\bibfnamefont {S.}~\bibnamefont
  {Shtrikman}},\ }\bibfield  {title} {\bibinfo {title} {Skyrmion crystals and
  their symmetries},\ }\href {https://doi.org/10.1103/PhysRevD.40.3421}
  {\bibfield  {journal} {\bibinfo  {journal} {Phys. Rev. D}\ }\textbf {\bibinfo
  {volume} {40}},\ \bibinfo {pages} {3421} (\bibinfo {year}
  {1989})}\BibitemShut {NoStop}%
\bibitem [{\citenamefont {Manton}\ and\ \citenamefont
  {Sutcliffe}(1995)}]{Manton_1995}%
  \BibitemOpen
  \bibfield  {author} {\bibinfo {author} {\bibfnamefont {N.~S.}\ \bibnamefont
  {Manton}}\ and\ \bibinfo {author} {\bibfnamefont {P.~M.}\ \bibnamefont
  {Sutcliffe}},\ }\bibfield  {title} {\bibinfo {title} {Skyrme crystal from a
  twisted instanton on a four-torus},\ }\href
  {https://doi.org/10.1016/0370-2693(94)01375-M} {\bibfield  {journal}
  {\bibinfo  {journal} {Phys. Lett. B}\ }\textbf {\bibinfo {volume} {342}},\
  \bibinfo {pages} {196} (\bibinfo {year} {1995})}\BibitemShut {NoStop}%
\bibitem [{\citenamefont {Castillejo}\ \emph {et~al.}(1989)\citenamefont
  {Castillejo}, \citenamefont {Jones}, \citenamefont {Jackson}, \citenamefont
  {Verbaarschot},\ and\ \citenamefont {Jackson}}]{Castillejo_1989}%
  \BibitemOpen
  \bibfield  {author} {\bibinfo {author} {\bibfnamefont {L.}~\bibnamefont
  {Castillejo}}, \bibinfo {author} {\bibfnamefont {P.~S.~J.}\ \bibnamefont
  {Jones}}, \bibinfo {author} {\bibfnamefont {A.~D.}\ \bibnamefont {Jackson}},
  \bibinfo {author} {\bibfnamefont {J.~J.~M.}\ \bibnamefont {Verbaarschot}},\
  and\ \bibinfo {author} {\bibfnamefont {A.}~\bibnamefont {Jackson}},\
  }\bibfield  {title} {\bibinfo {title} {Dense {Skyrmion} systems},\ }\href
  {https://doi.org/https://doi.org/10.1016/0375-9474(89)90161-9} {\bibfield
  {journal} {\bibinfo  {journal} {Nucl. Phys. A}\ }\textbf {\bibinfo {volume}
  {501}},\ \bibinfo {pages} {801} (\bibinfo {year} {1989})}\BibitemShut
  {NoStop}%
\bibitem [{\citenamefont {Goldhaber}\ and\ \citenamefont
  {Manton}(1987)}]{Goldhaber_1987}%
  \BibitemOpen
  \bibfield  {author} {\bibinfo {author} {\bibfnamefont {A.~S.}\ \bibnamefont
  {Goldhaber}}\ and\ \bibinfo {author} {\bibfnamefont {N.~S.}\ \bibnamefont
  {Manton}},\ }\bibfield  {title} {\bibinfo {title} {Maximal symmetry of the
  {Skyrme} crystal},\ }\href {https://doi.org/10.1016/0370-2693(87)91502-4}
  {\bibfield  {journal} {\bibinfo  {journal} {Phys. Lett. B}\ }\textbf
  {\bibinfo {volume} {198}},\ \bibinfo {pages} {231} (\bibinfo {year}
  {1987})}\BibitemShut {NoStop}%
\bibitem [{\citenamefont {Jackson}\ and\ \citenamefont
  {Verbaarschot}(1988)}]{Jackson_1988}%
  \BibitemOpen
  \bibfield  {author} {\bibinfo {author} {\bibfnamefont {A.~D.}\ \bibnamefont
  {Jackson}}\ and\ \bibinfo {author} {\bibfnamefont {J.~J.~M.}\ \bibnamefont
  {Verbaarschot}},\ }\bibfield  {title} {\bibinfo {title} {Phase structure of
  the {Skyrme} model},\ }\href {https://doi.org/10.1016/0375-9474(88)90302-8}
  {\bibfield  {journal} {\bibinfo  {journal} {Nucl. Phys. A}\ }\textbf
  {\bibinfo {volume} {484}},\ \bibinfo {pages} {419} (\bibinfo {year}
  {1988})}\BibitemShut {NoStop}%
\end{thebibliography}%

\end{document}